\documentclass[11pt]{ucthesis}

\usepackage{graphicx}
\usepackage{pstricks}
\usepackage{color,dcolumn}
\usepackage{amssymb,amsmath,amsbsy,bbold}
\usepackage{mathrsfs}
\usepackage{axodraw}
\usepackage{latexsym}

\newlength{\captsize}           \let\captsize=\small
\newlength{\captwidth}          
\newlength{\beforetableskip}    
\newcommand{\capt}[1]{\begin{minipage}{\captwidth}
              \let\normalsize=\captsize
              \caption[0]{#1}
              \end{minipage}\\ \vspace{\beforetableskip}}

\def\iso{\mathchoice{\cong}{\cong}{\isoS}{\cong}}
\def\isoS{\vbox{\baselineskip 0pt  \lineskip 0.5pt
    \ialign{$ \mathsurround=0pt  \scriptstyle \hfil ## \hfil $\crcr
        \sim \crcr = \crcr}}}

\newcommand{\mathbold}[1]{\mbox{\boldmath $\bf #1$}}
\newcommand\T{\rule{0pt}{2.6ex}}
\newcommand\Bot{\rule[-1.2ex]{0pt}{0pt}}
\def\ddel{\!\!\mathrel{\raise1.5ex\hbox{$\leftrightarrow$\kern-.85em
\lower1.7ex\hbox{$\partial$}}}}
\def\gev{~{\rm GeV}}

\def\Tr{{\rm Tr}}
\def\abar{{\bar a}}
\def\bbar{{\bar b}}
\def\cbar{{\bar c}}
\def\dbar{{\bar d}}
\def\ebar{{\bar e}}
\def\fbar{{\bar f}}
\def\gbar{{\bar g}}
\def\hb{{\bar h}}
\def\vcc{\left(\vt^* \vo + c.c.\right)}
\def\vo {{\widehat v_1}}
\def\vt {{\widehat v_2}}

\def\lam{\lambda}

\def\wtil{\widetilde}
\def\ben{\begin{enumerate}}
\def\een{\end{enumerate}}
\def\beq{\begin{equation}}
\def\eeq{\end{equation}}
\def\beqa{\begin{eqnarray}}
\def\eeqa{\end{eqnarray}}

\def\ifmath#1{\relax\ifmmode #1\else $#1$\fi}
\def\lsim{\mathrel{\raise.3ex\hbox{$<$\kern-.75em\lower1ex\hbox{$\sim$}}}}
\def\gsim{\mathrel{\raise.3ex\hbox{$>$\kern-.75em\lower1ex\hbox{$\sim$}}}}
\def\sect#1{section~\ref{#1}}
\def\sects#1#2{sections~\ref{#1} and \ref{#2}}

\def\eq#1{eq.~(\ref{#1})}
\def\Ref#1{ref.~\cite{#1}}
\def\Refs#1#2{refs.~\cite{#1} and \cite{#2}}

\def\eqs#1#2{eqs.~(\ref{#1}) and (\ref{#2})}

\def\eqst#1#2{eqs.~(\ref{#1})--(\ref{#2})}
\def\eqthree#1#2#3{eqs.~(\ref{#1}), (\ref{#2}) and (\ref{#3})}
\def\Eq#1{Eq.~(\ref{#1})}
\def\Eqst#1#2{Eqs.~(\ref{#1})--(\ref{#2})}
\def\Eqs#1#2{Eqs.~(\ref{#1}) and (\ref{#2})}
\def\Eqst#1#2{Eqs.~(\ref{#1})--(\ref{#2})}
\def\App#1{Appendix~\ref{#1}}

\def\anti{\overline}

\def\mud{M_U}
\def\mdd{M_D}
\def\vev#1{\langle #1 \rangle}
\def\qlo{Q^0_L}
\def\qlcal{\mathcal{Q}_L}
\def\uro{U^0_R}
\def\dro{D^0_R}

\def\ur{U_R}
\def\dr{D_R}
\def\eiuo{\eta_1^{U,0}}
\def\eiiuo{\eta_2^{U,0}}
\def\eido{\eta_1^{D,0}}
\def\eiido{\eta_2^{D,0}}

\def\eiuoa{\eta_a^{U,0}}
\def\eidoa{\eta_a^{D,0}}

\def\eiua{\eta_a^U}
\def\eida{\eta_a^D}

\def\cbma{c_{\beta-\alpha}}

\def\sbma{s_{\beta-\alpha}}

\def\phm{\phantom{-}}
\def\phaa{\phantom{AA}}

\def\beq{\begin{equation}}
\def\eeq{\end{equation}}
\def\ifmath#1{\relax\ifmmode #1\else $#1$\fi}

\def\call{\mathcal{L}}

\def\sb  {s_{\beta}}
\def\cb  {c_{\beta}}

\def\tanb{\tan\beta}

\def\hl{h^0}
\def\ha{A^0}
\def\hh{H^0}
\def\hpm{{H^\pm}}

\def\go{G^0}

\def\mha{m_{\ha}}
\def\mhl{m_{\hl}}
\def\mhh{m_{\hh}}
\def\mhpm{m_{\hpm}}

\def\ls#1{\ifmath{_{\lower1.5pt\hbox{$\scriptstyle #1$}}}}
\def\lss#1{\ifmath{^{\,\lower2.5pt\hbox{$\scriptstyle #1$}}}}
\def\lsup#1{^{\lower 6pt\hbox{$\scriptstyle#1$}}}
\def\llsup#1{^{\lower 3pt\hbox{$\scriptstyle#1$}}}
\def\lasup#1{^{\lower 2pt\hbox{$\scriptstyle#1$}}}
\def\nicefrac#1#2{\hbox{$\frac{#1}{#2}$}}

\def\half{\ifmath{{\textstyle{\frac{1}{2}}}}}

\def\quarter{\ifmath{{\textstyle{\frac{1}{4}}}}}
\def\eighth{\ifmath{{\textstyle{\frac{1}{8}}}}}

\def\ie{{\it i.e.}}

\def\th{e^{i \theta_{23}}}

\def\thet{e^{i \theta_{23}}}
\def\thetminus{e^{-i \theta_{23}}}
\def\thetdoub{e^{-2 i \theta_{23}}}
\def\thetdoubpl{e^{2 i \theta_{23}}}

\def\Tr{{\rm Tr}}
\def\LL{{\Lambda^2}}
\def\zsix{Z_6 e^{-i \theta_{23}}}
\def\zsixr{\Re(Z_6 e^{-i \theta_{23}})}
\def\zsixi{\Im(Z_6 e^{-i \theta_{23}})}
\def\zfive{Z_5 e^{-2 i \theta_{23}}}

\def\zfii{\Im(Z_5 \,e^{-2i\theta_{23}})}
\def\zseveni{\Im(Z_7 \,e^{-i\theta_{23}})}
\def\zsevenr{\Re(Z_7 \,e^{-i\theta_{23}})}
\def\zfiver{\Re(Z_5 \,e^{-2i\theta_{23}})}
\def\hi{h_i \hspace{2mm} (i=1,2,3)}
\def\gtwo{\frac{g^2}{16\pi^2}}

\def\gc{\frac{g^2}{16\pi^2c_W^2}}
\newcommand{\fourpoint}[4]{\fcolorbox{white}{white}{
  \begin{picture}(190,60) (78,-47)
    \SetWidth{0.5}
    \SetColor{Black}
    \DashCArc(105,-20)(18,117,477){4}
    \Photon(68,-42)(143,-42){4}{6.5}
    \Vertex(106,-38){1.6}
    \Text(69,-35)[lb]{\Black{$#1$}}
    \Text(130,-35)[lb]{\Black{$#2$}}
    \Text(100,0)[lb]{\Black{$#3$}}
 \Text(157,-20)[lb]{\Black{$#4$}}

\end{picture} }
}

\newcommand{\gaugeprop}[5]{\fcolorbox{white}{white}{
 
 \begin{picture}(194,50) (88,-2)
    \SetWidth{0.5}
    \SetColor{Black}
    \Photon(102,9)(138,9){4}{3}
    \DashCArc(123,12)(18,0,-180){4}
    \Photon(80,9)(102,9){4}{2.3}
    \Photon(138,9)(160,9){4}{2.3}

    \Text(80,16)[lb]{\Black{$#1$}}
    \Text(145,16)[lb]{\Black{$#2$}}
    \Text(115,-3)[lb]{\Black{$#3$}}
    \Text(115,30)[lb]{\Black{$#4$}}
 \Text(167,9)[lb]{\Black{$#5$}}

\end{picture} }}
\newcommand{\Gaugeprop}[5]{\fcolorbox{white}{white}{
 
 \begin{picture}(250,40) (86,-2)
    \SetWidth{0.5}
    \SetColor{Black}
    \Photon(102,9)(138,9){4}{3}
    \DashCArc(123,12)(18,0,-180){4}
    \Photon(80,9)(102,9){4}{2.3}
    \Photon(138,9)(160,9){4}{2.3}

    \Text(80,16)[lb]{\Black{$#1$}}
    \Text(148,16)[lb]{\Black{$#2$}}
    \Text(115,-3)[lb]{\Black{$#3$}}
    \Text(115,30)[lb]{\Black{$#4$}}
 \Text(167,9)[lb]{\Black{$#5$}}

\end{picture} }}

\newcommand{\loopgraph}[6]{\fcolorbox{white}{white}{
  \begin{picture}(200,53) (86,-12)
    \SetWidth{0.5}
    \SetColor{Black}
    \DashArrowArc(120,9)(18,0,-180){4}
    \DashArrowArc(120,9)(18,180,-0){4}
    \Photon(80,9)(102,9){-4}{2.3}
    \Photon(138,9)(160,9){4}{2.3}

    \Text(80,16)[lb]{\Black{$#1$}}
    \Text(148,16)[lb]{\Black{$#2$}}
    \Text(115,-6)[lb]{\Black{$#3$}}
    \Text(115,30)[lb]{\Black{$#4$}}
 \Text(170,9)[lb]{\Black{$#5$}}
 \Text(170,-9)[lb]{\Black{$#6$}}
\end{picture} }
}
\newcommand{\Loopgraph}[6]{\fcolorbox{white}{white}{
  \begin{picture}(250,58) (85,-12)
    \SetWidth{0.5}
    \SetColor{Black}
    \DashArrowArc(120,9)(18,0,-180){4}
    \DashArrowArc(120,9)(18,180,-0){4}
    \Photon(80,9)(102,9){-4}{2.3}
    \Photon(138,9)(160,9){4}{2.3}

    \Text(80,16)[lb]{\Black{$#1$}}
    \Text(148,16)[lb]{\Black{$#2$}}
    \Text(115,-6)[lb]{\Black{$#3$}}
    \Text(115,30)[lb]{\Black{$#4$}}
 \Text(170,9)[lb]{\Black{$#5$}}
 \Text(170,-9)[lb]{\Black{$#6$}}
\end{picture} }
}
\newcommand{\feynrule}[4]{\fcolorbox{white}{white}{
  \begin{picture}(180,48) (48,-42)

    \SetWidth{0.5}
    \SetColor{Black}
    \Photon(45,-20)(70,-20){3.5}{3}
    \DashArrowLine(90,-5)(70,-20){4}
    \DashArrowLine(70,-20)(90,-35){4}
  \Text(50,-12)[lb]{\Black{$#1^\mu$}}
 \Text(92,-7)[lb]{\Black{$#2$}}
 \Text(92,-42)[lb]{\Black{$#3$}}

 \Text(160,-25)[b]{$= #4 (p_{2} + p_{1})^\mu $}
  \end{picture}}
}
\newcommand{\feynruleVV}[4]{\fcolorbox{white}{white}{
  \begin{picture}(170,50) (31,-43)
    \SetWidth{0.5}
    \SetColor{Black}
    \Photon(40,-3)(70,-20){3}{3.5}
    \Photon(40,-37)(70,-20){-3}{3.5}
    \DashLine(100,-20)(70,-20){4}
  \Text(21,-5)[lb]{\Black{$#1^\mu$}}
  \Text(21,-45)[lb]{\Black{$#1^\mu$}}
 \Text(90,-18)[lb]{\Black{$#3$}}

 \Text(150,-25)[b]{$= #4 \hspace{1mm}g^{\mu\nu} $}
  \end{picture}}
}
\newcommand{\feynruletwo}[5]{\fcolorbox{white}{white}{
  \begin{picture}(185,48) (48,-42)
    \SetWidth{0.5}
    \SetColor{Black}
    \Photon(45,-20)(70,-20){4}{2.5}
    \DashArrowLine(90,-5)(70,-20){4}
    \DashArrowLine(70,-20)(90,-35){4}
  \Text(50,-12)[lb]{\Black{$#1^\mu$}}
 \Text(92,-7)[lb]{\Black{$#2$}}
 \Text(92,-42)[lb]{\Black{$#3$}}
 \Text(173,-20)[b]{$= #4(p_{2} + p_{1})^\mu $}
 \Text(160,-35)[b]{$= #5(p_{2} + p_{1})^\mu $}
  \end{picture}}
}
\newcommand{\feynrulefour}[5]{\fcolorbox{white}{white}{
  \begin{picture}(170,58) (42,-47)
    \SetWidth{0.5}
    \SetColor{Black}
    \Photon(40,-3)(70,-20){3}{3.5}
    \Photon(40,-37)(70,-20){-3}{3.5}
    \DashArrowLine(90,-3)(70,-20){4}
    \DashArrowLine(70,-20)(90,-37){4}
  \Text(35,0)[lb]{\Black{$#1^\mu$}}
  \Text(35,-50)[lb]{\Black{$#1^\mu$}}
 \Text(92,0)[lb]{\Black{$#2$}}
 \Text(92,-48)[lb]{\Black{$#3$}}

 \Text(160,-20)[b]{$= #4 g^{\mu\nu} $}
 \Text(140,-39)[b]{$#5$}
  \end{picture}}
}
\renewcommand{\Re}{{\rm Re}}
\renewcommand{\Im}{{\rm Im}}

\def\slashi#1{\setbox0=\hbox{$#1$}#1\hskip-\wd0\hbox to\wd0{\hss\sl/\/\hss}}
\begin{document}

\title{Phenomenology of the Basis-Independent CP-Violating Two-Higgs Doublet Model}
\author{Deva A. O'Neil}
\degreeyear{2009}
\degreemonth{June}
\degree{DOCTOR OF PHILOSOPHY}
\chair{Professor Howard Haber}
\committeememberone{Professor Michael Dine}
\committeemembertwo{Professor Jason Nielsen}
\numberofmembers{3} 
\deanlineone{Dean Lisa Sloan}
\deanlinetwo{Vice Provost and Dean of Graduate Studies}
\deanlinethree{}
\field{Physics}
\campus{Santa Cruz}

\begin{frontmatter}

\maketitle
\copyrightpage

\tableofcontents
\listoffigures
\listoftables

\begin{abstract}
The Two-Higgs Doublet Model (2HDM) is a model of low-energy particle interactions that is identical to the Standard Model except for the addition of an extra Higgs doublet. This extended Higgs sector would appear in experiments as the presence of multiple Higgs particles, both neutral and charged.  The neutral states may either be eigenstates of CP (in the CP-conserving 2HDM), or be mixtures of CP eigenstates (in the CP-violating 2HDM).  In order to understand how to measure the couplings of these new particles, this document presents the theory of the CP-violating 2HDM in a basis-independent formalism and explicitly identifies the physical parameters of the model, including a discussion of $\tan\beta$-like parameters.  The CP-conserving limit, decoupling limit, and the custodial limit of the model are presented.  

In addition, phenomenological constraints from the oblique parameters ($S$, $T$, and $U$) are discussed. A survey of the parameter space of this model shows that the 2HDM is consistent with a large range of possible values for $T$.  Our results also suggest that the 2HDM favors a slightly positive value of $S$ and a value of $U$ within $.02$ of zero, which is consistent with present data within the statistical error.  In a scenario in which the heaviest scalar particle is the charged Higgs boson, we find that the measured value of $T$ puts an upper limit on the mass difference between the charged Higgs boson and the heaviest neutral Higgs boson.
\end{abstract}

\begin{acknowledgements}
The text of this dissertation includes a reprint of the following previously published material:  

Howard E. Haber, Deva O'Neil.\emph{Basis-Independent Methods for the Two-Higgs-Doublet Model II: The Significance of $\tan\beta$.} Phys.Rev.D74:015018,2006. [hep-ph/0602242]
The co-author listed in this publication directed and supervised the research which forms the basis for the dissertation.

This effort could not have been completed without the strong support and guidance from the members of my committee, Howard Haber, Michael Dine, and Jason Nielsen.  In particular, I would especially like to thank Dr. Haber for serving as my advisor and for the time and effort he spent guiding this research, and Dr. Dine for his advice and mentoring.  In completing this dissertation, I have also benefited from conversations with Dr. Nielsen and Dr. John Mason.  I am deeply grateful to have had the privilege of working with these individuals, some of the foremost leaders in this field. 

\end{acknowledgements}

\end{frontmatter}

\chapter{Introduction} The nature of electroweak symmetry breaking is one of the most important remaining puzzles in particle physics today.  The Standard Model of particle physics contains a mechanism for electroweak symmetry breaking (EWSB), but experimental confirmation for it has not yet materialized.  Many other models of EWSB have also been proposed.  Since the Large Hadron Collider, which will start running in fall 2009, is optimized to make discoveries at the electroweak scale, refining experimental predictions and constraints for these models is a pressing research goal.
The purpose of this document is to explore electroweak symmetry breaking through the Two-Higgs Doublet Model (2HDM), an extension of the Standard Model.  The 2HDM presents interesting theoretical possibilities, such as CP-violation in the scalar sector of the theory.  It also presents challenges, since its phenomena are constrained by electroweak precision data. In this document a basis-independent version of the 2HDM will be presented in both CP-conserving and CP-violating scenarios.  In addition to the formalism of the model, identification of the observable parameters of the 2HDM will be emphasized.  New insights into custodial symmetry in the context of this model will also be discussed. Finally, the phenomenology of the CP-violating 2HDM will be explored, making use of the ``oblique'' parameters S, T, and U; and the well-known parameter $\tan\beta$. 
 
\section{Electroweak Symmetry Breaking in the Standard Model}
The Standard Model of particle physics is constructed by applying gauge symmetries to the interactions of fundamental particles. At high energies (above the electroweak scale) the gauge group is SU(3)$\times$SU(2)$\times $U(1). Here we will focus on the electroweak sector of the theory [SU(2)$\times $U(1)], introduced by Glashow, Weinberg, and Salam \cite{glashow,weinberg}.  A gauge symmetry alone requires a massless vector field corresponding to each generator of the symmetry group.  However, spontaneous breaking of the gauge symmetry produces masses for the vector fields.  The breaking of SU(2$)\times $U(1) to $\rm{U(1)}_{EM}$ leads to one gauge boson remaining massless (which we identify as the photon) and three gauge bosons acquiring mass, which reproduces the pattern observed in nature in the electroweak interactions.

This spontaneous symmetry breaking occurs when a scalar field acquires a non-zero vacuum expectation value (the ``Higgs Mechanism"). Once one postulates the existence of such a scalar field (the Higgs field), it may then be employed to give masses to fermions. The details of this mechanism are discussed below.

\subsection{The Higgs Mechanism}
Using the notation of \cite{peskinSchroeder}, we will represent this hypothetical scalar field as $\phi$, and take its U(1) charge to be $+\half$. It also has an SU(2) spinor structure. Then under the SU(2)$\times$U(1) gauge group, the field transforms as \beq\phi \rightarrow e^{i\alpha^a\tau^a}e^{i\beta/2}\phi,\eeq where $\tau^a$ are the generator matrices for SU(2), ie $\tau^a = \sigma^a/2$.  By convention, the vev of this field is taken to have the form \beq<\phi> = \frac{1}{\sqrt{2}}
\displaystyle \binom{0}{v}.\eeq
The kinetic term of the scalar Lagrangian is written \beq \mathcal{L}_{KE} = |D_\mu \phi |^2 \label{kin},\eeq where $D_\mu$ is the covariant derivative $\delta_\mu - ig A^a_\mu\tau^a-i\half g' B_\mu$.  The mass terms for the gauge bosons appear when the vacuum expectation value of \eq{kin} is taken:
\beqa \mathcal{L}_{KE} & = \half\Bigl( 0 \,\,\, v\Bigr)\left(gA_\mu^a \tau^a +\half g' B_\mu\rm{I}\right)\left(gA^{\mu b} \tau^b +\half g' B^\mu\rm{I}\right)\displaystyle \binom{0}{v}\nonumber\\
& = \half v^2\Bigl( 0 \,\,\, 1\Bigr)\left(\frac{g}{2}A_\mu^a \sigma^a +\frac{g'}{2} B_\mu \rm{I}\right)\left(\frac{g}{2}A^{\mu b} \sigma^b +\frac{g'}{2} B^\mu \rm{I}\right)\displaystyle \binom{0}{1}\nonumber\\
 & =  \frac{v^2}{8}\left[g^2(A_\mu^1)^2 + g^2 (A_\mu^2)^2 +\left(g' B_\mu - g A_\mu^3\right)^2\right].\label{masst}\eeqa
The first two terms in \eq{masst} gives the mass of the charged field $W_\mu^\pm = \frac{1}{\sqrt{2}}(A_\mu^1 \mp i A_\mu^2)$, and the second gives the mass of the neutral field $Z_\mu = \frac{1}{\sqrt{g^2 + g'^2}}(gA_\mu^3 - g' B_\mu)$.  The remaining orthogonal field is $A_\mu = \frac{1}{\sqrt{g^2 + g'^2}}(gA_\mu^3 + g' B_\mu)$. 
Thus, one can rewrite \eq{masst} in terms of observable fields as follows:
\beq \mathcal{L}_{KE} = \frac{v^2}{4}g^2 W_\mu^+ W^{\mu -} +\frac{v^2}{8}( g^2 + g'^2)Z_\mu Z^{\mu }. \eeq
Reading off the masses for the gauge bosons yields 
\beqa m_W &=& g\frac{v}{2}, ~ ~ m_Z = \sqrt{g^2 + g'^2}\frac{v}{2}, \nonumber\\
m_A &=& 0.  \eeqa
The massless field $A_\mu$ we identify as the photon.

\subsection{The Scalar Lagrangian \label{scsec}}
In order for the Higgs field to have a non-zero vev, we take its Lagrangian to have the form
\beq \mathcal{L} = |D_\mu \phi |^2 +\mu^2 \phi^\dagger \phi - \lambda( \phi^\dagger \phi )^2, \label{phipote}\eeq so that the potential energy has a minimum at \beq \label{hmas} v = \sqrt{\mu^2 / \lambda}.\eeq   It is conventional to expand $\phi$ around its vev:
\beq\phi = \frac{1}{\sqrt{2}}\displaystyle \binom{\sqrt{2}~G^+}{v +h(x)+ i G^0},\,\,\,\,\,\,\,\,\,\,\,\,
 	\phi^*= \frac{1}{\sqrt{2}}\binom{\sqrt{2}~G^-}{\\ v +h(x)- i G^0}.\eeq
The $G^0$, $G^\pm$ are Goldstone bosons; The field h(x) is a neutral scalar field of zero vev, whose excitations give rise to a scalar particle, the Higgs boson.  We will work in unitary gauge, absorbing $G^0$ and $G^\pm$ into the gauge potential terms, so that $h(x)$ is a real-valued field.  In this gauge, 
\beq  \phi = \frac{1}{\sqrt{2}}\displaystyle \binom{0}{v +h(x)}.\eeq
The Higgs boson's potential energy density can be found from the latter terms in \eq{phipote}:
\beq V = \mu^2 h^2 +\lambda v h^3 + \frac{1}{4}\lambda h^4. \eeq
One can then read off the mass of the Higgs boson, $m_h = \sqrt{2}\mu.$  Comparing to \eq{hmas}, one obtains 
\beq m_h =  \sqrt{2 \lambda}v .\label{higgsm}\eeq
Thus, without knowing the value of the Higgs self-coupling $\lambda$, one cannot predict the value of $m_h, $ even though $v$ is known from measurements of the W mass to be $246$ GeV.  

Although the Goldstone fields do not explicitly appear in the Lagrangian, and thus do not correspond to observable particles, the degrees of freedom that they represent are still present in the theory after electroweak symmetry breaking.  Three of the original four massless vector fields have acquired mass, which adds three degrees of freedom to the gauge boson sector. More precisely, the Goldstone bosons $G^0, G^\pm$ become the longitudinally polarized states of $Z^0$ and $ W^\pm,$ respectively, which is reflected in the equivalence theorem \cite{Bagger:1989fc},\cite{Chanowitz:1985hj},\cite{Cornwall:1974km},\cite{Lee:1977eg}:  A high energy process $(q \gg m_W)$ involving longitudinal gauge bosons has the same amplitude as one in which they are replaced by the corresponding Goldstone bosons [$Z^0\rightarrow G^0$ and $ W^\pm \rightarrow G^\pm$], up to $\mathcal{O}(m_W/q)$.  

The mixing between the $A_\mu^3$ and $B_\mu$ that produces the physical Z boson and photon fields may be represented in terms of the ``weak mixing angle" $\theta_W$, as follows:
\beq \left( \begin{array}{c}Z^0 \\ A \end{array}\right) = \left( \begin{array}{cc}\cos\theta_W & - \sin \theta_W \\\sin \theta_W & \cos\theta_W\end{array}\right)\left( \begin{array}{c}A^3 \\ B\end{array}\right),\eeq
where
$\cos\theta_W = \frac{g}{\sqrt{g^2 + g'^2}}$ and $\sin\theta_W = \frac{g'}{\sqrt{g^2 + g'^2}}.$  (Note that there is a simple relation between the masses of the Z and W, $m_W = m_Z \cos\theta_W$.)  Then one can rewrite the covariant derivative as 
\beq D_\mu = \delta_\mu - i \frac{g}{\sqrt{2}}(W_\mu^+ T^+ +W_\mu^- T^-) - i \frac{g}{\cos\theta_W}Z_\mu (T^3-\sin^2\theta_W Q) - ie A_\mu  Q, \eeq where the generators in this basis are $T^\pm = \half(\sigma^1 \pm i \sigma^2)$ and $Q = T^3 +Y$, and $e$ is the electric charge, related to $g$ via $g = \frac{e}{\sin\theta_W}$.  

We have thus shown how our original gauge symmetry $SU(2)\times U(1)$ with generators $T^a$ and $Y$ has been broken, leaving an unbroken symmetry $U(1)_{EM}$ whose generator is $Q$ and gauge boson is the photon.  The other 3 gauge bosons have gained mass through the Higgs mechanism.  In the Standard Model, this mechanism also implies the existence of a fundamental massive scalar, the Higgs boson.  

\subsection{Generating Fermion Masses \label{smyuk}} 
To explore the effect of electroweak symmetry breaking on the fermion sector, let us first concentrate on leptons.  We will later generalize this discussion to quarks.  In the Lagrangian, left-handed leptons appear in SU(2) doublets $E_L =\displaystyle \binom{\nu_e}{e^-}_L$.  For the first generation,
\beq \mathscr{L_f} = \bar{E}_L(i\slashi{D})E_L + \bar{e}_R(i\slashi{D})e_R + \mbox{quark terms}.\eeq
The right- and left-handed fermion fields are defined the usual way; $\psi_{R,L}\equiv P_{R,L}\psi$, where $P_{R,L}\equiv \half(1\pm\gamma_5)$.
Explicit lepton mass terms cannot be added to this model, since left-handed fermions appear as SU(2) doublets and the right-handed fermions are singlets.  Hence, terms such as $m_e \bar{e}_R e_L+m_e \bar{e}_L e_R$ are forbidden.  Thus, to generate mass terms, one requires a scalar SU(2) doublet $\phi$ that interacts with right- and left-handed leptons, which one can parametrize in terms of a dimensional coupling constant $\eta^E$ as 
\beq -\mathscr{L}_{Y} =  \eta^E \bar{E}_L\cdot \phi e_R + h.c.\label{yukelec} \eeq
After $\phi$ acquires a vev, a mass term is generated:
\beq -\mathscr{L}_{Y} =  \frac{v}{\sqrt{2}}\eta^E \bar{e}_L e_R + h.c.,\eeq
from which one can identify $m_e = \frac{1}{\sqrt{2}}\eta^E v$.

An analogous calculation can be made for quarks; in a one-generation model, masses would be generated in the form $m_d = \frac{v}{\sqrt{2}}\eta^D$ and  $m_u = \frac{v}{\sqrt{2}}\eta^U$ from Yukawa interactions similar to \eq{yukelec}.  However, when the model is expanded to 3 generations, additional complications arise from the mixing of quark generations.  Writing the SU(2) doublets as $Q_L^i =\displaystyle \binom{U^i}{D^i}_L$, one has

\beq \label{y0}
-\mathscr{L}_{\rm Y}
=\anti \qlo \cdot\wtil\phi\,\eta^U\,  \uro +\anti Q_L^0\cdot\phi\,(\eta^D)^\dagger
\,\dro  +{\rm h.c.}\,,\eeq
where $\wtil\phi\equiv i\sigma_2 \phi^*$, and the flavor indices have been suppressed.
Here, $\qlo $, $\uro $, $\dro $ denote the interaction basis quark fields, which
are vectors in the quark flavor space.  In this basis, the couplings  $\eta^{Q,0}$ ($Q=U\,,\,D$)
are two non-diagonal $3\times 3$ matrices.  To identify the quark mass-eigenstates, one applies unitary transformations of the left- and right-handed $U^0$ and $D^0$ fields to diagonalize the matrices $\eta^Q$, i.e.:
\beqa \label{biunitary}
&& P_L U=V_L^U P_L U^0\,,\qquad P_R U=V_R^U P_R U^0\,,\nonumber \\
&& P_L D=V_L^D P_L D^0\,,\qquad P_R D=V_R^D P_R D^0\,,
\eeqa
and the Cabibbo-Kobayashi-Maskawa (CKM) matrix is defined by $K\equiv V_L^U
V_L^{D\,\dagger}$.  The  Yukawa coupling matrices in this basis are as follows:
\beq\eta^U \equiv V_L^U \,\eta^{U,0}\, V_R^{U\,\dagger}\,,\qquad\qquad
    \eta^D\equiv V_R^D\, \eta^{D,0}\, V_L^{D\,\dagger}\,.\eeq
[Note the different ordering of $V_L^Q$ and $V_R^Q$ 
in the definitions of $\eta^Q$ for $Q=U$, $D$.] In terms of the mass-eigenstates, \eq{y0} becomes
\beq \label{ymod}
-\mathscr{L}_{\rm Y}=\anti U_L\cdot \wtil\phi^0\,\eta^U \ur
+\anti D_L K^\dagger\cdot\phi^-\, \eta^{U\,\dagger}\ur +\anti U_L K\cdot \wtil\phi^+\,\eta^{D,\dagger}\dr
+\anti D_L\cdot\phi^0\, \eta^{D\,\dagger}\dr+{\rm h.c.}\,\eeq
One could write \eq{ymod} more compactly by defining $\mathcal{U}\equiv K^\dagger U,$ and $\mathcal{Q}_L =\binom{\mathcal{U}}{D}_L$.  Then the Yukawa Lagrangian in the mass-eigenstate basis can be written
\beq \label{ycute}
-\mathscr{L}_{\rm Y}=\bar{\mathcal{Q}}_L\cdot \wtil\phi\,\eta^U \mathcal{U}_R
+\bar{\mathcal{Q}}_L\cdot\phi\, \eta^{D\,\dagger}\dr +{\rm h.c.}\,\eeq
One can now obtain the mass matrices $M_U$ and $M_D$ by taking the vev of $\phi$ in \eq{ymod}, which yields
\beqa
M_U=\frac{v}{\sqrt{2}}\eta^U&=&{\rm diag}(m_u\,,\,m_c\,,\,m_t)
= V_L^U M_U^{0} V_R^{U\,\dagger}\,, \label{diagumass}\\[6pt] 
M_D=\frac{v}{\sqrt{2}}\eta^{D\,\dagger}&=&{\rm diag}(m_d\,,\,m_s\,,\,m_b)
= V_L^D M_D^{0} V_R^{D\,\dagger}\,. \label{diagdmass}
\eeqa
By generating masses for the fermions and gauge bosons, the Higgs mechanism allows the Standard Model to reproduce the phenomena observed in nature.  In order to confirm that this is how electroweak symmetry breaking is implemented in real life, we would have to observe the production of the Higgs boson in particle accelerators.  Certain theoretical constraints may be used to indicate the likely range of the Higgs mass, as discussed in the following section.

\subsection{Constraints on the Standard Model Higgs Mass \label{constraints}}

\subsubsection{Bounds on the Higgs Mass from Finiteness and Vacuum Stability}
The Standard Model cannot be a valid description of nature at all energy scales.  It must be superseded by a theory that incorporates gravitational interactions near the Planck scale ($10^{19}$ GeV).  It may be that the scale at which new physics beyond the Standard Model emerges, $\Lambda$, is high (far above the TeV scale), in which case the mass of the SM Higgs must be fairly light, lest the Higgs self-coupling become divergent at a scale below $\Lambda$.  For $\Lambda \sim M_{Pl}$, the resulting upper bound from the two-loop renormalization group equation (RGE) is $m_h < 180$ GeV \cite{Hambye:1996wb}, in rough agreement with more recent (two-loop) calculations, which have placed the upper bound at $174$ GeV \cite{Hung:1996gj} and $161.3\pm20.6$ GeV \cite{Pirogov:1998tj}. Since the only case in which $\lambda$ would remain finite at all energy scales would be in the non-interacting  (or ``trivial") theory, in which $\lambda = 0$, this is sometimes called a ``triviality" argument.

On the other hand, if new physics enters at a low scale (on the order of 1 TeV), the Higgs boson can be heavier.   In a pure scalar ($\phi^4$) theory, the Higgs mass can be as much as 1 TeV before $\lambda$ is driven to infinity below the cut-off scale \cite{Lindner:1985uk}.  Cabbibo et al. \cite{Cabibbo:1979ay} derive a stringent upper bound on $m_h$ by extending this analysis to include Yukawa interactions with the top quark.  For the RGE of $\lambda$ they exhibit 
\beq  16 \pi^2 \frac{d\lambda}{dt} = 12 \lambda^2+ 6 \lambda y_t^2 -3 y_t^4 +\mathcal{O}(\alpha),\label{reg}\eeq
where $t = \ln(q^2/v^2)$ and $y_t$ is the Yukawa coupling for the quark, $y_t =\sqrt{2} m_t/v$. Requiring that the Higgs coupling $\lambda(q)$ be finite (up to some high energy scale where new physics sets in, such as $m_{\rm GUT}$), they numerically calculate the maximum value for $\lambda(v)$. (This upper bound for $\lambda(v)$ is dependent on the mass of the top quark, which was not then known.) The coupling $\lambda(v) $ can then be related to the Higgs mass via  \eq{higgsm}.  For $m_t = 175\, \rm{GeV}$, this corresponded roughly to \beq m_h \lsim 200 \,\rm{GeV} \label{cabbibobound}.\eeq  This calculation can be repeated for two loops, but theoretical uncertainties are significant \cite{Hambye:1996wb}.    

To bound the Higgs boson mass from below, one considers vacuum stability.  This condition specifies that $V(v)\leq V(\phi)$ for all $|\phi|< \Lambda$, where $V(v)$ is the value of the scalar potential at the electroweak minimum.  For small values of the running coupling $\lambda(v)$, the top quark contribution to the RGE can produce a negative value of $\lambda(q)$, so that the radiatively-corrected effective scalar potential would be unbounded from below \cite{Lindner:1988ww,Sher:1993mf,Altarelli:1994rb,Casas:1994qy}\footnote{This argument is controversial.  Branchina et al. \cite{branch1,branch2} argue that the region of instability in the scalar potential lies beyond the range of validity for the perturbative RGE.  This absence of vacuum instability is confirmed by lattice results \cite{Fodor:2007fn}.  The results of \Refs{branch1}{branch2} are disputed by \cite{einhornjones}.}. This occurs because while $V_{tree} \sim \lambda \phi^4$, for large $\phi$, $V_{eff} \sim \lambda(t) \phi^4,$ where $t = \ln(\phi^2/M^2)$ and $M$ is the renormalization scale \cite{Sher:1988mj}. Thus at some scale $t$, $\lambda(t)$ becomes negative.

\begin{figure}[!ht]
  \begin{center}
    \includegraphics[scale=.5]{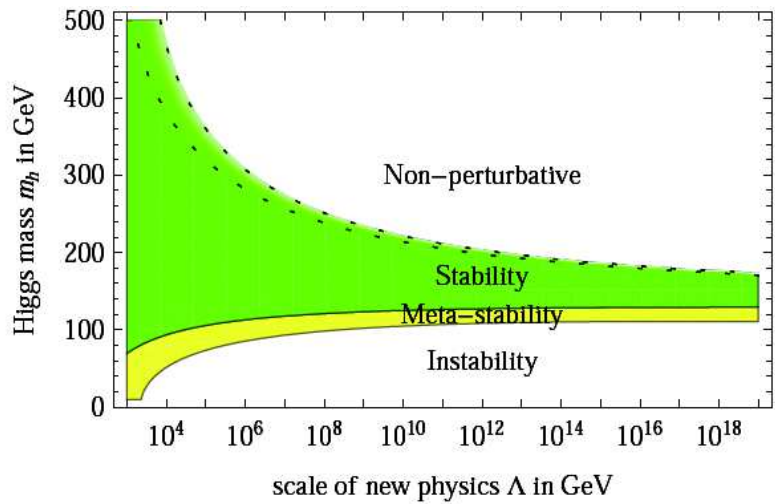}
  \end{center}
\caption{Bounds on the Standard Model Higgs mass from \Ref{isidori}, using $m_t = 173$ GeV and $\alpha_3(M_Z) = .118$. The lowest bound is from vacuum metastability, the middle bound from vacuum stability, and the upper (dotted) bounds are from perturbativity requiring $\lambda(\Lambda) < 3,6$ (see \cite{Hambye:1996wb}).}\label{isifig}
\end{figure}

One subtlety that appears in the literature on vacuum stability is that the scalar potential at large $|\phi|$ can have a region deeper than the electroweak vacuum provided that the decay of the ``false" electroweak vacuum is suppressed \cite{Anderson:1990aa,Isidori:2001bm,Arnold:1991cv,Espinosa:1995se,Sher:1988mj,Sher:1988db}.  By requiring that the lifetime of the metastable electroweak vacuum is less than the age of the universe, one derives a lower bound on $m_h$ that is not as strict as the one from stability. The most recent calculation of this metastable region is given in \Ref{isidori} and shown in Fig. \ref{isifig}.  

Because these RGEs arise from perturbative calculations, one might worry that they would not be valid for large $\lambda$.  Lattice calculations have been used in order to produce a non-perturbative analysis, which would be valid even for high values of the coupling.  The non-perturbative limit on the Higgs mass is found to be $m_h < 9\,m_W \approx 700$ GeV \cite{Langguth}.  This result is similar to that derived by Lunscher and Weisz \cite{lusch}, $m_h < 9.6 \,m_W$.  This consistency suggests that the perturbative calculations may be roughly accurate.

\subsubsection{Higgs Mass Bounds from Unitarity \label{unit}}
In a theory of the electroweak interactions without a Higgs boson, tree-level perturbative calculations of scattering amplitudes (such as $W^+ W^- \rightarrow W^+ W^-$) have terms proportional to $\frac{s}{m_W^2}$, which lead to amplitudes greater than unity at high values of the CM energy ($s \gg m_W^2$).  This violation of unitarity is fixed by the presence of the Higgs boson; once diagrams involving the Higgs boson are included, the terms of order $\frac{s}{m_W^2}$ in the tree-level scattering amplitudes cancel, leaving only $\mathcal{O}(1)$ terms.  Thus, the Higgs boson has the effect of ``unitarizing'' gauge boson scattering.  In principle, different mechanisms could unitarize $\bar{f}f\rightarrow V_L V_L$ and $V_L V_L\rightarrow V_L V_L$, respectively; in the Standard Model the Higgs boson unitarizes both \cite{Jager:1998va}.  

Lee et al. \cite{Lee:1977eg}\cite{Lee:1977yc} derived a critical value of the Higgs mass from tree-level unitarity of processes involving longitudinally polarized gauge bosons, with the condition $|a_0|\leq 1$, where $a_0$ is the amplitude of the zeroeth partial wave for $W^+ W^- \rightarrow W^+ W^-$. A more restrictive version of this condition, $|\Re~a_0| \leq \half$ ~\cite{akeroyd,Ginzburg:2003fe,luscher}, produces a stricter bound of \beq m_h^2 \leq \frac{4 \pi \sqrt{2}}{3 G_F} \approx (700 \,\rm{GeV})^2,\label{marcianobound}\eeq as derived in \Ref{marciano}.

This is a tree-level result, which represents the maximum value of $m_h$ for which a perturbative analysis of the scattering amplitude is reliable at all energy scales.  If the Higgs mass is above this value, it would indicate that the weak interactions become strongly coupled at high energies, or that additional scalar particles not described by the SM are present to stabilize the amplitudes.  

This analysis has been extended to two-loops by Durand et al. \cite{durand}.  Refinements of their calculation in \cite{rw} give results that are very similar to that of \eq{marcianobound}. 

Attempts have been made to analyze unitarity using non-perturbative approaches, which would be valid for large $\lambda$.  In particular, one can analyze SU(N)$\times$U(1) theories in the large N limit \cite{einhorn}. Although the numerical results would not necessarily be valid for $N = 2$, this approach may yield a conceptual understanding of the strong coupling regime.  Furthermore, it can be used to validate the perturbative approach in the small $\lambda$ limit.  To next-to-leading order in $1/N$, the non-perturbative analysis of $f\bar{f}\rightarrow h \rightarrow VV$ and $f\bar{f}\rightarrow h \rightarrow f\bar{f}$ scattering in the large N limit matches the NNLO perturbative results for $m_h < 800\,-\,900$ GeV \cite{ghin}.  It is also found that above $m_h = 1$ TeV, the Higgs mass no longer increases as the coupling increases, which violates \eq{higgsm}.  This analysis suggests that if the Standard Model is correct, the Higgs particle will be found below about 1 TeV even if the weak interaction becomes strongly coupled. 

\subsection{Results of Higgs Searches}  
The experimental lower bound on the Higgs mass is currently determined by data from LEP-II, the electron-positron collider at CERN which reached maximum CM energies of 209 GeV.  Higgs searches focused on the ``Higgsstrahlung" channel, $e^+ e^- \rightarrow ZH.$  No definite discovery of the Higgs was made by time LEP was decommissioned, which put a limit on the SM Higgs mass of $m_h > 114.4 $ GeV \cite{Barate:2003sz}.  

At the time of this writing, Higgs searches are proceeding at the upgraded Tevatron, a $p\bar{p}$ collider running at Fermilab with energy 1.96 TeV.  Based on non-observation of the Higgs boson by winter 2009, the combined results from both detectors (CDF and D0) were sufficient to exclude a SM Higgs in the mass range $160 \,\rm GeV < m_h < 170\, \rm GeV$ \cite{tevatron}.   The dominant channel for Higgs production at this energy scale is gluon fusion ($gg\rightarrow h \rightarrow W^+W^-,$ with final state $l \nu l \nu$, $l = e^-, \mu^-$), although $Wh\rightarrow WWW$ (Higgsstrahlung) is also being searched.  The most likely channels to produce a lower energy Higgs ($m_h \leq 140\, \rm GeV$) at the Tevatron are $Wh \rightarrow l\nu b \bar{b}$, $Zh \rightarrow llb \bar{b}$, and $Zh \rightarrow \bar{\nu}\nu b \bar{b}$.  It is expected that by the end of 2010, the Tevatron will either show evidence of Higgs producation (up to $3 \sigma$ level), or be able to exclude the existence of the Higgs over a large energy range ($145 \,\rm GeV < m_h < 185\, \rm GeV$) \cite{qian}.

\subsection{Future of Higgs Searches} 
With an energy of 14 TeV, the Large Hadron Collider at CERN, a proton-proton collider, will be powerful enough to produce hundreds of thousands of Higgs particles (if they are light) or tens of thousands of Higgs particles (if they are heavy) \cite{rainwater}. Most of these will be the result of gluon fusion, $gg \rightarrow h$, via a top quark loop \cite{Assamagan:2004mu}.  Higgs particles are also likely to be produced through $qq\rightarrow qqh$ processes (called weak boson fusion or vector boson fusion).  The Higgs would then most likely decay to $\bar{b}b$\footnote{Although decay rates to $\bar{b}b$ would be high, this channel is not useful for Higgs discovery due to high background rates.}, $\tau \tau$ and/or $W^+ W^-$, depending on its mass \cite{Aglietti:2006ne}. Below $\sim$ 150 GeV, the branching ratio for $h \rightarrow \gamma\gamma$ is also high enough to be observable. Despite having a smaller branching ratio than $\tau \tau$ or $W^+ W^-$, this channel is easier to distinguish from background ($q\bar{q},gg \rightarrow \gamma \gamma$) \cite{higgshunt}. 

Should an actual scalar particle be discovered in these searches, the question will be asked, ``Is this particle the Standard Model Higgs Boson?"  Although observation of the particle will yield measurements of its mass, electric charge, and color charge, to distinguish between different theoretical models, it is important to extract information about gauge and Yukawa couplings.  What might be discovered that is not predicted by the Standard Model is the subject of the next section. 

\section{Beyond the Standard Model}
Since the Higgs boson has not been discovered in experiments, it is not known whether the Higgs mechanism works as predicted by the Standard Model. It may be, for example, that breaking electroweak symmetry through one (or more) scalar fields is not what happens in nature.  However, precision tests of the SM suggest that the GWS description of gauge symmetry in the electroweak sector is the correct one at low energies, requiring some mechanism of electroweak symmetry breaking to generate masses for vector and fermion particles.  An example of the experimental evidence for postulating that the particles observed so far are part of a spontaneously broken gauge theory is the universality of coupling constants in gauge boson interactions.  In particular, the GWS theory predicts that the coupling constant $g$ is the same for cubic and quartic interactions of charged gauge bosons ($W^\pm$).  Description of the neutral gauge boson interactions requires only one additional parameter ($g'$ or $\sin\theta_W$). 

Another observable consequence of the GWS gauge symmetry is in the different behavior of left- and right-handed fields.  The SU(2) gauge group applies charges $T^3 = \pm \half$ to left-handed leptons (and their corresponding right-handed antiparticles) and $T^3 = 0$ to right-handed leptons (and left-handed antileptons). These charges appear in the $Z l^+ l^-$ couplings, which are proportional to $T^3 - \sin^2 \theta_W Q$, and thus can be measured in $Z \rightarrow  l^+ l^-$ decays. For example, the branching ratio of  $Z \rightarrow q\bar{q}$ ($q = u,d,c,s,b$) to $Z \rightarrow  l^+ l^-$  was calculated from Standard Model fits to be $R_\ell \equiv \Gamma_{had}/\Gamma_{\ell \ell} =  20.744$, which agrees with the value of $20.767 \pm 0.025$ measured at LEP \cite{Z-Pole}.  This asymmetry between left- and right-handed fields also results in a net polarization of the decay products in $Z \rightarrow f \bar{f}$ (ie, an excess of $f_L\bar{f}_R$ over $f_R \bar{f}_L$).  This ``polarization asymmetry" is parametrized in \beq \mathcal{A}_{LR} = \frac{\sigma_L - \sigma_R}{\sigma_L + \sigma_R}\,,\eeq
where $\sigma_L$ ($\sigma_R$) is the $e^+ e^-$ cross-section for Z production from left-handed (right-handed) electrons. 
The measured value of this asymmetry has been found to be $\mathcal{A}_{LR} = .15138 \pm .00216$  \cite{Abe:2000dq}, which is within 2 $\sigma$ of the Standard Model fit $\mathcal{A}_{LR} = .1473 \pm .0011$ \cite{Amsler:2008zzb}.

Although these (and other) precision tests of the electroweak sector confirm the GWS model of gauge interactions, and in particular the different SU(2) charge assignments for right- and left-handed fermion fields,  the electroweak symmetry is not necessarily broken through a single scalar doublet, as in the SM.  In constructing other theories, it is common to build on the Standard Model's particle content, so as to preserve the success of the SM fits to precision electroweak data.\footnote{There are also ``higgless'' theories such as models with extra dimensions, which provide a Goldstone boson through mechanisms other than interacting scalar fields.} One can extend the Standard Model's scalar sector by adding additional singlets, doublets, and/or higher multiplets, for example.  Although the exact content of the scalar sector is unknown, the relation $m_w = m_z\cos\theta_W$ must be preserved (up to loop corrections), which arises naturally from postulating an unbroken global SU(2) symmetry.  This ``custodial" symmetry will be discussed in detail in Chapter \ref{custchapter}. 
\chapter{The CP-Violating Two-Higgs Doublet Model}
\section{Motivation}
	The Standard Model by itself is not expected to be a complete description of nature.  As discussed in section \ref{constraints}, the SM is considered to be a low-energy effective theory, which ceases to be valid above some energy scale $\Lambda$.  Even at low energies, the SM with its single Higgs doublet may not be the correct theory.  Extensions of the SM have been proposed as solutions to observations such as dark matter and various theoretical problems (grand unification, the ``naturalness" problem, the strong CP problem, and insufficient CP violation to account for the matter/anti-matter imbalance).  The most popular of these, supersymmetry, produces both coupling constant unification and possible dark matter candidates.  The scalar sector of supersymmetry (in its simplest implementation, the Minimal Supersymmetric Standard Model, or MSSM) has two Higgs doublets.  Independently of supersymmetry, the Two-Higgs Doublet Model (2HDM) is an extension of the SM, which is identical to the SM except for the one extra Higgs doublet.   The 2HDM may be interesting on its own as a potential theory of nature, since the extended Higgs Sector allows for CP violation beyond what is produced by the SM.  It is also useful for gaining insight into the scalar sector of supersymmetry, and other models that contain similar scalar content. 

A theory with two Higgs doublets has the potential to produce dangerous Higgs-mediated flavor-changing neutral currents (FCNCs) unless the off-diagonal  couplings of the neutral Higgs bosons to quarks are absent (or sufficiently small).  It is common to apply a discrete symmetry that restricts the Higgs scalar potential so as to eliminate these off-diagonal couplings~\cite{desh,type2,type1,hallwise,lavoura2}.  The general 2HDM discussed here will not have any symmetries imposed, and thus the FCNCs will be assumed to be suppressed by fine-tuning or heavy scalar masses.  The degree to which the Higgs-fermion couplings are constrained by measurements of flavor-changing rates will be left to future work.  

If evidence of multiple Higgs bosons is discovered experimentally, it will be necessary to know how to connect the experimentally observed quantities with the physical parameters of the model.  Since one would not know in advance what symmetries are present that constrain the scalar sector, the definition of the physical parameters of the Higgs sector should be defined from the most generic implementation of the 2HDM.  In a generic 2HDM, an example of an ``unphysical'' parameter is the common construction
\beq \label{tanbdef}
\tan\beta\equiv\frac{\vev{\Phi_2^0}}{\vev{\Phi_1^0}}\,,
\eeq
where $\Phi_1^0$ and $\Phi_2^0$ are the neutral components of the two Higgs doublets \cite{davidson}.  As defined, this parameter is ambiguous in a general 2HDM because it depends on the choice of basis for the Higgs fields. The two identical hypercharge-one fields can be redefined by a global $2\times 2$ unitary transformation.  The goal of this work is to construct \emph{physical} parameters, which must be basis-independent.  The physical parameter that replaces \eq{tanbdef} was developed in \Ref{davidson} for the CP-conserving 2HDM.  The analog for the CP-violating 2HDM will be discussed in Chapter \ref{tanbsec}.  

The goal of this chapter is to derive the scalar, gauge boson and Yukawa couplings of the CP-violating 2HDM.  (These were presented for the CP-conserving case in \Ref{davidson}.)  A Two-Higgs Doublet Model is CP-violating if there is no basis in which the couplings of the scalar Lagrangian are real-valued.  Complex-valued Higgs couplings can lead to the mixing of the CP-even and CP-odd eigenstates to produce Higgs fields that have indefinite CP.  To develop this model in the basis-independent formalism, I will start by presenting my work with H. Haber in \Ref{haberoneil}.  We begin by reviewing basis-independence in \sect{sec:two}.   In \sect{sec:three}, we introduce the Higgs basis (defined to be a basis
in which one of the two neutral scalar fields has zero vacuum
expectation value), which possesses some invariant
features.  We review the construction of the Higgs basis and use the basis-independent
formalism to highlight the invariant qualities of this basis choice.
Ultimately, we are interested in the Higgs mass-eigenstates.
In the most general CP-violating 2HDM, three neutral Higgs states
mix to form mass-eigenstates that are not eigenstates of CP.  
In \sect{sec:four}, we demonstrate how
to define basis-independent Higgs mixing parameters that are crucial for
deriving an invariant form for the Higgs couplings. In \sect{sec:five}
and \sect{sec:six} we provide the explicit basis-independent forms for
the Higgs couplings to bosons (gauge bosons and Higgs boson self-couplings)
and fermions (quarks and leptons), respectively.  

\section{The Basis-Independent Formalism}
\label{sec:two}
The fields of the two-Higgs-doublet model (2HDM) consist of two
identical complex hypercharge-one, SU(2)$\ls{\rm L}$ doublet scalar fields
$\Phi_a(x)\equiv (\Phi^+_a(x)\,,\,\Phi^0_a(x))$, 
where $a=1,2$ labels the two Higgs doublet fields, and will 
be referred to as the Higgs ``flavor'' index.   
The Higgs doublet fields can always be redefined
by an arbitrary non-singular complex transformation
$\Phi_a\to B_{ab}\Phi_b$, where the matrix $B$ depends on eight real
parameters.  However, four of these parameters can be used
to transform the scalar field kinetic energy terms 
into canonical form.\footnote{That is, starting from
$\mathscr{L}_{\rm KE}=a\,(D_\mu\Phi_1)^\dagger(D_\mu\Phi_1)+
b\,(D_\mu\Phi_2)^\dagger(D_\mu\Phi_2)
+\bigl[c\,(D_\mu\Phi_1)^\dagger(D_\mu\Phi_2)+{\rm h.c.}\bigr]$, 
where $a$ and $b$ are real and $c$ is complex, one
can always find a (non-unitary) transformation
$B$ that removes the four real degrees of freedom
corresponding to $a$, $b$ and $c$ and sets $a=b=1$ and $c=0$.
Mathematically, such a transformation is an element of the coset space
GL$(2,\mathbb{C})/$U(2).}
The most general redefinition of the scalar fields [which leaves invariant
the form of the canonical kinetic 
energy terms 
$\mathscr{L}_{\rm KE}=(D_\mu\Phi)^\dagger_{\abar} (D^\mu\Phi)_a$] 
corresponds to a global U(2) transformation,
$\Phi_a\to U_{a\bbar}\Phi_b$ [and $\Phi_\abar^\dagger\to\Phi_\bbar^\dagger
U^\dagger_{b\abar}$], where the $2\times 2$ unitary matrix $U$ satisfies
$U^\dagger_{b\abar}U_{a\cbar}=\delta_{b\cbar}$.  In our index
conventions, replacing an unbarred index with a barred index is
equivalent to complex conjugation.   We only allow sums over 
barred--unbarred index pairs, which are
performed by employing
the U(2)-invariant tensor $\delta_{a\bbar}$.  
The basis-independent formalism consists of writing all equations
involving the Higgs sector fields in a U(2)-covariant form.
Basis-independent quantities can then be identified as 
U(2)-invariant scalars, which are easily identified as products of
tensor quantities with
all barred--unbarred index pairs summed with no Higgs
flavor indices left over.

We begin with the most general 2HDM scalar potential.  An explicit
form for the scalar potential in a generic basis is given in 
Appendix~\ref{app:one}.
Following \Refs{branco}{davidson}, the scalar potential can be written
in U(2)-covariant form:
\beq \label{genericpot}
\mathcal{V}=Y_{a\bbar}\Phi_\abar^\dagger\Phi_b
+\half Z_{a\bbar c\dbar}(\Phi_\abar^\dagger\Phi_b)
(\Phi_\cbar^\dagger\Phi_d)\,,
\eeq
where the indices $a$, $\bbar$, $c$ and $\dbar$
are labels with respect to the two-dimensional Higgs
flavor space and $Z_{a\bbar c\dbar}=Z_{c\dbar a\bbar}$.
The hermiticity of $\mathcal{V}$ yields
$Y_{a \bbar}= (Y_{b \abar})^\ast$ and
$Z_{a\bbar c\dbar}= (Z_{b\abar d\cbar})^\ast$.
Under a U(2) transformation, the tensors $Y_{a\bbar}$ and
$Z_{a\bbar c\dbar}$ transform covariantly:
$Y_{a\bbar}\to U_{a\cbar}Y_{c\dbar}U^\dagger_{d\bbar}$
and $Z_{a\bbar c\dbar}\to U_{a\ebar}U^\dagger_{f\bbar}U_{c\gbar}
U^\dagger_{h\dbar} Z_{e\fbar g\hb}$.  Thus, the scalar potential
$\mathcal{V}$ is a U(2)-scalar.  The interpretation of these results
is simple.  Global U(2)-flavor
transformations of the two Higgs doublet fields do not change the
functional form of the scalar potential.  However, the coefficients of
each term of the potential depends on the choice of basis.  The
transformation of these coefficients under a U(2) basis change are
precisely the transformation laws of $Y$ and $Z$ given above.

We shall assume that the vacuum of the theory respects the electromagnetic
U(1)$_{\rm EM}$ gauge symmetry.  In this case, the non-zero vacuum
expectation values of $\Phi_a$ must be aligned.  The standard
convention is to make a gauge-SU(2)$\ls{\rm L}$ 
transformation (if necessary) such
that the lower (or second) component of the doublet fields correspond
to electric charge $Q=0$.  In this case, the most general
U(1)$_{\rm EM}$-conserving vacuum expectation values are:
\beq \label{emvev}
\langle \Phi_a
\rangle={\frac{v}{\sqrt{2}}} \left(
\begin{array}{c} 0\\ \widehat v_a \end{array}\right)\,,\qquad
{\rm with}\qquad
\widehat v_a \equiv e^{i\eta}\left(
\begin{array}{c} \cb\,\\ \sb\,e^{i\xi} \end{array}\right)
\,,
\eeq
where $v\equiv 2m_W/g=246$~GeV and $\widehat v_a$ is a vector of unit norm.
The overall phase $\eta$ is arbitrary.  By convention, we take
$0\leq\beta\leq\pi/2$ and $0\leq\xi<2\pi$.
Taking the derivative of \eq{genericpot} with respect to $\Phi_b$, and
setting $\vev{\Phi^0_a}=v_a/\sqrt{2}$, we find the covariant form for
the scalar potential minimum conditions:
\beq \label{potmingeneric}
v\,\widehat v_\abar^\ast\,
[Y_{a\bar b}+\half v^2 Z_{a\bbar c\dbar}\, \widehat v_\cbar^\ast\,
\widehat v_d]=0 \,.
\eeq

Before proceeding, let us consider the most general global-U(2)
transformation~(see p.~5 of \Ref{murnaghan}):
\beq \label{utransform}
U=e^{i\psi}
\left(\begin{array}{cc} e^{i\gamma}\cos\theta &
\quad e^{-i\zeta}\,\sin\theta  \\
 -e^{i\zeta}\,\sin\theta & \quad e^{-i\gamma}\,\cos\theta
\end{array}\right)\,,
\eeq
where $-\pi\leq \theta\,,\,\psi <\pi$ and
$-\pi/2\leq \zeta\,,\,\gamma\leq\pi/2$ defines the closed and bounded
U(2) parameter space.  The matrices $U$ with $\psi = 0$ span an SU(2)
matrix subgroup of U(2).  The factor of $\{e^{i\psi}\}$ constitutes a U(1)
subgroup of U(2). More precisely, 
U(2)~$\iso$~SU(2)$\times$U(1)$/\mathbb{Z}_2$.
In the scalar sector, this U(1) coincides with
global hypercharge U(1)$_{\rm Y}$.   However,
the former U(1) is distinguished from hypercharge by the fact
that it has no effect on the other fields of the Standard Model.

Because the scalar potential is \textit{invariant}
under U(1)$_{\rm Y}$ hypercharge 
transformations,\footnote{The SU(2)$\ls{\rm L}\times$U(1)$_{\rm Y}$
gauge transformations act on the fields of the Standard Model, but
do \textit{not} transform the coefficients of 
the terms appearing in the Lagrangian.}
it follows that $Y$ and $Z$ are invariant under U(1)-flavor
transformations.  Thus, from the standpoint of the Lagrangian, only
SU(2)-flavor transformations correspond to a change of basis.
Nevertheless, the vacuum expectation value $\widehat v$ does change by
an overall phase under flavor-U(1) transformations.  Thus, it is
convenient to expand our definition of the basis to include the phase
of $\widehat v$.  In this convention, all
U(2)-flavor transformations correspond to a change of basis.
The reason for this choice is that it permits us to expand our
potential list of basis-independent quantities to include quantities
that depend on $\widehat v$. Since $\Phi_a\to U_{a\bbar}\Phi_b$ it
follows that $\widehat v_a\to U_{a\bbar} \widehat v_b$, and the
covariance properties of quantities that depend on $\widehat v$ are
easily discerned.

The unit vector $\widehat v_a$ can also be regarded as an
eigenvector of unit norm of the Hermitian matrix
$V_{a\bbar}\equiv \widehat v_a \widehat v_{\bbar}^*$.  The overall phase of
$\hat v_a$ is not determined in this definition, but as noted above
different phase choices are related by U(1)-flavor transformations.
Since $V_{a\bbar}$ is hermitian, it possesses a second eigenvector of unit norm
that is orthogonal to $\widehat v_a$.  We denote this eigenvector by
$\widehat w_a$, which satisfies:
\beq \label{vw}
\widehat v_{\bbar}^* \widehat w_b=0\,.
\eeq
The most general solution to \eq{vw}, up to an overall multiplicative
phase factor, is:
\beq \label{wdef}
\widehat w_b \equiv \widehat
v_\abar^\ast\epsilon_{ab}=e^{-i\eta}\left(
\begin{array}{c} -\sb\,e^{-i\xi}\\ \cb\end{array}\right)\,.\
\eeq
That is,
we have chosen a convention in which $\widehat w_b \equiv e^{i\chi}\widehat
v_\abar^\ast\epsilon_{ab}$, where $\chi=0$.
Of course, $\chi$ is not fixed by \eq{vw};
the existence of this
phase choice is reflected in the non-uniqueness of 
the Higgs basis, as discussed in \sect{sec:three}.

The inverse relation to \eq{wdef} is
easily obtained: $\widehat v^\ast_{\abar}=
\epsilon_{\abar\bbar}\,\widehat w_b$.
Above, we have introduced two Levi-Civita tensors
with $\epsilon_{12}=-\epsilon_{21}=1$ and $\epsilon_{11}=\epsilon_{22}=0$.
However, $\epsilon_{ab}$ and $\epsilon_{\abar\bbar}$ are not proper
tensors with respect to the full flavor-U(2) group (although these are
invariant SU(2)-tensors).  Consequently, $\widehat w_a$ does not
transform covariantly with respect to the full flavor-U(2) group. If
we write $U=e^{i\psi}\widehat U$, with $\det\widehat U=1$ (and
$\det U=e^{2i\psi}$), it is simple to check that under a
U(2) transformation
\beq \label{uvtransform}
\widehat v_a\to U_{a\bbar}\widehat v_b\qquad {\rm implies~that}\qquad
\widehat w_a\to ({\rm det}~U)^{-1}\,U_{a\bbar\,} \widehat w_b\,.
\eeq

Henceforth, we shall define a pseudotensor\footnote{In tensor calculus,
analogous quantities are usually referred to as
tensor densities or relative tensors~\cite{synge}.}
as a tensor
that transform covariantly with respect to the flavor-SU(2)
subgroup but whose transformation law
with respect to the full flavor-U(2)
group is only covariant modulo an overall nontrivial phase equal to
some integer power of $\det U$.  Thus, $\widehat w_a$ is a pseudovector.
However, we can use $\widehat w_a$ to construct proper tensors.  For
example, the Hermitian matrix $W_{a\bbar}\equiv\widehat w_a
\widehat w^*_{\bbar}=\delta_{a\bbar}-V_{a\bbar}$ is a proper
second-ranked tensor.

Likewise, a pseudoscalar (henceforth referred to as 
a pseudo-invariant) is defined
as a quantity that transforms
under U(2) by multiplication by
some integer power of $\det U$.  We reiterate that pseudo-invariants
\textit{cannot} be physical observables as the latter must be true
U(2)-invariants.

\section{The Higgs Bases}
\label{sec:three}
Once the scalar potential minimum is determined, which defines
$\widehat v_a$, one class of basis choices is uniquely selected.
Suppose we begin in a generic $\Phi_1$--$\Phi_2$ basis.
We define new Higgs doublet fields:
\beq \label{hbasisdef}
H_1=(H_1^+\,,\,H_1^0)\equiv \widehat v_{\abar}^*\Phi_a\,,\qquad\qquad
H_2=(H_2^+\,,\,H_2^0)\equiv \widehat w_{\abar}^*\Phi_a= \epsilon_{\bbar\abar}
\widehat v_b\Phi_a \,.
\eeq
The transformation between
the generic basis and the Higgs basis, $H_a = \widehat U_{a \bar{b}} \Phi_b$, is given by the following
flavor-SU(2) matrix:
\beq \label{ugenhiggs}
\widehat U=\left(\begin{array}{cc}\widehat v_1^* &\quad \widehat v_2^*\\
\widehat w_1^* &\quad \widehat w_2^*\end{array}\right)=
\left(\begin{array}{cc}\phm\widehat v_1^* &\quad \widehat v_2^*\\
-\widehat v_2 &\quad \widehat v_1\end{array}\right)\,.
\eeq
This defines a particular Higgs basis.

Inverting \eq{hbasisdef} yields:
\beq \label{hbasis}
\Phi_a=H_1 \widehat v_a+ H_2 \widehat w_a = H_1 \widehat v_a+ H_2 \widehat v^*_{\bbar}\epsilon_{ba}\,.
\eeq
The definitions of $H_1$ and $H_2$ imply that
\beq \label{higgsvevs}
\vev{H_1^0}=\frac{v}{\sqrt{2}}\,,\qquad\qquad \vev{H_2^0}=0\,,
\eeq
where we have used \eq{vw} and the fact that
$\widehat v^{\,*}_{\abar}\,\widehat v_a=1$.

The Higgs basis is not unique.  Suppose one begins in a generic
$\Phi'_1$--$\Phi'_2$ basis, where $\Phi'_a=V_{a\bbar}\Phi_b$ and
$\det V\equiv e^{i\chi}\neq 1$.  If we now define:
\beq \label{hbasispdef}
H'_1\equiv \widehat v_{\abar}^*\Phi'_a\,,\qquad\qquad
H'_2\equiv \widehat w_{\abar}^*\Phi'_a\,,
\eeq
then
\beq \label{hbasischi}
H'_1=H_1\,,\qquad\qquad H'_2=(\det V)H_2= e^{i\chi} H_2\,.
\eeq
That is, $H_1$ is an invariant field, whereas $H_2$ is pseudo-invariant
with respect to arbitrary U(2) transformations.
In particular, the unitary matrix
\beq \label{ud}
U_D\equiv\left(\begin{array}{cc} 1 &\quad 0\\ 0 &\quad e^{i\chi}\end{array}
\right)
\eeq
transforms from the unprimed Higgs basis to the primed Higgs basis.
The phase angle
$\chi$ parameterizes the class of Higgs bases.  From the definition of
$H_2$ given in \eq{hbasisdef}, this phase freedom can be attributed
to the choice of an overall phase in the definition of $\widehat
w$ as discussed in \sect{sec:two}.  This phase freedom will be
reflected by the appearance of pseudo-invariants in the study of the
Higgs basis.  However, pseudo-invariants are useful in that they can
be combined to create true invariants, which are candidates for
observable quantities.

It is now a simple matter to insert
\eq{hbasis} into \eq{genericpot} to obtain:
\beqa \label{hbasispot}
\mathcal{V}&=& Y_1 H_1^\dagger H_1+ Y_2 H_2^\dagger H_2
+[Y_3 H_1^\dagger H_2+{\rm h.c.}]\nonumber \\[5pt]
&&\quad
+\half Z_1(H_1^\dagger H_1)^2 +\half Z_2(H_2^\dagger H_2)^2
+Z_3(H_1^\dagger H_1)(H_2^\dagger H_2)
+Z_4( H_1^\dagger H_2)(H_2^\dagger H_1) \nonumber \\[5pt]
&&\quad +\left\{\half Z_5 (H_1^\dagger H_2)^2
+\big[Z_6 (H_1^\dagger H_1)
+Z_7 (H_2^\dagger H_2)\big]
H_1^\dagger H_2+{\rm h.c.}\right\}\,,
\eeqa
where $Y_1$, $Y_2$ and $Z_{1,2,3,4}$ are U(2)-invariant quantities and
$Y_3$ and $Z_{5,6,7}$ are pseudo-invariants.
The explicit forms for the Higgs basis coefficients have been given in
\Ref{davidson}.  The invariant coefficients are conveniently expressed
in terms of the second-ranked tensors $V_{a\bbar}$ and $W_{a\bbar}$
introduced in section \ref{sec:two}:
\beqa \label{invariants}
Y_1 &\equiv& \Tr(YV)\,,\qquad\qquad\qquad\,\, Y_2 \equiv \Tr(YW)\,,\nonumber\\
Z_1 &\equiv& Z_{a\bbar c\dbar}\,V_{b\abar}V_{d\cbar}\,,\qquad\qquad\,\,\,\,
Z_2 \equiv Z_{a\bbar c\dbar}\,W_{b\abar}W_{d\cbar}\,,\qquad\qquad \nonumber\\
Z_3 &\equiv& Z_{a\bbar c\dbar}\,V_{b\abar}W_{d\cbar}\,,\qquad\qquad\,\,\,
Z_4 \equiv Z_{a\bbar c\dbar}\,V_{b\cbar}W_{d\abar}\,,
\eeqa
whereas the pseudo-invariant coefficients are given by:
\beqa \label{pseudoinvariants}
\hspace{-0.3in} Y_3 &\equiv&
Y_{a\bbar}\,\widehat v_\abar^\ast\, \widehat w_b\,,\qquad\qquad\qquad
Z_5 \equiv
Z_{a\bbar c\dbar}\,\widehat v_\abar^\ast\, \widehat w_b\,
\widehat v_\cbar^\ast\, \widehat w_d\,,\nonumber \\ \hspace{-0.3in}
Z_6 &\equiv&  Z_{a\bbar c\dbar}\,\widehat v_\abar^\ast\,\widehat v_b\,
\widehat v_\cbar^\ast\, \widehat w_d\,,\qquad\quad
Z_7 \equiv
     Z_{a\bbar c\dbar}\,\widehat v_\abar^\ast\, \widehat w_b\,
\widehat w_\cbar^\ast\,\widehat w_d\,.
\eeqa
The invariant coefficients are manifestly real, whereas the
pseudo-invariant coefficients are potentially complex.

Using \eq{uvtransform}, it follows that under a flavor-U(2)
transformation specified by the matrix $U$, the pseudo-invariants
transform as:
\beq \label{tpseudo}
[Y_3, Z_6, Z_7]\to (\det U)^{-1}[Y_3, Z_6, Z_7]\qquad {\rm and} \qquad
Z_5\to  (\det U)^{-2} Z_5 \,.
\eeq
One can also deduce \eq{tpseudo} from \eq{hbasispot} by
noting that $\mathcal{V}$ and $H_1$ are invariant
whereas $H_2$ is pseudo-invariant field that is transforms as:
\beq \label{h2pseudo}
H_2\to (\det U)H_2\,.
\eeq

In the class of Higgs bases defined by \eq{hbasischi}, $\widehat{v}=(1,0)$
and $\widehat{w}=(0,1)$, which are independent of the angle $\chi$
that distinguishes among different Higgs bases.  That is,
under the phase transformation specified by \eq{ud}, both $\widehat v$
and $\widehat w$ are unchanged.
Inserting these values of $\widehat v$ and $\widehat w$ into
\eqs{invariants}{pseudoinvariants} yields the coefficients
of the Higgs basis scalar potential.  For example, the coefficient of
$H_1^\dagger H_2$ is given by $Y_{12}=Y_3$ in the unprimed Higgs basis
and $Y'_{12}=Y'_3$ in the primed Higgs basis.  Using \eq{tpseudo}, it
follows that
$Y'_{12}=Y_{12} e^{-i\chi}$, which is consistent with the matrix
transformation law $Y'=U_D Y U_D^\dagger$.

From the four complex pseudo-invariant coefficients, one can form
four independent real invariants $|Y_3|$, $|Z_{5,6,7}|$ and three
invariant relative phases $\arg(Y_3^2 Z_5^*)$, 
$\arg(Y_3 Z_6^*)$ and $\arg(Y_3 Z_7^*)$.
Including the six invariants of \eq{invariants}, we have 
therefore identified thirteen
independent invariant real degrees of freedom prior to imposing the
scalar potential minimum conditions.  \Eq{potmingeneric} then imposes three
additional conditions on the set of thirteen invariants\footnote{The 
second condition of
\eq{hbasismincond} is a complex equation that can be rewritten in
terms of invariants: $|Y_3|=\half|Z_6|v^2$ and $Y_3
Z_6^*=-\half|Z_6|^2 v^2$.}
\beq \label{hbasismincond}
Y_1=-\half Z_1 v^2\,,\qquad\qquad\qquad
Y_3=-\half Z_6 v^2\,.
\eeq
This leaves eleven independent real degrees of freedom 
(one of which is the vacuum expectation
value~$v=246$~GeV) that specify the 2HDM parameter space.

The doublet of scalar fields in the Higgs basis
can be parameterized as follows:
\beq
\label{hbasisfields}
H_1=\left(\begin{array}{c}
G^+ \\ {\frac{1}{\sqrt{2}}}\left(v+\varphi_1^0+iG^0\right)\end{array}
\right)\,,\qquad
H_2=\left(\begin{array}{c}
H^+ \\ {\frac{1}{\sqrt{2}}}\left(\varphi_2^0+ia^0\right)\end{array}
\right)\,,
\eeq
and the corresponding hermitian conjugated fields are likewise defined.
We identify $G^\pm$ as a charged Goldstone boson pair
and $G^0$ as the CP-odd neutral Goldstone boson.\footnote{The definite CP
property of the neutral Goldstone boson persists even if the Higgs Lagrangian
is CP-violating (either explicitly or spontaneously), 
as shown in Chapter \ref{custchapter}.}
In particular, the identification of $G^0=\sqrt{2}\,\Im\, H_1^0$
follows from the fact that we have
defined the Higgs basis [see \eqs{hbasisdef}{higgsvevs}]
such that $\vev{H_1^0}$ is real and non-negative.  Of the remaining fields,
$\varphi_1^0$ is a CP-even neutral scalar field,  
$\varphi_2^0$ and $a^0$ are states of indefinite CP quantum 
numbers,\footnote{The CP-properties of 
the neutral scalar fields (in the Higgs basis) can be determined
by studying the pattern of gauge boson/scalar boson couplings and the 
scalar self-couplings in the interaction Lagrangian (see \sect{sec:five}).
If the scalar potential is CP-conserving, then two orthogonal
linear combinations of $\varphi_2^0$ and $a^0$ can be found that are
eigenstates of CP.  
By an appropriate rephasing of $H_2$ 
(which corresponds to some particular choice among the possible Higgs
bases) such that all the coefficients of the scalar potential in the
Higgs basis are real, one can then identify $\varphi_2^0$ as a CP-even 
scalar field and $a^0$ as a CP-odd scalar field.  See Chapter \ref{custchapter}
for further details.}
and $H^\pm$ is the physical charged Higgs boson pair.
If the Higgs sector is CP-violating, then
$\varphi_1^0$, $\varphi_2^0$, and $a^0$ all mix to produce three
physical neutral Higgs mass-eigenstates of indefinite CP quantum numbers. 
 
\section{The Physical Higgs Mass-Eigenstates}
\label{sec:four}
To determine the Higgs mass-eigenstates, one must examine the terms of
the scalar potential that are quadratic in the scalar fields (after
minimizing the scalar potential and defining shifted scalar
fields with zero vacuum expectation values).  
This procedure is carried out in Appendix~\ref{app:two} starting 
from a generic basis.  However, there is an advantage in     
performing the computation in the Higgs basis
since the corresponding scalar potential coefficients 
are
invariant or pseudo-invariant quantities [\eqst{hbasispot}{pseudoinvariants}].
This will allow us to identify U(2)-invariants
in the Higgs mass diagonalization procedure.

Thus, we proceed by inserting \eq{hbasis}
into \eq{genericpot} and examining the terms linear and quadratic in
the scalar fields.  The requirement that the coefficient of the linear term
vanishes corresponds to the scalar potential minimum conditions
[\eq{hbasismincond}].  These conditions are then used in the evaluation
of the coefficients of the terms quadratic in the fields.  One can easily
check that no quadratic terms involving the Goldstone boson fields survive
(as expected, since the Goldstone bosons are massless).  This confirms
our identification of the Goldstone fields in \eq{hbasisfields}.
The charged Higgs boson mass is also easily determined:
\beq \label{hplus}
m_{H^\pm}^2=Y_{2}+\half Z_3 v^2\,.
\eeq
The three remaining neutral fields mix, and
the resulting neutral Higgs
squared-mass matrix in the $\varphi_1^0$--$\varphi_2^0$--$a^0$ basis is:
\beq \label{matrix33}
\mathcal{M}=v^2\left( \begin{array}{ccc}
\hspace{-.1 in}Z_1&\,\, \Re(Z_6) &\,\, -\Im(Z_6)\\
\hspace{-.1 in}\Re(Z_6)  &\,\, \half\left[Z_3+Z_4+\Re(Z_5)\right]+Y_2/v^2 & \,\,
- \half \Im(Z_5)\\\hspace{-.1 in} -\Im(Z_6) &\,\, - \half \Im(Z_5) &\,\,
 \half\left[Z_3+Z_4-\Re(Z_5)\right]+Y_2/v^2\end{array}\right).
\eeq
Note that $\mathcal{M}$ depends implicitly on the choice of Higgs
basis [\eq{hbasischi}] via the
$\chi$-dependence of the pseudo-invariants $Z_5$ and $Z_6$.
Moreover, the real and imaginary parts of these pseudo-invariants mix
if $\chi$ is changed.  Thus, $\mathcal{M}$ does not
possess simple transformation
properties under arbitrary flavor-U(2) transformations.
Nevertheless, we demonstrate
below that the eigenvalues and normalized eigenvectors are
U(2)-invariant.  First, we compute the characteristic equation:
\beq \label{charpoly}
\det(\mathcal{M}-xI)=-x^3+\Tr(\mathcal{M})\,x^2
-\half\left[(\Tr\mathcal{M})^2-\Tr(\mathcal{M}^2)\right]x+
\det(\mathcal{M})\,,
\eeq
where $I$ is the $3\times 3$ identity matrix. [The coefficient of
$x$ in \eq{charpoly} is particular to $3\times 3$
matrices (see Fact 4.9.3 of \Ref{matrixref}).]
Explicitly,
\beqa
\Tr(\mathcal{M}) &=& 2Y_2+(Z_1+Z_3+Z_4)v^2\,,\nonumber\\
\Tr(\mathcal{M}^2) &=& Z_1^2 v^4
+\half v^4\left[(Z_3+Z_4)^2+|Z_5|^2+4|Z_6|^2\right]
+2Y_2[Y_2+(Z_3+Z_4)v^2]\,,\nonumber \\
\det(\mathcal{M}) &=& \quarter\left\{Z_1 v^6[(Z_3+Z_4)^2-|Z_5|^2]
-2v^4[2Y_2+(Z_3+Z_4)v^2]|Z_6|^2\right. \nonumber \\
&&\qquad\qquad\qquad \left. +4Y_2 Z_1 v^2[Y_2+(Z_3+Z_4)v^2]
+2v^6\Re(Z_5^* Z_6^2)\right\}\,.
\eeqa
Clearly, all the coefficients of the characteristic polynomial are
U(2)-invariant.  Since the roots of this polynomial are the
squared-masses of the physical Higgs bosons, it follows that the
physical Higgs masses are basis-independent as required.
Since $\mathcal{M}$ is a real symmetric matrix, the eigenvalues of
$\mathcal{M}$ are real.  However, if any of these eigenvalues are
negative, then the extremal solution of \eq{potmingeneric} with $v\neq 0$ 
is \textit{not} a minimum of the scalar potential.  
The requirements that
$\mhpm^2>0$ [\eq{hplus}] and the positivity of the squared-mass
eigenvalues of $\mathcal{M}$ provide basis-independent conditions for
the desired spontaneous symmetry breaking pattern specified by \eq{emvev}.

The real symmetric squared-mass matrix $\mathcal{M}$ can be diagonalized by
an orthogonal transformation
\beq \label{rmrt}
R\mathcal{M} R^T=\mathcal{M}_D\equiv {\rm diag}~(m_1^2\,,\,m_2^2\,,\,m_3^2)\,,
\eeq
where $RR^T=I$ and the $m_k^2$ are the eigenvalues of $\mathcal{M}$
[\textit{i.e.}, the roots of \eq{charpoly}].
A~convenient form for $R$ is:
\beqa \label{rmatrix}
R=R_{12}R_{13}R_{23} &=&\left( \begin{array}{ccc}
c_{12}\,\, &-s_{12}\quad &0\\
s_{12}\,\, &\phm c_{12}\quad &0\\
0\,\, &\phm 0\quad &1\end{array}\right)\left( \begin{array}{ccc}
c_{13}\quad &0\,\, &-s_{13}\\
0\quad & 1\,\,&\phm 0\\
s_{13}\quad &0\,\, &\phm c_{13}\end{array}\right) \left( \begin{array}{ccc}
1\quad &0\,\, &\phm 0\\
0\quad &c_{23}\,\, &-s_{23}\\
0\quad &s_{23}\,\, &\phm c_{23}\end{array}\right) \nonumber \\[10pt]
&=&
\left( \begin{array}{ccc}
c_{13}c_{12}\quad &-c_{23}s_{12}-c_{12}s_{13}s_{23}\quad &-c_{12}c_{23}s_{13}
+s_{12}s_{23}\\[6pt]
c_{13}s_{12}\quad &c_{12}c_{23}-s_{12}s_{13}s_{23}\quad
& -c_{23}s_{12}s_{13}-c_{12}s_{23}\\
s_{13}\quad &c_{13}s_{23}\quad &c_{13}c_{23}\end{array}\right)\,,
\eeqa
where $c_{ij}\equiv \cos\theta_{ij}$ and $s_{ij}\equiv\sin\theta_{ij}$.
Note that $\det R=1$, although we could have chosen an orthogonal
matrix with determinant equal to $-1$ by choosing $-R$ in place
of $R$.  In addition, if we take the range of the angles to be
$-\pi\leq\theta_{12}$, $\theta_{23}<\pi$ and $|\theta_{13}|\leq\pi/2$,
then we cover the complete parameter space of SO(3) matrices
(see p.~11 of \Ref{murnaghan}).  That is, we work in a convention
where $c_{13}\geq 0$.
However, this parameter space includes points that simply correspond
to the redefinition of two of the Higgs mass-eigenstate fields by
their negatives.  Thus, we may reduce the parameter
space further and define all Higgs mixing angles modulo~$\pi$.
We shall verify this assertion at the end of this section.

The neutral Higgs mass-eigenstates are denoted by $h_1$, $h_2$ and
$h_3$:
\beq \label{rotated}
\left( \begin{array}{c}
h_1\\ h_2\\h_3 \end{array}\right)=R \left(\begin{array}{c} \varphi_1^0\\
\varphi_2^0\\ a^0\end{array}\right)\,.
\eeq
It is often convenient to choose a convention for the mass
ordering of the $h_k$ such that $m_1\leq m_2\leq m_3$.

Since the mass-eigenstates $h_k$ do not depend on
the initial basis choice, they must be U(2)-invariant fields.
In order to present a formal proof of this assertion, we need to
determine the transformation properties of the elements of $R$ under
an arbitrary U(2) transformation.  In principle, these can be
determined from \eq{rmrt}, using the fact that the $m_k^2$ are
invariant quantities.  However, the form of
$\mathcal{M}$ is not especially convenient for this purpose as noted
below \eq{matrix33}.  This can be ameliorated by introducing 
the unitary matrix:
\beq
W=\left(\begin{array}{ccc} 1&\qquad 0&\qquad 0\\
0 &\qquad 1/\sqrt{2} &\qquad 1/\sqrt{2} \\
0  &\qquad -i/\sqrt{2} &\qquad i/\sqrt{2}\end{array}\right)\,,
\eeq
and rewriting \eq{rmrt} as
\beq \label{rwmrw}
(RW)(W^\dagger \mathcal{M}W)(RW)^\dagger=\mathcal{M}_D
={\rm diag}~(m_1^2\,,\,m_2^2\,,\,m_3^2)\,.
\eeq
A straightforward calculation yields:
\beqa \hspace{-.2 in}
W^\dagger \mathcal{M}W \!&=& \!v^2\left(\begin{array}{ccc} Z_1 &\quad
\nicefrac{1}{\sqrt{2}}Z_6  &\quad \nicefrac{1}{\sqrt{2}}Z_6^* \\
 \nicefrac{1}{\sqrt{2}}Z_6^*  &\quad \half(Z_3+Z_4)+Y_2/v^2 &\quad
\half Z_5^*\\  \nicefrac{1}{\sqrt{2}}Z_6  &\quad \half Z_5 &\quad
 \half(Z_3+Z_4)+Y_2/v^2\end{array}\right),\,\,\,\label{WMW}\\[12pt]
RW&=&\left(\begin{array}{ccc}q_{11} 
&\qquad  \nicefrac{1}{\sqrt{2}}q^*_{12}\,e^{i\theta_{23}}
&\qquad \nicefrac{1}{\sqrt{2}} q_{12}\,e^{-i\theta_{23}} \\[4pt]
q_{21} &\qquad  \nicefrac{1}{\sqrt{2}}q^*_{22}\,e^{i\theta_{23}}
&\qquad \nicefrac{1}{\sqrt{2}}q_{22}\,e^{-i\theta_{23}} \\[4pt]
q_{31} &\qquad  \nicefrac{1}{\sqrt{2}}q^*_{32}\,e^{i\theta_{23}}
&\qquad \nicefrac{1}{\sqrt{2}} q_{32}\,e^{-i\theta_{23}}
\end{array}\right),\label{RW}
\eeqa
where
\beqa \label{qkldef}
\!\!\!\!\! q_{11}&=&c_{13}c_{12}\,,\qquad\qquad\qquad\,\,\,
q_{21}=c_{13}s_{12}\,,\qquad\qquad\qquad
q_{31}=s_{13}\,,\nonumber \\
\!\!\!\!\! q_{12}&=&-s_{12}-ic_{12}s_{13}\,,\qquad\quad\!
q_{22}= c_{12}-is_{12}s_{13}\,,\qquad\quad\,
q_{32}= ic_{13}\,.
\eeqa

The matrix $RW$ defined in \eq{RW} is unitary
and satisfies $\det RW=i$. Evaluating this determinant yields:
\beq \label{detrw}
\half\sum_{j,k,\ell=1}^3\,\epsilon_{jk\ell} q_{j1}\Im(q^*_{k2} q_{\ell 2})=1\,,
\eeq
while unitarity implies:
\beqa 
&&\qquad\qquad\qquad\quad
\Re\left(q_{k1} q_{\ell 1}^* + q_{k2}q_{\ell 2}^* \right)=\delta_{k\ell}\,,
\label{unitarity1}\\[6pt]
&&\sum_{k=1}^3\,|q_{k1}|^2=\half\sum_{k=1}^3\,|q_{k2}|^2=1\,,\qquad\qquad
\sum_{k=1}^3\,q_{k2}^{\,2}=\sum_{k=1}^3\,q_{k1}q_{k2}=0\,.\label{unitarity2}
\eeqa
These results can be used to prove the identity~\cite{cpcarlos}:
\beq \label{epsid}
q_{j1}=\half\sum_{k,\ell=1}^3\,\epsilon\ls{jk\ell}\,\Im(q^*_{k2}q_{\ell 2})\,.
\eeq

Since the matrix elements of
$W^\dagger\mathcal{M}W$ only involve invariants and pseudo-invariants,
we may use \eq{rwmrw} to determine the flavor-U(2)
transformation properties of $q_{k\ell}$ and $e^{i\theta_{23}}$.  
The resulting transformation laws are:
\beq \label{qtrans}
q_{k\ell}\to q_{k\ell}\,,\qquad {\rm and}\qquad
e^{i\theta_{23}}\to (\det U)^{-1}e^{i\theta_{23}}\,,
\eeq
under a U(2) transformation $U$.  That is, the $q_{k\ell}$ are
invariants, or equivalently
$\theta_{12}$ and $\theta_{13}$ (modulo $\pi$)
are U(2)-invariant angles, whereas
$e^{i\theta_{23}}$ is a pseudo-invariant.
\Eq{qtrans} is critical for the rest of the paper.  Finally, to show
that the Higgs mass-eigenstates are invariant fields, we rewrite
\eq{rotated} as
\beq \label{rotated2}
\left( \begin{array}{c}
h_1\\ h_2\\h_3 \end{array}\right)=RW \left(\begin{array}{c}
\sqrt{2}\,\Re H_1^0-v\\ H_2^0\\H_2^{0\,\dagger} \end{array}\right)\,.
\eeq
Since the $q_{k\ell}$, $H_1$ and 
the product $e^{i\theta_{23}}H_2$ are U(2)-invariant quantities, 
it follows that the $h_k$ are invariant fields.

The transformation laws given in 
\eqs{tpseudo}{qtrans} imply that the quantities 
$Z_5\,e^{-2i\theta_{23}}$, 
$Z_6\, e^{-i\theta_{23}}$ and $Z_7\, e^{-i\theta_{23}}$ 
are U(2)-invariant.  These combinations will appear
in the physical Higgs boson self-couplings of \sect{sec:five} and in the
expressions for the invariant mixing angles 
given in Appendix~\ref{app:three}.  With this in
mind, it is useful to rewrite the neutral Higgs mass diagonalization equation
[\eq{rmrt}] as follows.  With $R\equiv R_{12}R_{13}R_{23}$ given by
\eq{rmatrix}, 
\beq \label{mtilmatrix}
\widetilde{\mathcal{M}}\equiv R_{23}\mathcal{M}R_{23}^T=\!
v^2\!\left( \begin{array}{ccc}
\hspace{-.12 in}Z_1&\,\, \Re(Z_6 \, e^{-i\theta_{23}}) &\,\, -\Im(Z_6 \, e^{-i\theta_{23}})\\
\hspace{-.12 in}\Re(Z_6 e^{-i\theta_{23}}) &\,\,\Re(Z_5 \,e^{-2i\theta_{23}})+ A^2/v^2 & \,\,
\hspace{-.12 in}- \half \Im(Z_5 \,e^{-2i\theta_{23}})\\ -\Im(Z_6 \,e^{-i\theta_{23}})
 &\,\, - \half \Im(Z_5\, e^{-2i\theta_{23}}) &\,\, A^2/v^2\end{array}\right)\!.
\eeq
where 
${A}^2$ is defined by
\beq \label{madef}
{A}^2\equiv Y_2+\half[Z_3+Z_4-\Re(Z_5 e^{-2i\theta_{23}})]v^2\,.
\eeq
The diagonal neutral Higgs squared-mass matrix is then given by the following:
\beq \label{diagtil}
\widetilde{R}\,\widetilde{\mathcal{M}}\,\widetilde{R}^T=\mathcal{M}_D={\rm
 diag}(m_1^2\,,\,m_2^2\,,\,m_3^2)\,, 
\eeq
where the diagonalizing matrix $\widetilde{R}\equiv R_{12}R_{13}$ 
depends only on $\theta_{12}$ and $\theta_{13}$:
\beq \label{rtil}
\widetilde R=\left(\begin{array}{ccc}c_{12}c_{13} & \quad -s_{12} &
\quad -c_{12}s_{13} \\ c_{13}s_{12} & \quad c_{12} & \quad-s_{12}s_{13}\\
s_{13} & \quad 0 & \quad c_{13}\end{array}\right)\,.
\eeq
\Eqst{mtilmatrix}{rtil} provide a manifestly U(2)-invariant 
squared-mass matrix
diagonalization, since the elements of  $\widetilde{R}$
and $\widetilde{\mathcal{M}}$ are invariant quantities.

\Eq{rotated2} can be conviently written as
\beq h_k = \frac{1}{\sqrt{2}}\left[\overline H_1\lsup{0\,\dagger} q_{k1} + H_2\lsup{0\,\dagger} q_{k2}\thetminus + \overline H_1^0 q_{k1}^* + H_2^0 q_{k2}^* \thet\right]\,,\label{convenient}\eeq
where $\overline H_1^0 \equiv H_1^0 -v\widehat v_a/\sqrt{2}$. The $q_{k\ell}$, defined for $k=1,2,3$ and $\ell=1,2$ by \eq{qkldef}, are displayed in Table~\ref{tab1}.
To account for the Goldstone boson ($k=4$) we have also introduced:
$q_{41}=i$ and $q_{42}=0$ Note that the $q_{k\ell}$ are invariant, and $\overline H_1^0$ and $H_2^0 \thet$ are invariant fields.

In this section, all computations were carried out by first
transforming to the Higgs basis.  
The advantage of this procedure is that one can readily
identify the relevant invariant and pseudo-invariant quantities
involved in the determination of the Higgs mass-eigenstates.  We may now
combine \eqs{hbasisdef}{convenient} to obtain explicit expressions for the 
Higgs mass-eigenstate fields $h_k$ in terms of the scalar fields in
the generic basis $\Phi_a$.  
Since these expressions do not depend
on the Higgs basis, one could have obtained the 
results for the Higgs mass-eigenstates directly without reference
to Higgs basis quantities.  In Appendix~\ref{app:two}, we present a
derivation starting from the generic basis, which produces the following
expressions for the Higgs mass-eigenstates (and the
Goldstone boson) in terms of the generic basis fields:
\beq \label{hmassinv}
h_k=\frac{1}{\sqrt{2}}\left[\overline\Phi_{\abar}\lsup{0\,\dagger}
(q_{k1} \widehat v_a+q_{k2}\widehat w_a e^{-i\theta_{23}})
+(q^*_{k1}\widehat v^*_{\abar}+q^*_{k2}\widehat w^*_{\abar}e^{i\theta_{23}})
\overline\Phi_a\lsup{0}\right]\,,
\eeq
for $k=1,\ldots,4$, where $h_4=G^0$.  The shifted neutral fields are defined
by $\overline\Phi_a\lsup{0}\equiv \Phi_a^0-v\widehat v_a/\sqrt{2}$.
\begin{table}[h!]
\centering
\caption{The U(2)-invariant quantities $q_{k\ell}$ are functions of the
the neutral Higgs mixing angles $\theta_{12}$ and $\theta_{13}$, where
$c_{ij}\equiv\cos\theta_{ij}$ and $s_{ij}\equiv\sin\theta_{ij}$.\label{tab1}}
\begin{tabular}{|c||c|c|}\hline
$\phaa k\phaa $ &\phaa $q_{k1}\phaa $ & \phaa $q_{k2} \phaa $ \\ \hline
$1$ & $c_{12} c_{13}$ & $-s_{12}-ic_{12}s_{13}$ \\
$2$ & $s_{12} c_{13}$ & $c_{12}-is_{12}s_{13}$ \\
$3$ & $s_{13}$ & $ic_{13}$ \\
$4$ & $i$ & $0$ \\ \hline
\end{tabular}
\end{table}

\noindent
Since the $q_{k\ell}$ are U(2)-invariant and
$\widehat w_a e^{-i\theta_{23}}$ is a \textit{proper} vector under
U(2) transformations, it follows that
\eq{hmassinv} provides a
U(2)-invariant expression for the Higgs mass-eigenstates.
It is now a simple matter to invert \eq{hmassinv} to obtain
\beq \label{master}
\Phi_a=\left(\begin{array}{c}G^+\widehat v_a+H^+ \widehat w_a\\[6pt]
\displaystyle
\frac{v}{\sqrt{2}}\widehat v_a+\frac{1}{\sqrt{2}}\sum_{k=1}^4
\left(q_{k1}\widehat v_a+q_{k2}e^{-i\theta_{23}}\widehat w_a\right)h_k
\end{array}\right)\,,
\eeq
where $h_4\equiv G^0$.  The form of the charged upper component of $\Phi_a$
is a consequence of \eq{hbasis}.
The U(2)-covariant expression for $\Phi_a$ in terms of the 
Higgs mass-eigenstate scalar fields given by \eq{master}
is one of the central results of
this paper.
In sections \ref{sec:five} and \ref{sec:six}, we shall employ 
this result for $\Phi_a$ in the computation of
the Higgs couplings of the 2HDM.

Finally, we return to the question of the domains of the angles
$\theta_{ij}$.  We assume that $Z_6\equiv |Z_6|e^{i\theta_6}\neq 0$
(the special case of $Z_6=0$ is treated at the end of \App{app:three}).
Since $e^{-i\theta_{23}}$ is a pseudo-invariant, we prefer to deal with
the invariant angle $\phi$:
\beq \label{invang}
\phi\equiv\theta_6-\theta_{23}\,,\qquad {\rm where}
\qquad \theta_6\equiv\arg Z_6\,.
\eeq
As shown in Appendix~\ref{app:three}, the invariant angles
$\theta_{12}$, $\theta_{13}$ and
$\phi$ are determined modulo $\pi$ in terms of invariant combinations of
the scalar potential parameters.
This domain is smaller than the one defined by
$-\pi\leq\theta_{12}$, $\theta_{23}<\pi$ and $|\theta_{13}|\leq\pi/2$,
which covers the parameter space of SO(3) matrices.
Since the U(2)-invariant mass-eigenstate fields $h_k$ are real, one
can always choose to redefine any one of the $h_k$ by its negative.
Redefining two of the three Higgs fields $h_1$, $h_2$ and $h_3$
by their negatives\footnote{In order
to have an odd number of Higgs mass-eigenstates
redefined by their negatives, one would have to employ an orthogonal
Higgs mixing matrix with $\det~R=-1$.} is
equivalent to multiplying two of the rows of $R$ by $-1$.  
In particular,
\beqa
\hspace{-0.25in} \theta_{12}\to\theta_{12}\pm\pi &\Longrightarrow&
h_1\to -h_1~{\rm and}~ h_2\to -h_2  
\,, \label{signflip12} \\
\hspace{-0.25in} \phi\to\phi\pm\pi\,,
\quad\!\! \theta_{13}\to -\theta_{13}\,,\quad\!\!  
\theta_{12}\to \pm\pi-\theta_{12}
&\Longrightarrow&
 h_1\to -h_1~{\rm and}~h_3\to -h_3
\,,\label{signflip13}\\
\hspace{-0.25in} \theta_{13}\to \theta_{13}\pm\pi\,,\quad\!\!  
\theta_{12}\to -\theta_{12}
&\Longrightarrow&
 h_1\to -h_1~{\rm and}~h_3\to -h_3
\,,\label{signflip13p}\\
\hspace{-0.25in}\phi\to\phi\pm\pi\,,
\quad\!\! \theta_{13}\to -\theta_{13}\,,
\quad\!\! \theta_{12}\to -\theta_{12} &\Longrightarrow&
 h_2\to -h_2~{\rm and}~h_3\to -h_3
\,,\label{signflip23}\\
\hspace{-0.25in} \theta_{13}\to \theta_{13}\pm\pi\,,
\quad\!\! \theta_{12}\to \pm\pi-\theta_{12} &\Longrightarrow&
 h_2\to -h_2~{\rm and}~h_3\to -h_3
\,.\label{signflip23p}
\eeqa
This means that if we adopt a convention in which $c_{12}$,
$c_{13}$ and $\sin\phi$ are 
non-negative, with the angles defined modulo $\pi$,
then the sign of the Higgs mass-eigenstate fields
will be fixed.  

Given a choice of the overall sign
conventions of the neutral Higgs fields, the number of solutions for
the invariant angles $\theta_{12}$, $\theta_{13}$ and $\phi$ 
modulo $\pi$ are in one-to-one correspondence with the 
possible mass orderings of the $m_k$ (except at 
certain singular points of the parameter space\footnote{At singular
points of the parameter space corresponding to two (or three) mass-degenerate 
neutral Higgs bosons,
some (or all) of the invariant Higgs mixing angles are indeterminate.
An indeterminate invariant angle also arises in the case of $Z_6=0$
and $c_{13}=0$ as explained at the end of \App{app:three}.\label{fnmass}}).
For example,  note that
\beq \label{signfliph1h2} 
\hspace{-0.25in} 
\theta_{12}\to\theta_{12}\pm\pi/2 \Longrightarrow
 h_1\to \mp h_2~{\rm and}~ h_2\to \pm h_1\,.
\eeq
That is, two solutions for $\theta_{12}$ exist modulo $\pi$.
If $m_1< m_2$, then \eq{m2m1} implies that 
the solutions for $\theta_{12}$ and $\phi$
are correlated such that $s_{12}\cos\phi\geq 0$, and (for fixed $\phi$) 
only one $\theta_{12}$ solution
modulo~$\pi$ survives. 
The corresponding effects on the invariant angles that result from swapping
other pairs of neutral Higgs fields are highly
non-linear and cannot be simply exhibited in closed form.
Nevertheless, we can use the results of Appendix~\ref{app:three}
to conclude that for $m_{1,2}<m_3$
(in a convention where $\sin\phi\geq 0$),  \eq{phieq} yields
$s_{13}\leq 0$, and for $m_1<m_2<m_3$,
\eq{imz56f} implies that $\sin 2\theta_{56}\cos\phi\geq 0$, where
$\theta_{56}\equiv -\half\arg(Z_5^* Z_6^2)$.

The sign of the neutral Goldstone field is conventional, but is not
affected by the choice of Higgs mixing angles.  Finally, we note that
the charged fields $G^\pm$ and $H^\pm$ are complex.  \Eq{master}
implies that $G^\pm$ is an invariant field and $H^\pm$
is a pseudo-invariant field that transforms as:
\beq \label{hplustrans}
H^\pm\to (\det U)^{\pm 1}\,H^\pm
\eeq
with respect to U(2) transformations.
That is, once the Higgs Lagrangian is written in terms
the Higgs mass-eigenstates and the Goldstone bosons, one is still free
to rephase the charged fields.  By convention, we shall fix this phase
according to \eq{master}.

\section{Higgs Couplings to Bosons}
\label{sec:five}
We begin by computing the Higgs self-couplings in terms of
U(2)-invariant quantities.   First, we use \eq{master} to obtain:
\beqa \label{phiphi}
\Phi^\dagger_{\abar}\Phi_b &=& \half v^2 V_{b\abar}+
vh_k\left[V_{b\abar}\,\Re~q_{k1}+\half\left(\widehat v_b\widehat w^*_{\abar}
q^*_{k2}e^{i\theta_{23}}+\widehat v^*_{\abar}\widehat w_b
q_{k2}e^{-i\theta_{23}}\right)\right]\nonumber \\
&&+\half h_j h_k\left[V_{b\abar}\Re(q_{j1}^* q_{k1})+
W_{b\abar}\Re(q_{j2}^* q_{k2})\right.\nonumber\\
&&\quad\left.
+\widehat v_b\widehat w^*_{\abar} q^*_{j_2}q_{k1}
e^{i\theta_{23}}+\widehat v^*_{\abar}\widehat w_b q^*_{j1}
q_{k2}e^{-i\theta_{23}}\right] \nonumber \\
&& +G^+G^- V_{b\abar}+H^+H^- W_{b\abar}+G^- H^+ \widehat
v^*_{\abar}\widehat w_b+ G^+ H^- \widehat w^*_{\abar}\widehat v_b\,,
\eeqa
where repeated indices are summed over and $j,k=1,\ldots,4$.
We then insert \eq{phiphi} into \eq{genericpot}, and expand out the
resulting expression.  We shall write:
\beq
\mathcal{V}=\mathcal{V}_0+\mathcal{V}_2+\mathcal{V}_3+\mathcal{V}_4\,,
\eeq
where the subscript indicates the overall degree of the fields that appears in
the polynomial expression.  $\mathcal{V}_0$ is a constant of no
significance and $\mathcal{V}_1=0$ by the scalar potential minimum condition.
$\mathcal{V}_2$ is obtained in Appendix~\ref{app:three}.  In this
section, we focus on the cubic Higgs self-couplings that reside in
$\mathcal{V}_3$ and the quartic Higgs self-couplings that reside in
$\mathcal{V}_4$.

Using \eqs{invariants}{pseudoinvariants}, one can
express $\mathcal{V}_3$ and $\mathcal{V}_4$ in terms of the invariants
($Y_1$, $Y_2$ and $Z_{1,2,3,4}$) and pseudo-invariants ($Y_3$, $Z_{5,6,7}$).
In the resulting expressions, we have eliminated $Y_1$ and $Y_3$ by the
scalar potential minimum conditions [\eq{hbasismincond}].
The cubic Higgs couplings are governed by the following terms of the
scalar potential:
\beqa \label{hcubic}
\mathcal{V}_3&=&\half v\, h_j h_k h_\ell
\biggl[q_{j1}q^*_{k1}\Re(q_{\ell 1}) Z_1
+q_{j2}q^*_{k2}\,\Re(q_{\ell 1})(Z_3+Z_4) +
\Re(q^*_{j1} q_{k2}q_{\ell 2}Z_5\,
e^{-2i\theta_{23}}) \nonumber \\
&&\qquad\qquad\qquad\quad
+\Re\left([2q_{j1}+q^*_{j1}]q^*_{k1}q_{\ell 2}Z_6\,e^{-i\theta_{23}}\right)
+\Re (q_{j2}^*q_{k2}q_{\ell 2}Z_7\,e^{-i\theta_{23}})
\biggr]\nonumber \\
&& \hspace{-0.2in} +v\,h_k
G^+G^-\biggl[\Re(q_{k1})Z_1+\Re(q_{k2}\,e^{-i\theta_{23}}Z_6)\biggr]\nonumber\\
&&\hspace{-0.2in}+v\,h_k H^+H^-\biggl[\Re(q_{k1})Z_3+\Re(q_{k2}\,e^{-i\theta_{23}}Z_7)\biggr]
\nonumber \\
&& \hspace{-0.2in} +\half v \,h_k\biggl\{G^-H^+\,e^{i\theta_{23}}
\left[q^*_{k2} Z_4
+q_{k2}\,e^{-2i\theta_{23}}Z_5+2\Re(q_{k1})Z_6
\,e^{-i\theta_{23}}\right]+{\rm h.c.}\biggr\}\,,
\eeqa
where there is an implicit sum over the repeated indices\footnote{Note that the
sum over repeated indices can be rewritten by appropriately symmetrizing
the relevant coefficients.  For example, $\sum_{jk\ell} g_{jk\ell}\,h_j h_k
h_\ell= \sum_{j\leq k\leq\ell} h_j h_k h_\ell\,[g_{jk\ell}+{\rm perm}]$,
where ``perm'' is an instruction to add additional terms (as needed) 
such that the indices $j$, $k$ and $\ell$ appear in all possible 
\textit{distinct} permutations.\label{fn1}}
$j$, $k$, $\ell=1,2,3,4$.  Since the neutral Goldstone boson field is
denoted by $h_4\equiv G^0$, we can extract the cubic couplings of
$G^0$ by using $q_{41}=i$ and $q_{42}=0$.  The only cubic
Higgs--$G^0$ couplings
that survive are:
\beqa \label{G3}
\mathcal{V}_{3G}&=&
\half v\,\sum_{k=1}^3\sum_{\ell =1}^3\,G^0
h_k h_\ell\biggl[\Im( q_{k2} q_{\ell 2}Z_5\,e^{-2i\theta_{23}})+
2q_{k1}\,\Im\left(q_{\ell 2} Z_6\,e^{-i\theta_{23}}\right)\biggr]
\nonumber \\
&& +\half v\,\sum_{\ell=1}^3\,G^0 G^0 h_\ell\biggl[q_{\ell 1}
Z_1+\Re(q_{\ell 2}Z_6e^{-i\theta_{23}})\biggr]\,,
\eeqa
where we have used the fact that $q_{j1}$ is real for $j=1,2,3$.

At the end of the \sect{sec:four}, we noted that $H^+$ is a
pseudo-invariant field.  However $e^{i\theta_{23}}H^+$ is a U(2)-invariant
field [see \eqs{qtrans}{hplustrans}], and it is precisely this
combination that shows up in \eq{hcubic}.  Moreover, 
as shown in \sect{sec:four}, the $q_{k\ell}$
and the quantities $Z_5\,e^{-2i\theta_{23}}$, $Z_6\,e^{-i\theta_{23}}$
and $Z_7\,e^{-i\theta_{23}}$ are also invariant with respect
to flavor-U(2) transformations.  Thus, we conclude that
\eq{hcubic} is U(2)-invariant as required.

The quartic Higgs couplings are governed by the following terms of the
scalar potential:
\beqa
&& \hspace{-0.18in}
\mathcal{V}_4=\eighth h_j h_k h_l h_m
\biggl[q_{j1}q_{k1}q^*_{\ell 1}q^*_{m1}Z_1
+q_{j2}q_{k2}q^*_{\ell 2}q^*_{m2}Z_2\nonumber \\[5pt]
&&\quad\,\,+2q_{j1}q^*_{k1}q_{\ell 2}q^*_{m2}(Z_3+Z_4)+2\Re(q^*_{j1}q^*_{k1}q_{\ell 2}q_{m2}Z_5\,e^{-2i\theta_{23}})\nonumber \\[5pt]
&&\quad\,\,+4\Re(q_{j1}q^*_{k1}q^*_{\ell 1}q_{m2}Z_6\,e^{-i\theta_{23}})
+4\Re(q^*_{j1}q_{k2}q_{\ell
  2}q^*_{m2}Z_7\,e^{-i\theta_{23}})\biggr]\nonumber \\[5pt]
&&  +\half h_j h_k G^+ G^-\biggl[q_{j1}q^*_{k1} Z_1 + q_{j2}q^*_{k2}Z_3
+2\Re(q_{j1}q_{k2}Z_6\,e^{-i\theta_{23}})\biggr]
 \nonumber \\[5pt]
&&  +\half h_j h_k H^+ H^-\biggl[q_{j2}q^*_{k2} Z_2 + q_{j1}q^*_{k1}Z_3
+2\Re(q_{j1}q_{k2}Z_7\,e^{-i\theta_{23}})\biggr] \nonumber \\
&& +\half h_j h_k\biggl\{G^- H^+\,e^{i\theta_{23}} \left[q_{j1}q^*_{k2}Z_4
+ q^*_{j1}q_{k2}Z_5\,e^{-2i\theta_{23}}+q_{j1}q^*_{k1}Z_6
\,e^{-i\theta_{23}}\right.\nonumber\\
&&\quad\,\,\left.+q_{j2}q^*_{k2}Z_7\,e^{-i\theta_{23}}\right]+{\rm h.c.}
\biggr\}+\half Z_1 G^+ G^- G^+ G^- +\half Z_2 H^+H^- H^+ H^-\nonumber \\[5pt]
&&+ (Z_3+Z_4)G^+ G^- H^+ H^- +\half Z_5 H^+H^+G^-G^- +\half Z_5^* H^-H^- G^+G^+\nonumber \\[5pt]
&&+ G^+G^-(Z_6 H^+ G^-\! + Z_6^* H^- G^+) + H^+H^-(Z_7 H^+ G^-\! + Z_7^* H^- G^+)
\,,\label{scalpot}\eeqa
where there is an implicit sum over the repeated indices
$j$, $k$, $\ell$, $m=1,2,3,4$.
One can check the U(2)-invariance of
$\mathcal{V}_4$ by noting that $Z_5 H^+ H^+$, $Z_6 H^+$ and $Z_7 H^+$ are
U(2)-invariant combinations.\footnote{It is instructive to
write, \textit{e.g.}, $Z_6 H^+ = (Z_6\,e^{-i\theta_{23}})
(H^+\,e^{i\theta_{23}})$, \textit{etc.} to exhibit the well-known
U(2)-invariant combinations.}
It is again straightforward to isolate the
quartic couplings of the neutral Goldstone boson ($h_4\equiv G^0$):
\beqa
&&\mathcal{V}_{4G}=\eighth q_{11}^4 Z_1 G^0 G^0 G^0 G^0
+\half \Im(q_{m2}Z_6\,e^{-i\theta_{23}})\,G^0 G^0 G^0 h_m
 \nonumber\\[5pt]
&&\,\,\,
+\quarter G^0 G^0 h_\ell h_m\biggl[q_{\ell 1}q_{m1}Z_1
+q_{\ell 2}q^*_{m2}(Z_3+Z_4)-\Re(q_{\ell 2} q_{m2} Z_5 e^{-2i\theta_{23}})\biggr.\nonumber\\[5pt]
&&\,\,\,\quad\,\,\biggl.
+2q_{\ell 1}\Re(q_{m2} Z_6 e^{-i\theta_{23}})\biggr]+\half G^0 h_k h_\ell h_m
\biggl[q_{k1}\Re(q_{\ell 2} q_{m2} Z_5 e^{-2i\theta_{23}})\biggr.\nonumber \\[5pt]
&&\,\,\,\quad\,\,\biggl.
+q_{k1} q_{\ell 1}\Re(q_{m2} Z_6 e^{-i\theta_{23}})
+\Re(q_{k2} q_{\ell 2}q^*_{m2} Z_7 e^{-i\theta_{23}})\biggr] \nonumber \\[5pt]
&&\,\,\,
- \Im(q_{m2} Z_6 \,e^{-i\theta_{23}})\,G^+G^-G^0 h_m
- \Im(q_{m2} Z_7 \,e^{-i\theta_{23}})\,H^+H^-G^0 h_m
 \nonumber \\[5pt]
&&\,\,\,
+\half i\,G^0 h_m \biggl\{G^-H^+e^{i\theta_{23}}\left[q^*_{m2}Z_4-
q_{m2}Z_5 e^{-2i\theta_{23}}\right]+{\rm h.c.}\biggr\}
\nonumber \\[5pt]
&&\,\,\,
+\half Z_1 G^0G^0G^+G^- +\half Z_3 G^0G^0H^+H^- \nonumber\\
&&\,\,\,+\half Z_6 G^0 G^0G^-H^+
+\half Z_6^* G^0 G^0G^+H^-,
\eeqa
where the repeated indices $k$, $\ell$, $m=1,2,3$ are summed over.

The Feynman rules are obtained by multiplying the relevant terms of
the scalar potential by $-iS$, where the symmetry factor 
$S=\prod_i n_i!$ for an the interaction term that possesses $n_i$ 
identical particles of type $i$.
Explicit forms for the $q_{k\ell}$ in terms of the invariant mixing angles
$\theta_{12}$ and $\theta_{13}$ are displayed in Table~\ref{tab1}.
For example, the Feynman rule for the cubic self-coupling of the lightest
neutral Higgs boson is given by $ig(h_1 h_1 h_1)$ where
\beqa
g(h_1 h_1 h_1)&=& -3v\biggl[Z_1 c_{12}^3 c_{13}^3
+(Z_3+Z_4)c_{12} c_{13} |s_{123}|^2
+c_{12} c_{13}\,\Re(s_{123}^2 Z_5\,e^{-2i\theta_{23}})
\nonumber \\ && \qquad
-3 c_{12}^2 c_{13}^2\,\Re(s_{123} Z_6\,e^{-i\theta_{23}})
-|s_{123}|^2\,\Re(s_{123} Z_7\, e^{-i\theta_{23}})
\biggr]\,,
\eeqa
where $s_{123}\equiv s_{12}+ic_{12}s_{13}$.
Similarly, the Feynman rule for the quartic self-coupling of the lightest
neutral Higgs boson is given by $ig(h_1 h_1 h_1 h_1)$ where
\beqa
g(h_1 h_1 h_1 h_1)&=& -3\biggl[Z_1 c_{12}^4 c_{13}^4 + Z_2|s_{123}|^4
+2(Z_3+Z_4)c_{12}^2 c_{13}^2 |s_{123}|^2\nonumber\\
&&\qquad + 2c_{12}^2 c_{13}^2\,\Re(s_{123}^2 Z_5\, e^{-2i\theta_{23}})
- 4 c_{12}^3 c_{13}^3\,\Re(s_{123} Z_6\, e^{-i\theta_{23}}) \nonumber \\
&& \qquad -4 c_{12}c_{13}|s_{123}|^2\,\Re(s_{123}Z_7\, e^{-i\theta_{23}})
\biggr]\,.
\eeqa

We turn next to the coupling of the Higgs bosons to the gauge bosons.
These arise from the Higgs boson kinetic energy terms when the
partial derivatives are replaced by the gauge covariant derivatives:
$\mathscr{L}_{\rm KE}=D^\mu\Phi_{\abar}^\dagger D_\mu\Phi_a$.
In the SU(2)$\ls{\rm L}\times$U(1) electroweak gauge theory,
\beq \label{covder}
D_\mu\Phi_a=\left(\begin{array}{c} \displaystyle
\partial_\mu\Phi^+_a+\left[\frac{ig}{c_W}\left(\half-s_W^2\right)Z_\mu
+ieA_\mu\right]\Phi^+_a+\frac{ig}{\sqrt{2}}W_\mu^+\Phi^0_a \\[8pt]
\displaystyle \partial_\mu\Phi^0_a-\frac{ig}{2c_W}Z_\mu\Phi_a^0+
\frac{ig}{\sqrt{2}}W_\mu^-\Phi^+_a\end{array}\right)\,,
\eeq
where $s_W\equiv\sin\theta_W$ and $c_W\equiv\cos\theta_W$.  Inserting
\eq{covder} into $\mathscr{L}_{\rm KE}$ yields the Higgs boson--gauge boson
interactions in the generic basis.  Finally, we use \eq{master} to
obtain the interaction Lagrangian of the gauge bosons with the
physical Higgs boson mass-eigenstates.  The resulting interaction
terms are:
\beqa
\mathscr{L}_{VVH}&=&\left(gm_W W_\mu^+W^{\mu\,-}+\frac{g}{2c_W}
m_Z Z_\mu Z^\mu\right)\Re(q_{k1}) h_k \nonumber \\[5pt]
&&
+em_WA^\mu(W_\mu^+G^-+W_\mu^-G^+)
-gm_Zs_W^2 Z^\mu(W_\mu^+G^-+W_\mu^-G^+)
\,,\label{VVH} \\[10pt]
\mathscr{L}_{VVHH}&=&\left[\quarter g^2  W_\mu^+W^{\mu\,-}
+\frac{g^2}{8c_W^2}Z_\mu Z^\mu\right]
\Re(q_{j1}^* q_{k1}+q_{j2}^* q_{k2})\,h_j h_k \nonumber \\[5pt]
&& +\biggl[\half g^2 W_\mu^+ W^{\mu\,-}+e^2A_\mu A^\mu+\frac{g^2}{c_W^2}\left(\half -s_W^2\right)^2Z_\mu Z^\mu \biggr.\nonumber \\[5pt]
&& \qquad \biggl.
+\frac{2ge}{c_W}\left(\half -s_W^2\right)A_\mu Z^\mu\biggr](G^+G^-+H^+H^-)
\nonumber \\[5pt]
&&
+\biggl\{
\left(\half eg A^\mu W_\mu^+ -\frac{g^2s_W^2}{2c_W}Z^\mu W_\mu^+\right)
(q_{k1}G^-+q_{k2}\,e^{-i\theta_{23}}H^-)h_k +{\rm h.c.}\biggr\}
\,, \nonumber\\
&&\phantom{a}\label{VVHH}
\eeqa
and
\beqa
\hspace{-1.3in}
\mathscr{L}_{VHH}=\frac{g}{4c_W}\,\Im(q_{j1}q^*_{k1}+q_{j2}q^*_{k2})
Z^\mu h_j\ddel_\mu h_k &&\nonumber \\[5pt]
&& \hspace{-2.5in}
-\half g\biggl\{iW_\mu^+\left[q_{k1} G^-\ddel\lsup{\,\mu} h_k+
q_{k2}e^{-i\theta_{23}}H^-\ddel\lsup{\,\mu} h_k\right]
+{\rm h.c.}\biggr\}\nonumber \\[5pt]
&& \hspace{-2.5in}
+\left[ieA^\mu+\frac{ig}{c_W}\left(\half -s_W^2\right)
Z^\mu\right](G^+\ddel_\mu G^-+H^+\ddel_\mu H^-)\,,\label{VHH}
\eeqa
where the repeated indices $j,k=1,\ldots,4$ are summed over.  The
neutral Goldstone boson interaction terms can be ascertained by taking
$h_4\equiv G^0$:
\beqa
\!\!\mathscr{L}_{VG}&\!=\!&\left[\quarter g^2  W_\mu^+W^{\mu\,-}
+\frac{g^2}{8c_W^2}Z_\mu Z^\mu\right]G^0 G^0 \nonumber \\[6pt]
&&
+\biggl\{
\half ieg A^\mu W_\mu^+ G^- G^0
-\frac{ig^2s_W^2}{2c_W}Z^\mu W_\mu^+
G^-G^0 +{\rm h.c.}\!\biggr\}
\nonumber \\[6pt]
&& 
+\frac{g}{2c_W} \Re(q_{k1}) Z^\mu G^0\ddel_\mu h_k
+\half g\left(W_\mu^+G^-\ddel\lsup{\,\mu}G^0+W_\mu^-G^+\ddel\lsup{\,\mu}G^0
\right)\,.
\eeqa

Once again, we can verify by inspection that the Higgs boson--vector
boson interactions are U(2)-invariant.  Moreover, one can derive
numerous relations among these couplings using the properties of the
$q_{k\ell}$.  In particular,  
\eqst{unitarity1}{epsid} imply the following relations among the 
Higgs boson--vector boson 
couplings~\cite{Gunion:1997aq,Grzadkowski:1999ye,cpcarlos}:
\beqa
&& g(ZZh_j)= m_Z\sum_{k,\ell=1}^3
\,\epsilon\ls{jk\ell}\, g(Zh_k h_\ell)\,,\qquad (j=1,2,3)\,,
\label{id1}\\[6pt]
&&\sum_{k=1}^3\,[g(VVh_k)]^2 = \frac{g^2 m_V^4}{m^2_W}\,,
\qquad\qquad\quad V=W^\pm~{\rm or}~Z\,,\label{id2}
\\[6pt]
&&\!\!\!\!\sum_{1\leq j<k\leq 3}
\,[g(Zh_j h_k)]^2 = \frac{g^2}{4c_W^2}\,,\label{id3}
\eeqa
\beq
g(ZZh_j)g(ZZh_k)+4m_Z^2\sum_{\ell=1}^3\,g(Zh_j h_\ell)g(Zh_k h_\ell)=
\frac{g^2 m_Z^2}{c_W^2}\,\delta_{jk}\,,\label{id4}
\eeq
where the Feynman rules for the $VVh_k$ and $Zh_j h_k$ vertices
are given by $ig^{\mu\nu}\,g(VVh_k)$ and $(p_k-p_j)^\mu\,g(Zh_j h_k)$,
respectively,
and the four-momenta $p_j$, $p_k$ of the neutral 
Higgs bosons $h_j$, $h_k$ point into the
vertex.\footnote{The Feynman rule for the $ZZh_k$ vertex 
includes a factor of two relative to the coefficient of the
corresponding term in $i\mathscr{L}_{VVH}$ due to the identical
$Z$ bosons. The Feynman rule for the
$Zh_j h_k$ vertex is given by $\half(g/c_W)\Im[q_{j1}q_{k1}^*+
q_{j2}q_{k2}^*](p_k-p_j)^\mu$.  Here, the factor of two relative to
the corresponding term in \eq{VHH} arises from the implicit double sum over
$j$ and $k$ in the Lagrangian.  Note that the rule for the $Zh_j h_k$
vertex does not depend on the ordering of $j$ and $k$.}
Note that \eq{id4} holds for $j,k=1,2,3,4$.

\section{Higgs Couplings to Fermions}
\label{sec:six}
The most general Yukawa
couplings of Higgs bosons to fermions yield neutral Higgs-mediated
flavor-changing neutral currents at
tree-level~\cite{Glashow:1976nt,Georgi,Paschos:1976ay}.  Typically, these
couplings are in conflict with the experimental bounds on FCNC
processes.  Thus, most model builders impose restrictions on the
structure of the Higgs fermion couplings to avoid the potential for
phenomenological disaster.  However, even in the case of the most general
Higgs-fermion couplings, parameter regimes exist where FCNC effects
are sufficiently under control.  In the absence of new physics beyond
the 2HDM, such parameter regimes are unnatural (but can be arranged
with fine-tuning).  In models such as the minimal supersymmetric
extension of the Standard Model (MSSM), supersymmetry-breaking effects
generate all possible  Higgs-fermion Yukawa couplings allowed by
electroweak gauge invariance.  Nevertheless, the FCNC effects are
one-loop suppressed and hence phenomenologically acceptable.

In this section, we will study the basis-independent description of
the Higgs-fermion interaction.  In a generic basis,
the so-called type-III model~\cite{typeiii2,typeiii1,davidson} of Higgs fermion
interactions is governed by the following interaction Lagrangian:
\beq \label{ymodeliii0}
-\mathscr{L}_{\rm Y}
=\anti \qlo\, \wtil\Phi_1\eiuo\,  \uro +\anti Q_L^0\,\Phi_1(\eido)^\dagger
\,\dro + \anti \qlo\, \wtil\Phi_2\eiiuo\, \uro 
+\anti \qlo\, \Phi_2(\eiido)^\dagger \,\dro +{\rm h.c.}\,,
\eeq
where $\Phi_{1,2}$ are the Higgs doublets and $\wtil\Phi_i\equiv
i\sigma_2 \Phi^*_i$.
As in section \ref{smyuk}, $\qlo $, $\uro $,
$\dro $ denote the interaction basis quark fields, which
are vectors in the quark
flavor space, and $\eta_1^{Q,0}$ and $\eta_2^{Q,0}$ ($Q=U\,,\,D$)
are four $3\times 3$ matrices in quark flavor space.
We have omitted the leptonic couplings in \eq{ymodeliii0};
these are obtained from \eq{ymodeliii0}
with the obvious substitutions $Q_L^0\to L_L^0$ and $D_R^0\to E_R^0$.  
(In the absence of right-handed neutrinos, there is no analog of $U_R^0$.)

The derivation of the couplings of the physical Higgs bosons with the
quark mass-eigenstates was given in \Ref{davidson} in the case of a
CP-conserving Higgs sector.  Here, we generalize that discussion to
the more general case of a CP-violating Higgs sector.  The first step
is to identify the quark mass-eigenstates.  This is accomplished by
setting the scalar fields to their vacuum expectation values and
performing unitary transformations of the left and right-handed up and
down quark multiplets such that the resulting quark mass matrices
are diagonal with non-negative entries.  In more detail, we define
left-handed and right-handed quark mass-eigenstate fields according to \eq{biunitary}, with the CKM matrix $K\equiv V_L^U
V_L^{D\,\dagger}$ as before.  In addition, we introduce ``rotated'' Yukawa coupling
matrices:
\beq
\eiua\equiv V_L^U \,\eiuoa\, V_R^{U\,\dagger}\,,\qquad\qquad
{\eida}\equiv V_R^D\, {\eidoa}\, V_L^{D\,\dagger}\,.
\eeq
We then rewrite \eq{ymodeliii0} in terms of the quark mass-eigenstate
fields and the transformed couplings:
\beq \label{ymodeliii}
-\mathscr{L}_{\rm Y}=\overline U_L \widetilde\Phi_{\bar{a}}\eta^U_a U_R+\overline D_L K^\dagger\widetilde\Phi_{\bar{a}}\eta^U_a U_R
+\overline U_L K\Phi_a \eta^{D\,\dagger}_{\bar{a}}D_R +\overline D_L\Phi_a \eta^{D\,\dagger}_{\bar{a}}D_R +{\rm h.c.}\,,
\eeq
where $\eta^Q_a\equiv (\eta^Q_1\,,\,\eta^Q_2)$ is a (basis-dependent) vector in U(2) space. If we assume that there is no basis in which can write either $\eta^U_2=\eta^D_2=0$ (Type I) or
$\eta^U_1=\eta^D_2=0$ (Type II), then this ``Type III'' 2HDM Yukawa Lagrangian.  One could also write it in a more compact form:
\beq \label{ymodeliiicute}
-\mathscr{L}_{\rm Y}=\overline\qlcal \widetilde\Phi_{\bar{a}}\eta^U_a \mathcal{U}_R
+\overline \qlcal\Phi_a \eta^{D\,\dagger}_{\bar{a}}D_R +{\rm h.c.}\,,
\eeq
where $\mathcal{U}\equiv K^\dagger U,$ and $\mathcal{Q}_L =\binom{\mathcal{U}}{D}_L$.

Under a U(2)-transformation of the scalar fields, $\eta^Q_a\to U_{a\bbar}\eta^Q_b$ and 
$\eta^{Q\,\dagger}_\abar\to\eta^{Q\,\dagger}_\bbar U^\dagger_{b\abar}$. 
Hence, the Higgs--quark Lagrangian is
U(2)-invariant.  We can construct basis-independent
couplings following the strategy of \sect{sec:three}
by transforming to the Higgs basis.  Using \eq{hbasis}, we can rewrite
\eq{ymodeliii} in terms of Higgs basis scalar fields:
\beq \label{hbasisymodeliii}
-\mathscr{L}_{\rm Y}=\overline \qlcal (\tilde H_1\kappa^U+\tilde H_2\rho^U) \mathcal{U}_R
+\overline \qlcal (H_1\kappa^{D\,\dagger}+ H_2\rho^{D\,\dagger}) D_R +{\rm h.c.}\,,
\eeq
where
\beq \label{kapparho}
\kappa^{Q}\equiv \widehat v^*_{\abar}\,\eta^{Q}_a\,,\qquad\qquad
\rho^{Q}\equiv \widehat w^*_{\abar}\,\eta^{Q}_a\,.
\eeq
Inverting \eq{kapparho} yields:
\beq \label{kapparhoinv}
\eta^Q_a=\kappa^Q\widehat v_a+\rho^Q\widehat w_a\,.
\eeq
Under a U(2) transformation, $\kappa^Q$ is invariant, whereas $\rho^Q$
is a pseudo-invariant that transforms as:
\beq \label{rhotrans}
\rho^Q\to (\det U)\rho^Q\,.
\eeq
By construction, $\kappa^U$ and $\kappa^D$ are proportional to the
(real non-negative) diagonal quark mass matrices $M_U$ and $M_D$,
respectively.  In particular, the $M_Q$ 
are obtained by inserting \eq{higgsvevs} into
\eq{hbasisymodeliii}.  As in the SM (see \eqs{diagumass}{diagdmass}), we find:
\beqa
M_U=\frac{v}{\sqrt{2}}\kappa^U&=&{\rm diag}(m_u\,,\,m_c\,,\,m_t)
= V_L^U M_U^{0} V_R^{U\,\dagger}\,, \\[6pt] 
M_D=\frac{v}{\sqrt{2}}\kappa^{D\,\dagger}&=&{\rm diag}(m_d\,,\,m_s\,,\,m_b)
= V_L^D M_D^{0} V_R^{D\,\dagger}\,,
\eeqa
where $M_U^0\equiv (v/\sqrt{2}) \widehat v^*_{\abar}\,\eta^{U,0}_a$
and $M_D^0\equiv (v/\sqrt{2}) \widehat v_a\,\eta^{D,0\,\dagger}_{\abar}$.
That is, we have chosen the unitary matrices $V^U_L$, $V^U_R$,
$V^D_L$ and $V^D_R$ 
such that $M_D$ and $M_U$ are diagonal matrices with
real non-negative entries.\footnote{This can be
accomplished by the singular-value decompositions of the
complex matrices $M_U^0$ and $M_D^0$~\cite{horn}.}
In contrast, the $\rho^Q$ are independent complex $3\times 3$
matrices.

In order to obtain the interactions of the physical Higgs bosons with
the quark mass-eigenstates, we do not require the intermediate step
involving the Higgs basis.  Instead, we insert \eq{master} into \eq{ymodeliii} and obtain:
\beqa \label{hffu2}
&& \hspace{-0.5in}
-\mathscr{L}_Y = \frac{1}{v}\overline D
\biggl\{M_D (q_{k1} P_R + q^*_{k1} P_L)+\frac{v}{\sqrt{2}}
\left[q_{k2}\,[e^{i\theta_{23}}\rho^D]^\dagger P_R+
q^*_{k2}\,e^{i\theta_{23}}\rho^D P_L\right]\biggr\}Dh_k \nonumber \\[5pt]
&&\quad  \hspace{-0.2in}
+\frac{1}{v}\overline U
\biggl\{M_U (q_{k1} P_L + q^*_{k1} P_R)+\frac{v}{\sqrt{2}}
\left[q^*_{k2}\,e^{i\theta_{23}}\rho^U P_R+
q_{k2}\,[e^{i\theta_{23}}\rho^U]^\dagger P_L\right]\biggr\}U h_k
\nonumber \\[5pt]
&&\quad \hspace{-0.3in}
+\biggl\{\overline U\left[K[\rho^D]^\dagger 
P_R-[\rho^U]^\dagger KP_L\right] DH^+ \biggr.\nonumber\\[5pt]
&&\qquad \biggl.
+\frac{\sqrt{2}}{v}\,\overline U\left[K\mdd P_R-\mud KP_L\right] DG^+
+{\rm h.c.}\biggr\}\,,
\eeqa
where $k=1,\ldots\,4$.  
Since $e^{i\theta_{23}}\rho^Q$ and $[\rho^Q]^\dagger H^+$ are U(2)-invariant,
it follows that
\eq{hffu2} is a basis-independent representation of the Higgs--quark
interactions.

The neutral Goldstone boson interactions ($h_4\equiv G^0$)
are easily isolated:
\beq \label{YG}
-\mathscr{L}_{YG}=\frac{i}{v}\left[\overline DM_D\gamma\ls{5}D
-\overline UM_U\gamma\ls{5}U\right]G^0\,.
\eeq
In addition, since the $q_{k1}$ are real for $k=1,2,3$, it 
follows that the piece of 
the neutral Higgs--quark couplings proportional to the quark mass
matrix is of the form $v^{-1}\overline Q \,M_Q \,q_{k1}\, Q\,h_k$.

The couplings of the neutral Higgs bosons to quark pairs
are generically CP-violating as a result
of the complexity of the $q_{k2}$ and the fact that the matrices
$e^{i\theta_{23}}\rho^Q$ are not generally purely real orpurely imaginary.  (Invariant conditions for the CP-invariance of these couplings
are given in Chapter \ref{cpcoch}).  
\Eq{hffu2} also exhibits Higgs-mediated
FCNCs at tree-level due to the $\rho^Q$ not being
flavor-diagonal.  Thus, for a phenomenologically
acceptable theory, the off-diagonal elements of $\rho^Q$ must be
small.

\chapter{Special Limits:  CP Conservation, The 2HDM with $Z_6 =0 $, and Custodial Symmetry\label{custchapter}}
In the limit of $g' \rightarrow 0$, the bosonic sector of the Standard Model has a global $SU(2)_L\times SU(2)_R$ symmetry. After electroweak symmetry breaking, this symmetry reduces to $SU(2)_{L+R} \equiv SU(2)_V$, known as the ``custodial symmetry.''  Violations of custodial symmetry lead to corrections to the relation $m_W^2 = m_Z^2 cos^2\theta_W.$  This custodial limit is a more restrictive case of CP conservation. In the 3-generation model, we know that CP violation must exist in the CKM matrix, so a 2HDM with no CP violation is unrealistic.  Thus, for the purposes of this chapter, when we refer to the ``CP conserving" model, we mean that the bosonic sector and the neutral Higgs-Quark couplings conserve CP. 

The CP-conserving limit of the basis-independent 2HDM has been analyzed in \cite{davidson} and \cite{haberoneil}\footnote{By CP conservation, we mean that the neutral scalars are CP eigenstates.  The complex phase in the CKM matrix induces CP violation in the charged Higgs interactions.}. In this chapter, we build on the results of previous work and define two cases of CP conservation that are distinguishable based on the parameter $\zsix$. In section \ref{indepsec}, we derive basis-independent expressions for the masses and mixing angles of the Higgs particles in the CP-conserving limit. In section \ref{realsec}, we make contact with the existing literature by working in the real basis and relating the Higgs fields to the parameters $\cbma$ and $\sbma$.  

We then analyze the scenario in which $Z_6 = 0$.  We will start by deriving the mass matrix and invariant expressions for the mixing angles in section \ref{zsixzero}, and then discuss the effects of CP violation or conservation in the scalar couplings.  We end with a discussion of the special case $Z_6 = Z_7 =0$ in section \ref{sec:herquetcpcons}.

In section \ref{custsec}, we apply these results to analyse the custodial limit of the 2HDM.  The use of the basis-independent formalism allows us to clarify aspects of custodial symmetry which have been made unnecessarily complicated in the literature.  We derive unambiguous conditions for custodial symmetry in the scalar sector and the Higgs-Quark sector and discuss the resulting implications for the scalar masses.
 
\section{The CP-Conserving 2HDM \label{cpcoch}}
\subsection{Basis-Independent Analysis of the CP-Conserving Limit\label{indepsec}}
In the CP-conserving limit, we impose CP-invariance on 
all bosonic couplings of the Higgs bosons and the fermionic couplings of the neutral Higgs bosons.  (We will ignore, for now, the CP violation in charged Higgs-quark interactions that arises from the complexity of the CKM matrix.)
The requirement of a CP-conserving bosonic sector is
equivalent to the requirement that the scalar potential is explicitly
CP-conserving and that the Higgs vacuum is CP-invariant
(\textit{i.e.}, there is no spontaneous CP-violation).  
Basis-independent conditions for a CP-conserving bosonic sector have been
given in refs.~\cite{cpx2,davidson,cpbasis,cpx}.  In \Ref{davidson}, these
conditions were recast into the following form.  The bosonic sector is
CP-conserving if and only if:\footnote{Since the scalar potential
minimum conditions imply that $Y_3=-\half Z_6 v^2$, no separate
condition involving $Y_3$ is required.}  
\beq \label{cpoddinv}
\Im[Z_6 Z_7^\ast]=\Im[Z_5^*Z_6^2]=\Im[Z_5^*(Z_6+Z_7)^2]=0\,.  
\eeq 
\Eq{cpoddinv} is equivalent to the requirement that
\beq \label{cpang}
\sin 2(\theta_5-\theta_6)=
\sin 2(\theta_5-\theta_7)=\sin(\theta_6-\theta_7)=0\,,
\eeq
where $\theta_5$ and $\theta_6$ are defined in \eq{app:invang} 
and $\theta_7\equiv\arg Z_7$ 
(note that $\theta_5$ is defined modulo
$\pi$ and $\theta_6$ and $\theta_7$ are defined modulo $2\pi$).

One can explore the consequences of CP-invariance by studying the
pattern of Higgs couplings and the structure of the neutral Higgs
boson squared-mass matrix [\eq{matrix33}].  
The tree-level couplings of $\go$ are
CP-conserving, even in the general CP-violating 2HDM.  In particular,
the couplings $\go\go\go$, $\go G^+G^-$, $\go H^+ H^-$
and $ZZ\go$ are absent.  Moreover, \eq{YG} implies that $\go$
possesses purely pseudoscalar couplings to the fermions.  Hence,
$\go$ is a CP-odd scalar, independently of the 
structure of the scalar potential.
We can therefore use the couplings of $\go$ to the neutral Higgs bosons as a
probe of the CP-quantum numbers of these states.  
The analysis of the neutral
Higgs boson squared-mass matrix (which does not depend on $Z_7$)
simplifies significantly when
$\Im[Z_5^* Z_6^2]=0$.  One can then choose a basis
where $Z_5$ and $Z_6$ are simultaneously real, in which case
the scalar squared-mass matrix decomposes into 
diagonal block form.  The upper 
$2\times 2$ block can be diagonalized analytically and yields the
mass-eigenstates $\hl$ and $\hh$ (with $\mhl\leq\mhh$).  The lower $1\times 1$
block yields the mass-eigenstate $\ha$.  If all the conditions of
\eqs{cpoddinv}{zrho} are satisfied, then the neutral Higgs boson
mass-eigenstates are also states of definite CP quantum number.
We shall demonstrate below that 
$\hl$ and $\hh$ are CP-even scalars and $\ha$ is a CP-odd scalar.

Since $Z_6\neq 0$ by assumption, \eq{tan2phi} yields
$\sin\phi\cos\phi=0$, and
\eq{imz56f} implies that either some of the neutral
Higgs boson masses are degenerate
or $s_{13}s_{12}c_{12}=0$.\footnote{Since
$Z_6\neq 0$, one can use \eqs{tan13}{tan213} to show
that $c_{13}\neq 0$.}  
In the case of degenerate masses, some of 
the invariant angles are not well defined, since any linear
combination of the degenerate states is also a mass-eigenstate.
Hence, the degenerate case must be treated separately.  In what
follows, we shall assume that all three neutral Higgs boson masses are
non-degenerate.  Note that if
$\sin\phi=0$, then \eq{phieq} yields
$s_{13}=0$, whereas if $\cos\phi=0$, then \eq{m2m1} yields
$\sin 2\theta_{12}=0$.%
\footnote{The same constraints are obtained by imposing the
requirement of CP-conserving Higgs couplings.  In particular,
the existence of a $\go h_k h_k$ coupling would imply that
$h_k$ is a state of mixed CP-even and CP-odd components.  All such
couplings must therefore be absent in the CP-conserving limit.
Using the results of \eqs{G3}{cpang} one can easily check that
at least one of these CP-violating couplings is present unless
$s_{13}=\sin\phi=0$ or $\cos\phi=\sin 2\theta_{12}=0$.}
Thus, we shall consider separately the two cases: \beqa
{\rm CP~Case~I:}\phantom{I}\quad \sin\phi=0   &\Longrightarrow& 
\Im(Z_5 e^{-2i\theta_{23}})=\Im(Z_6 e^{-i\theta_{23}})=0\,,\label{sp0}\\
{\rm CP~Case~II:}\quad \cos\phi=0   &\Longrightarrow& 
\Im(Z_5 e^{-2i\theta_{23}})=\Re(Z_6 e^{-i\theta_{23}})=0\,,\label{cp0}
\eeqa
where $\phi \equiv \arg(Z_6 \thetminus)$ is an invariant quantity.
The first case corresponds to the mass ordering $m_{A^0} > m_{H^0}$; the second case to the reverse. By comparing equations (\ref{sp0}) and (\ref{cp0}) with the mass matrix in \eq{mtilmatrix}, one identifies the CP-odd field as the following:
\beq \label{cpodd}
A^0=\begin{cases}
\,\, \Im(\thet H_2^0) \quad& \text{[CP Case I]}\,, \\ \,\,\Re(\thet H_2^0)\quad& \text{[CP Case II]}\,.\end{cases}
\eeq
This is equivalent to the statement that the fields in the Higgs basis transform as follows:
\beqa \label{tran}
\left( \begin{array}{c}H_1 \\ \th H_2 \end{array}\right)&\rightarrow&   \left( \begin{array}{c}H_1^* \\ e^{-i \theta_{23}} H_2^* \end{array}\right) \quad \text{[Case I]}\,,\nonumber\\
 \quad \left( \begin{array}{c}H_1 \\ i\th H_2 \end{array}\right)&\rightarrow&   \left( \begin{array}{c}H_1^* \\ (i\th H_2)^* \end{array}\right)\quad \text{[Case II]}\,.
\eeqa
Alternatively, one can define a CP-transformation on fields in the generic basis using the following \cite{haberoneil}:
\beq \left( \begin{array}{c}H_1 \\  H_2 \end{array}\right) = \left( \begin{array}{cc}\widehat{v}_1^* & \widehat{v}_2^*\\ - \widehat{v}_2 &  \widehat{v}_1  \end{array}\right) \left( \begin{array}{c}\Phi_1 \\ \Phi_2 \end{array}\right).\label{eqnten}\eeq  Substituting \eq{eqnten} into \eq{tran} yields
\beq  \left( \begin{array}{cc}\widehat{v}_1^* & \widehat{v}_2^* \\ - \widehat{v}_2 \thet&  \widehat{v}_1 \thet \end{array}\right) \left( \begin{array}{c}\Phi_1 \\ \Phi_2 \end{array}\right) \rightarrow \left( \begin{array}{cc}\widehat{v}_1 & \widehat{v}_2 \\ \mp \widehat{v}_2^* \thetminus& \pm \widehat{v}_1^* \thetminus \end{array}\right) \left( \begin{array}{c}\Phi_1^* \\ \Phi_2^*\end{array}\right),\eeq
or
\beqa   \left( \begin{array}{c}\Phi_1 \\ \Phi_2 \end{array}\right) &\rightarrow& \left( \begin{array}{cc}\widehat{v}_1 & -\widehat{v}_2^* \thetminus\\  \widehat{v}_2 &  \widehat{v}_1^*\thetminus \end{array}\right) \left( \begin{array}{cc}\widehat{v}_1 & \widehat{v}_2 \\ \mp \widehat{v}_2^* \thetminus&\pm  \widehat{v}_1^* \thetminus \end{array}\right) \left( \begin{array}{c}\Phi_1^* \\ \Phi_2^*\end{array}\right) \nonumber\\
&=& \left( \begin{array}{cc}\widehat{v}_1^2 \pm \widehat{v}_2^{*2}\thetdoub & \widehat{v}_1 \widehat{v}_2  \mp\widehat{v}_2^* \widehat{v}_1^*\thetdoub\\ \widehat{v}_1 \widehat{v}_2 \mp\widehat{v}_2^* \widehat{v}_1^*\thetdoub& \widehat{v}_2^2 \pm \widehat{v}_1^{*2}\thetdoub\end{array}\right)  \left( \begin{array}{c}\Phi_1^* \\ \Phi_2^*\end{array}\right).\eeqa Hence, the ``covariant'' form of \eq{tran} is 
\beq \Phi_a (\vec{x},t) \rightarrow (\widehat{v}_a \widehat{v}_b \pm\thetdoub \widehat{w}_a \widehat{w}_b)\Phi_\bbar^*(-\vec{x},t)  \label{covcons}, \eeq
with the positive (negative) solution corresponding to CP Case I (II). 

As a consistency check, we note that a CP-transformation of the Higgs doublets in the generic basis takes the following form (in the notation of \cite{cpbasis}):
\beq \mathcal{CP} ~\Phi_a(\vec{x},t) ~\mathcal{CP}^{-1} = (U^{CP})_{ab} \Phi^*_\bbar(-\vec{x},t).\eeq
Invariance of the vacuum under $CP$ requires \cite{branco}:
\beq  <\Phi_a>= (U^{CP})_{ab} <\Phi_\bbar>^*. \label{vac}\eeq
Eq.~(\ref{covcons}) indeed satisfies \eq{vac}, with 
\beq (U^{CP})_{ab} = \widehat{v}_a \widehat{v}_b \pm\thetdoub \widehat{w}_a \widehat{w}_b.\eeq 

When the scalar potential is CP-conserving, there always exists a basis in which the $Y$ and $Z$ parameters and the scalar vacuum expectation values are all real-valued. In a generic basis, the quantities  $Y_3$, $Z_5$, $Z_6$ and $Z_7$ are complex.  
For the CP-transformation given in \eq{covcons}, requiring $ \mathcal{CP} ~\mathcal{V} ~\mathcal{CP}^{-1}=  \mathcal{V}$ reproduces the relations in \eq{cpoddinv}.
If all three physical neutral fields couple to CP-even states, eg, $H^+ H^- h_k \neq 0$ and $W^+ W^- h_k \neq 0~\forall~k$, then the scalar sector violates CP.  Otherwise, the field $h_k$ for which all such couplings vanish is CP-odd, and the remaining two fields are CP-even.  In our basis-independent notation, the quantity $q_{k1}$ will be non-zero for the CP-even states and zero for the CP-odd state.  

Assuming that the masses are non-degenerate\footnote{One can investigate separately cases of degenerate masses, in which case not all of the mixing angles are well-defined, which we will not do here for the general case of $Z_6 \neq 0$.  We analyze the degenerate cases for $Z_6 = 0$ in section \ref{zsixzerodeg}.} and that $Z_6$ is non-zero, we find that
\beqa
{\rm If}~h_3~{\rm is~CP~odd}\,,\qquad
s_{13}&=&\Im(Z_6\,e^{-i\theta_{23}})=\sin\phi=0\,\,\,~{\rm [CP~Case~I]},\\
{\rm If}~h_2~{\rm is~CP~odd}\,,\qquad
s_{12}&=&\Re(Z_6\,e^{-i\theta_{23}})=\cos\phi=0\,\,\,~{\rm [CP~Case~IIa]},\\
{\rm If}~h_1~{\rm is~CP~odd}\,,\qquad
c_{12}&=&\Re(Z_6\,e^{-i\theta_{23}})=\cos\phi=0\,\,\,~{\rm [CP~Case~IIb]}.
\eeqa
The values of the $q_{k\ell}$ corresponding to cases I, IIa and IIb
are given in Tables~\ref{tab2}---\ref{tab4}.

\begin{table}[ht!]
\centering
\caption{The U(2)-invariant quantities $q_{k\ell}$ in the
CP-conserving limit.  Case I: $s_{13}=\Im(Z_6\,e^{-i\theta_{23}})=\sin\phi=0$. $G^0$ and $h_3$ are CP-odd; $h_1$ and $h_2$ are CP-even. } \vskip 0.08in
\label{tab2}
\begin{tabular}{|c||c|c|}\hline
$\phaa k\phaa $ &\phaa $q_{k1}\phaa $ & \phaa $\phm q_{k2} \phaa $ \\ \hline
$1$ & $c_{12}$ & $-s_{12}$ \\
$2$ & $s_{12}$ & $\phm c_{12}$ \\
$3$ & $0$ & $\phm i$ \\
$4$ & $i$ & $\phm 0$ \\ \hline 
\end{tabular}
\end{table}
\begin{table}[ht!]
\centering
\caption{The U(2)-invariant quantities $q_{k\ell}$ in the
CP-conserving limit. Case IIa: $s_{12}=\Re(Z_6\,e^{-i\theta_{23}})=\cos\phi=0$. $G^0$ and $h_2$ are CP-odd; $h_1$ and $h_3$ are CP-even. }
\vskip 0.08in
\label{tab3}
\begin{tabular}{|c||c|c|}\hline
$\phaa k\phaa $ &\phaa $q_{k1}\phaa $ & \phaa $\phm q_{k2} \phaa $ \\ \hline
$1$ & $c_{13}$ & $ -is_{13} $\\
$2$ & $0$ & $\phm 1$ \\
$3$ & $s_{13}$ & $\phm i c_{13}$ \\
$4$ & $i$ & $\phm 0$ \\ \hline 
\end{tabular}
\end{table}
\begin{table}[ht!]
\centering
\caption{The U(2)-invariant quantities $q_{k\ell}$ in the
CP-conserving limit. Case IIb: $c_{12}=\Re(Z_6\,e^{-i\theta_{23}})=\cos\phi=0$.  $G^0$ and $h_1$ are CP-odd; $h_2$ and $h_3$ are CP-even. } 
\vskip 0.08in
\label{tab4}
\begin{tabular}{|c||c|c|}\hline
$\phaa k\phaa $ &\phaa $\phm q_{k1}\phaa $ & \phaa $q_{k2} \phaa $ \\ \hline
$1$ & $\phm 0$ & $1$ \\
$2$ & $-c_{13}$ & $is_{13}$ \\
$3$ & $\phm s_{13}$ & $i c_{13}$ \\
$4$ & $\phm i$ & $0$ \\ \hline
\end{tabular}
\end{table}

In both Case I and Case II, $\widetilde{\mathcal{M}}$ assumes a block
diagonal form consisting of a $2\times 2$ block (corresponding to the
the CP-even Higgs bosons) and a $1\times 1$ block (corresponding to the
CP-odd Higgs boson). The CP-odd field has mass
\beqa
m_{A^0}^2&=&\half v^2\left[Y_2/v^2+Z_1+\half(Z_3+Z_4-\zfiver)\right]\quad\quad\mbox{\text{[Case I] }}\,,\nonumber\\
m_{A^0}^2&=&\half v^2\left[Y_2/v^2+Z_1+\half(Z_3+Z_4+\zfiver)\right]\quad\quad\mbox{\text{[Case II] }}\,.\label{evenmas}
\eeqa
It is possible to eliminate the explicit dependence on $\theta_{23}$ by defining a quantity $\varepsilon_{56}$ as follows: \beq 
\Re(Z_5^* Z_6^2)=\varepsilon_{56} |Z_5|\,|Z_6|^2\,,\qquad
\varepsilon_{56}\equiv\pm 1\,.
\eeq
Note that in the CP-conserving limit, \beq\Re[Z_5^* Z_6^2] = \zfiver\,\Re[(\zsix)^2] = \pm \zfiver |Z_6|^2\,,\eeq
where the upper (lower) sign corresponds to case I (II).  Then can write $\zfiver = \pm \varepsilon_{56}|Z_5|,$ and the masses of the neutral Higgs fields become
\beqa
m^2_{h^0,H^0}&=&\half v^2\left[Y_2/v^2+Z_1+\half(Z_3+Z_4+\varepsilon_{56}|Z_5|)\right.\nonumber\\
&&
\left.\mp\sqrt{\left[Y_2/v^2-Z_1+\half(Z_3+Z_4+\varepsilon_{56}|Z_5|)\right]^2
+4|Z_6|^2}\right],\nonumber \\
m_{A^0}^2&=&Y_2+\half(Z_3+Z_4-\varepsilon_{56}|Z_5|)v^2\,,\label{mha1}
\eeqa
where is defined by the relation
$\varepsilon_{56}$ is an invariant quantity, its value must be determined from experiment.

Additional constraints for a CP-conserving 2HDM arise when the Higgs-fermion couplings are included. Let us write the transformation of the fields $h_k$ under CP as $h_k \rightarrow \eta_k h_k$, where $\eta_k = \pm 1$. The Higgs-Quark Lagrangian [\eq{hffu2}] contains the term
\beq \frac{1}{\sqrt{2}}q_{k2}^*\,\overline{D}
e^{i\theta_{23}}\rho^D P_L Dh_k + h.c.\,\label{yukterm}\eeq
(and a similar term with $U$ in place of $D$).  If CP is conserved, the quark mass-eigenstates $D_i$ and $U_i$ transform under CP with some phase, 
\beq D_i \rightarrow e^{i \theta_{i}^D} D_i^*,\qquad U_i \rightarrow e^{i \theta_{i}^U} U_i^*\,.\eeq
However, one can rephase $D_i$ and $U_i$ such that $D_i \rightarrow D_i^*$ and $U_i \rightarrow U_i^*$ under a CP transformation.  Let us assume that we have done such a rephasing, defined by $D_i \rightarrow \eta_i^D D_i$ (and similarly for $U_i$), with $|\eta_i^Q|=1$.  This rephasing also transforms the Yukawa matrices, with the result $\rho_{ij}^Q \rightarrow \eta_i^Q \eta_j^{Q*} \rho_{ij}^Q$. Reinstituting the flavor indices, the CP transformation on \eq{yukterm} (having suitably rephased the quark fields) gives
\beq q_{k2}^*\,\overline{D}_i[e^{i\theta_{23}}\rho^D]_{ij} P_L D_jh_k \rightarrow 
\eta_k q_{k2}^*\,\overline{D}_i[e^{i\theta_{23}}\rho^D]_{ji} P_RD_jh_k\,,\eeq
where $\rho^D$ has been appropriately rephased, as described above.
Comparing this to the hermitian conjugate of \eq{yukterm}, 
\beq q_{k2}\,\overline{D}_i[e^{i\theta_{23}}\rho^D]_{ij}^\dagger P_RD_jh_k\,, \eeq
we obtain a condition for CP invariance of the neutral Higgs bosons, and with the analogous result for $\rho^U$.  Both conditions can be summarized as
\beq \eta_k q_{k2}^*[e^{i\theta_{23}}\rho^Q]_{ij}=[e^{i\theta_{23}}\rho^Q]_{ij}^*q_{k2}\,.\label{neut}\eeq

The values of $q_{k2}$ and $\eta_k$ can be obtained from
Tables~\ref{tab2}---\ref{tab4}.  One finds that \eq{neut} is equivalent to the following:
\beq \label{antihermitian}
e^{i\theta_{23}} \rho^Q\,\,{\rm is}\quad \begin{cases}
\mbox{\text{real}} &
\qquad\text{in~Case~I}\,,\\
\mbox{\text{imaginary}} &
\qquad\text{in~Cases~IIa~and~IIb}\,.\end{cases}
\eeq
In both Cases I and II, 
the results of \eqthree{sp0}{cp0}{antihermitian} imply that 
\beqa \Im(Z_6 \rho^Q)&=&\Im(Z_6 e^{-i\theta_{23}}e^{i\theta_{23}}\rho^Q)\nonumber\\
&=&\Re(Z_6 e^{-i\theta_{23}})\Im(e^{i\theta_{23}}\rho^Q)) - \Im(Z_6 e^{-i\theta_{23}})\Re(e^{i\theta_{23}}\rho^Q))\nonumber\\
&=&0\,.\eeqa
One can prove similar conditions involving $Z_7$ and $Z_5$, so that the complete set of conditions for CP-invariance of the couplings of the neutral Higgs bosons to fermion pairs is the following:
\beq \label{zrho}
\Im[Z_6 \rho^Q]=\Im[Z_7 \rho^Q]=\Im[Z_5 (\rho^Q)^2]=0\,.  
\eeq 
Thus, if \eqs{cpoddinv}{zrho} are satisfied, then
the neutral Higgs bosons are eigenstates of CP, and the
only possible source of CP-violation in the 2HDM is the 
unremovable phase in the CKM matrix $K$
that enters via the charged current interactions mediated by either
$W^\pm$ or $H^\pm$ exchange\footnote{One can also formulate a
basis-independent condition 
(that is invariant with respect to
separate redefinitions of the Higgs doublet fields and the quark
fields) for the absence of CP-violation in the charged current
interactions.  This condition involves the Jarlskog 
invariant~\cite{Jarlskog:1985cw,Jarlskog:1985ht}, and can also be written 
as~\cite{cpinvariants,branco}: 
$\Tr\ls{\rm f}\,\bigl[H^{U,0},\,H^{D,0}\bigr]\lasup{\,3}=0$
(summed over three quark generations), where
$H^{Q,0}\equiv M^{Q,0} M^{Q,0\,\dagger}$ and the $M^{Q,0}$ are defined below \eq{diagdmass}. Since CP-violating phenomena in the charged current interactions are observed
and well described by the CKM matrix, we shall not impose 
this latter condition here.} [see \eq{hffu2}].

Invariant techniques for describing the constraints on the Higgs-fermion
interaction due to CP-invariance have also been
considered in \Refs{cpx2}{branco}.  In these works, the authors
construct invariant expressions that are both U(2)-invariant and
invariant with respect to the redefinition of the quark fields.  For
example, the invariants denoted by $J_a$ and $J_b$ in \Ref{cpx2}
are given by $J_a\equiv \Im\, J^D$ and $J_b\equiv \Im\, J^U$ where
\beq \label{jqdef}
J^Q= \Tr(VYT^Q)\,,\qquad T^Q_{a\bbar}\equiv\Tr\ls{\rm f}(\eta^{Q,0}_a
\eta^{Q,0\,\dagger}_{\bbar}) =\Tr\ls{\rm f}(\eta^{Q}_a
\eta^{Q\,\dagger}_{\bbar})\,,
\eeq
and the trace $\Tr\ls{\rm f}$ sums over the diagonal quark
generation indices.  Note that the trace over generation indices
ensures that the resulting expression is invariant with respect to
unitary redefinitions of the quark fields [\eq{biunitary}].
Using \eq{aab} [with $A=Y$], it is straightforward to re-express
\eq{jqdef} as:
\beq
J^Q=Y_1\Tr\ls{\rm f}[(\kappa^Q)^2]+Y_3\Tr\ls{\rm f}[\kappa^Q\rho^Q]\,,
\eeq
after using \eqthree{invariants}{pseudoinvariants}{kapparho}.
Indeed, $J^Q$ is invariant with respect to U(2) transformations
since the product of pseudo-invariants $Y_3\,\rho^Q$ is a U(2)-invariant
quantity.  Moreover, taking the trace over the quark generation 
indices ensures that $J^Q$ is invariant with respect to 
unitary redefinitions of the quark fields.
In \Ref{cpx2}, a proof is given 
that $\Im\, J^Q=0$ is one of the invariant conditions 
for CP-invariance of the Higgs-fermion interactions. 
In our formalism, this result is easily verified. Using the 
scalar potential minimum conditions [\eq{hbasismincond}], we obtain:
\beqa
\Im\, J^Q&=&-\frac{v}{\sqrt{2}}\, \Im\bigl[Z_6
\Tr\ls{\rm f}(M_Q \rho^Q)\bigr]\,,\nonumber\\
&=& -\frac{v}{\sqrt{2}}\, \Tr\ls{\rm f}\Im\bigl[
M_Q Z_6\rho^Q\bigr]\,.\eeqa
But, CP-invariance requires [by \eq{zrho}] that
$Z_6\rho ^Q$ is real.  Since $M_Q$ is a real diagonal
matrix, it then immediately follows that
$\Im\,J^Q=0$.

\subsection{The CP-Conserving Limit in the Real Basis \label{realsec}}
For completeness, we will now analyse the CP-conserving 2HDM in a specific basis, in order to express the masses and physical Higgs states in terms of more traditional parameters.  In the standard notation of the CP-conserving 2HDM,
one considers only real basis choices, in which the Higgs Lagrangian
parameters and the scalar vacuum expectation values are 
real.  We can therefore restrict basis changes to O(2) 
transformations~\cite{davidson}.\footnote{If $Z_6=Z_7=\rho^Q=0$, then
the possible transformations among real bases are elements of
O(2)$\times\mathbb{Z}_2$.  In particular, the sign of $Z_5$ changes when
when the Higgs basis field
$H_2\to iH_2$. In this case, $Z_5$ is an O(2)-invariant but it is
a pseudo-invariant with respect to $\mathbb{Z}_2$.}
In this context, pseudo-invariants are SO(2)-invariant quantities that
change sign under an O(2) transformation with determinant
equal to $-1$.  Note that
$Z_5$ is now an invariant with respect to O(2) transformations, but
$Z_6$, $Z_7$ and $e^{-i\theta_{23}}$ are pseudo-invariants.  
In particular, for $Z_6\neq 0$ in the convention where $0\leq\phi<\pi$,
\beq \label{vareps}
e^{-i\theta_{23}}=e^{i\phi}e^{-i\theta_6}=\begin{cases}
\,\,\varepsilon_6 \quad & \text{[Case I]}\,, \\ \,\,i\varepsilon_6 \quad
& \text{[Case II]}\,,\end{cases}
\eeq
where $Z_6\equiv\varepsilon_6|Z_6|$ in the real basis.  That is,
$\varepsilon_6$ is a pseudo-invariant quantity (in contrast, the sign of $Z_5$
is invariant) with respect to O(2) transformations. Note that in the real basis, \beq \varepsilon_{56} = e^{2i\theta_5}e^{-2i\theta_6} = \rm{sgn}(Z_5) \varepsilon_6^2 = \rm{sgn}(Z_5)\,,\eeq
so that \eq{mha1} can be written in terms of the real-basis parameters:
\beq
m_{\ha}^2=Y_2+\half v^2\left(Z_3+Z_4-Z_5\right)\,.
\eeq

The generic real basis fields can be expressed in
terms of 
the two neutral CP-even scalar mass-eigenstates $\hl$, $\hh$
(with $\mhl\leq\mhh$) and the CP-odd scalar mass-eigenstate $\ha$,
$\go$ as follows~\cite{ghhiggs1,ghhiggs2,higgshunt}:
\beqa
\Phi_1^0&=&\frac{1}{\sqrt{2}}\left[v\widehat v_1-\hl s_\alpha
+\hh c_\alpha+i(\go\cb-\ha\sb)\right]\,,\\
\Phi_2^0&=&\frac{1}{\sqrt{2}}\left[v\widehat v_2+\hl c_\alpha
+\hh s_\alpha+i(\go\sb+\ha\cb)\right]\,,
\eeqa
with $\mhl\leq\mhh$,
where $\widehat v_a=(\cb\,,\,\sb)$,
$s_\alpha\equiv\sin\alpha$, $c_\alpha\equiv\cos\alpha$, and
$\alpha$ is the CP-even neutral Higgs boson mixing angle.  These
equations can be written more compactly as
\beq \label{cpchdm}
\Phi^0_a=\frac{1}{\sqrt{2}}\left[(v+\hl\sbma+\hh\cbma+i\go)\widehat v_a
+(\hl\cbma-\hh\sbma+i\ha)\widehat w_a\right]\,,
\eeq
where $\sbma\equiv\sin(\beta-\alpha)$ and
$\cbma\equiv\cos(\beta-\alpha)$.

Using the results of Tables~\ref{tab2}---\ref{tab4}
and comparing \eq{cpchdm} to \eq{master} [with
$e^{-i\theta_{23}}$ determined from \eq{vareps}],
one can identify the neutral Higgs fields $h_k$ with the eigenstates
of definite CP quantum numbers, $\hl$, $\hh$ and $\ha$, and relate
the angular factor $\beta-\alpha$ with the
appropriate invariant angle:\footnote{The extra minus signs 
in the identification of $h_2=-\varepsilon_6\hh$ 
in Case I and $h_2=-\hl$ in Case IIb arise due to
the fact that the standard conventions of the CP-conserving 2HDM
correspond to $\det R=-1$ (whereas $\det R=+1$ in Case~IIa).}
\beqa \label{cases12}
&&\hspace{-0.5in}{\rm Case~I:}\phm\quad h_1=\hl\,,\,\,
h_2=-\varepsilon_6\hh\,, \,\, h_3=\varepsilon_6\ha\,,\,\,
c_{12}=\sbma\,\,\, {\rm and}\,\, s_{12}=-\varepsilon_6\cbma\,,\nonumber \\
&&\hspace{-0.5in}{\rm Case~IIa:}\quad h_1=\hl\,,\,\,
h_2=\varepsilon_6\ha\,,\,\,h_3=\varepsilon_6\hh\,,\,\,
c_{13}=\sbma\,\,\, {\rm and}\,\, s_{13}=\varepsilon_6\cbma\,,\nonumber \\
&&\hspace{-0.5in}{\rm Case~IIb:}\quad h_1=\varepsilon_6\ha\,,\,\,
h_2=-\hl\,,\,\,h_3=\varepsilon_6\hh\,,\,\,
c_{13}=\sbma\,\,\, {\rm and}\,\, s_{13}=\varepsilon_6\cbma\,.
\eeqa
In the convention for the angular domain given by \eq{app:domains},
$c_{12}$ and $c_{13}$ are non-negative and therefore
$\sbma\geq 0$. 
The appearance of the pseudo-invariant quantity $\varepsilon_6$
in \eq{cases12}
implies that $\hh$, $\ha$ (and $H^\pm$) are pseudo-invariant fields,
and $\cbma$ is a pseudo-invariant with respect to O(2) 
transformations.\footnote{Note that $\sbma$ is invariant with respect to O(2)
transformations, which is consistent with our convention that
$\sbma\geq 0$.   The analogous results have
been obtained in \Ref{davidson} in a convention where $\cbma\geq 0$.}
In contrast, $\hl$ is an invariant field.

At this stage, we have not imposed any mass ordering of the three
neutral scalar states.  Since one can distinguish between the CP-odd
and the CP-even neutral scalars, it is sufficient to require that
$\mhl\leq\mhh$.  (If one does not care about the mass ordering
of $\ha$ relative to the CP-even states, then
Cases IIa and IIb can be discarded without loss of generality.)  
We can compute the masses of the CP-even scalars and
the angle $\beta-\alpha$~\cite{ghdecoupling} in any of the three cases:
\beqa
\mhl^2 &=& \mha^2\,\cbma^2+v^2\left[Z_1\sbma^2+Z_5\cbma^2+2\sbma\cbma
  Z_6\right]\,,\label{mhl2}\\
\mhh^2 &=& \mha^2\,\sbma^2+v^2\left[Z_1\cbma^2+Z_5\sbma^2-2\sbma\cbma
  Z_6\right]\,,\label{mhh2}
\eeqa
and
\beq \label{bma}
\tan[2(\beta-\alpha)]=\frac{2Z_6 v^2}{m_{\ha}^2+(Z_5-Z_1) v^2}\,,\qquad
\sin [2(\beta-\alpha)]=\frac{-2Z_6 v^2}{\mhh^2-\mhl^2}\,.
\eeq
Note that
\eqst{mhl2}{bma} are covariant with respect to O(2) transformations,
since $Z_6$ and $\cbma$ are both pseudo-invariant quantities.

We end this section with a very brief outline of the tree-level MSSM
Higgs sector.  Since this model is CP-conserving, it is conventional
to choose the phase conventions of the Higgs fields that yield
a real basis.  In the natural
supersymmetric basis, the $\lambda_i$ of \eq{pot}
are given by:
\beq
\lambda_1=\lambda_2=\quarter(g^2+g^{\prime\,2})\,,\quad
\lambda_3=\quarter(g^2-g^{\prime\,2})\,,\quad
\lambda_4=-\half g^2\,,\quad
\lambda_5=\lambda_6=\lambda_7=0\,,
\eeq
where $g$ and $g'$ are the usual electroweak couplings [with
$m_Z^2=\quarter(g^2+g^{\prime\,2})v^2$].
From these results, one can compute the (pseudo)invariants:
\beqa \label{mssmz}
Z_1=Z_2 &=& \quarter (g^2+g^{\prime\,2})\cos^2
2\beta\,,\quad
\! Z_3= Z_5+\quarter(g^2-g^{\prime\,2})\,,\quad
Z_4= Z_5-\half g^2\,,\nonumber \\ \hspace{-0.5in}
Z_5&=& \quarter (g^2+g^{\prime\,2})\sin^2 2\beta\,,\quad
Z_6=-Z_7= -\quarter (g^2+g^{\prime\,2})\sin 2\beta\cos2\beta\,.
\eeqa
The standard MSSM 
tree level Higgs sector formulae~\cite{ghhiggs1,ghhiggs2} for the Higgs masses and 
$\beta-\alpha$ are easily reproduced 
using \eq{mssmz} and the results of this section.
\section{The 2HDM with $Z_6 = 0$ \label{zsixzero}}
In this section we discuss the case where $Z_6=0$. For now, we will assume that all three neutral Higgs
squared-masses are non-degenerate.  Therefore, we require that
$Z_5\equiv |Z_5|e^{2i\theta_5}\neq 0$ in what follows,\footnote{If
$Z_5=Z_6=0$, then the neutral Higgs squared-mass matrix is diagonal in
the Higgs basis, with two degenerate Higgs boson mass-eigenstates for $A^2 \neq Z_1 v^2$.  If $A^2 = Z_1 v^2$ there are three degenerate Higgs boson mass-eigenstates.}
and define the invariant angle
$\phi_5\equiv \theta_5-\theta_{23}$.  Once the sign conventions of
the neutral Higgs fields are fixed, the invariant angles 
$\theta_{12}$, $\theta_{13}$ and $\phi_5$ are defined modulo $\pi$.
We first note that \eqthree{c23c12a}{c23c12b}{tan12}
are valid when $Z_6=0$.  Thus, setting \eq{c23c12b} to zero
implies that $\sin 2\theta_{13}=0$,\footnote{If $Z_6=0$ and 
$A^2=Z_1 v^2$, then Eq. (C7) is automatically
equal to zero.  We will see in section \ref{zsixzerodeg} that the neutral Higgs masses are degenerate in this scenario.}  
which yields two possible solutions, $s_{13}=0$ or $c_{13}=0$. 
In the former case, \eq{c23c12a} yields $\Im(Z_5
e^{-2i\theta_{23}})=0$, \textit{i.e.}, $\sin 2\phi_5=0$, and
\eq{tan12} implies that $\sin 2\theta_{12}=0$.  In the latter case, we can use \eq{tan12} to write
\beq \tan(2\theta_{12}) = - \tan(2 \phi_5).\eeq
Thus, we can define three cases:
\beqa \label{zsixz}\begin{array}{ccc}
s_{13}=0, & \Im(Z_5 \,e^{-2i\theta_{23}}) = c_{12} = 0 &\qquad\rm{~Case}~(\it{i})\,,\\
s_{13}=0, & \Im(Z_5 \,e^{-2i\theta_{23}}) = s_{12} = 0 &\qquad\rm{~Case}~(\it{ii})\,,\\
c_{13}=0, & \Im[Z_5 \,e^{2i(\theta_{12}- \theta_{23})}]=0 & \qquad\rm{~Case}~(\it{iii})\,.\end{array}
\eeqa
Let us start with $(i)$.  For $Z_6 = 0$, the mass matrix $\widetilde{\mathcal{M}}$ has the form [\eq{mtilmatrix}]
\beq \label{mtilmatrixcp}
\widetilde{\mathcal{M}}\equiv R_{23}\mathcal{M}R_{23}^T=
v^2\left( \begin{array}{ccc}
Z_1&\,\, 0 &\,\, -\half\Im(Z_5 \,e^{-2i\theta_{23}})\\
0 &\,\,\Re(Z_5 \,e^{-2i\theta_{23}})+ A^2/v^2 & \,\,
0\\ 0&\,\, -\half\Im(Z_5 \,e^{-2i\theta_{23}}) &\,\, A^2/v^2\end{array}\right).
\eeq
Applying the mass mixing matrix $\widetilde{R}\,\widetilde{\mathcal{M}}\,\widetilde{R}^T$ and setting  $s_{13} = c_{12}=0$ in $\widetilde{R}$ produces the diagonalized masses for case ($i$):
\beqa
m_1^2&=&  A^2 +v^2~\Re(Z_5 \,e^{-2i\theta_{23}})= Y_2 +\half[Z_3 + Z_4 + \Re(Z_5 \,e^{-2i\theta_{23}})]~v^2 \,, \nonumber \\
m_2^2 &=&Z_1 v^2\,,\nonumber\\
m_3^2 &=&A^2= Y_2 +\half[Z_3 + Z_4 - \Re(Z_5 \,e^{-2i\theta_{23}})]~v^2\,,\label{threemass}
\eeqa
where we have used the convention $c_{13} = + 1$ and $s_{12} = -1$ as required by \eq{app:domains}, and eliminated $A^2$ using \eq{madef}.  Note that since $\zfii = 0$, $\zfiver = \rm{sgn}(Z_5) |Z_5|^2$.  The ambiguity in sign will be clarified in section \ref{zsixzerocp}.

For case ($ii$), one uses instead $s_{13} = s_{12}=0$, with $c_{12} = c_{13} = +1$.  Then taking $\widetilde{R}\,\widetilde{\mathcal{M}}\,\widetilde{R}^T$ yields
\beqa
m_1^2&=&Z_1 v^2\,,\nonumber\\ 
m_2^2 &=&A^2 +v^2~\Re(Z_5 \,e^{-2i\theta_{23}})= Y_2 +\half[Z_3 + Z_4 + \Re(Z_5 \,e^{-2i\theta_{23}})]~v^2 \,, \nonumber \\
m_3^2 &=&A^2=Y_2 +\half[Z_3 + Z_4 - \Re(Z_5 \,e^{-2i\theta_{23}})]~v^2\,.\label{threemass2}
\eeqa

The final (non-degenerate) scenario for $Z_6 = 0$ is case ($iii$), in which $q_{12} = i e^{i \theta_{12}}$, $q_{22v}= e^{i \theta_{12}}$, $q_{41}=i$, $q_{31}=-1$, and all other $q_{k\ell}$ vanish.  Diagonalizing the mass matrix in \eq{mtilmatrixcp} yields
\beqa
m_1^2 &=&A^2 +v^2\Re(ie^{i\theta_{12}})\Re(ie^{i\theta_{12}}Z_5 \,e^{-2i\theta_{23}})\,,\nonumber\\ 
m_2^2 &=&A^2 +v^2\Re(e^{i\theta_{12}})\Re(e^{i\theta_{12}}Z_5 \,e^{-2i\theta_{23}}) \,, \nonumber \\
m_3^2 &=&Z_1 v^2\,.
\eeqa
Using $\Im(Z_5 \,e^{2i(\theta_{12} - \theta_{23})})= 0$ and $\Re[z_1,z_2] = \Re z_1 \Re z_2 - \Im z_1 \Im z_2$, the masses take the form
\beqa
m_1^2 &=& Y_2 +\half[Z_3 + Z_4 - \Re(Z_5 \,e^{2i(\theta_{12} - \theta_{23})})]~v^2\,,\nonumber\\ 
m_2^2 &=& Y_2 +\half[Z_3 + Z_4 + \Re(Z_5 \,e^{2i(\theta_{12} - \theta_{23})})]~v^2 \,, \nonumber \\
m_3^2 &=& Z_1 v^2\,.\label{threemass3}
\eeqa
Since $ \Im(Z_5 \,e^{2i(\theta_{12} - \theta_{23})})] = 0$, $ \Re(Z_5 \,e^{2i(\theta_{12} - \theta_{23})})] = \rm{sgn}(Z_5) |Z_5|^2$.  Thus, up to a reordering of the three fields, all three cases exhibit the same masses,
\beqa
m_1^2 &=& Y_2 +\half[Z_3 + Z_4 - \rm{sgn}(Z_5)|Z_5|]~v^2\,,\nonumber\\ 
m_2^2 &=& Y_2 +\half[Z_3 + Z_4 + \rm{sgn}(Z_5)|Z_5|]~v^2 \,, \nonumber \\
m_3^2 &=& Z_1 v^2\,.\label{massthree}
\eeqa

\subsection{CP Conservation with $Z_6 = 0$}\label{zsixzerocp}
Since the mass matrix for $Z_6 = 0$ is broken up into $2\times 2$ and $1\times 1$ blocks [see \eq{mtilmatrixcp}], one might expect that the scalar sector is automatically CP-conserving.  Indeed, the field whose mass is given by $Z_1 v^2$ in the three cases ($i$)--($iii$) is CP even, as one can deduce from the couplings $G^0 G^0 h_k$. However, the coupling $H^+ H^- h_k$ contains the expression $\Re(q_{k2}Z_7 \,e^{-i\theta_{23}})$ and the $\bar{Q}{Q}h_k$ interactions [\eq{hffu2}] contain $q_{k2}^*e^{i\theta_{23}}\rho^Q$ ($Q = U,D$). Thus, for CP to be conserved, one requires 
\beq\Im(Z_5^* Z_7^2) = 0,\qquad\Im[Z_5 (\rho^Q)^2]=0,\,\,\mbox{\text{and}}\,\,\Im[Z_7\rho^Q]=0.\label{conditionsz}\eeq   
One observes from \eqthree{threemass}{threemass2}{threemass3} that in case ($iii$), the angle $\theta_{23}-\theta_{12}$ plays the same role as $\theta_{23}$ in cases ($i$) and ($ii$).  Thus, it is convenient to define
\beq \label{zsixcpcond}
\bar{\theta}_{23} \equiv\quad \begin{cases}
 \quad\theta_{23} \quad\quad\quad\mbox{\text{cases (\emph{i}),(\emph{ii})}}\,;\\
 \quad\theta_{23}-\theta_{12} \quad\mbox{\text{case (\emph{iii})}}\,.\end{cases}
\eeq
Since $\Im(Z_5 \,e^{-2i\bar\theta_{23}}) =0$ [see \eq{zsixz}] and $\Im[Z_7\rho^Q]=0$, one can write
\beqa \Im[Z_5^*Z_7^2]& =& \Re(Z_5 \,e^{-2i\bar{\theta}_{23}})\Im(Z_7^2 \,e^{-2i\bar{\theta}_{23}})\nonumber\\
&=& \Re(Z_5 \,e^{-2i\bar{\theta}_{23}})2\Im(Z_7 \,e^{-i\bar{\theta}_{23}})\Re(Z_7 \,e^{-i\bar{\theta}_{23}})=0\,,\label{z57}\\
\Im[Z_7\,e^{-i\bar{\theta}_{23}} e^{i\bar{\theta}_{23}}\rho^Q]&=&  \Re(Z_7 \,e^{-i\bar{\theta}_{23}})\Im(\rho^Q \,e^{i\bar{\theta}_{23}})\nonumber\\
 &&+\,\Im(Z_7 \,e^{-i\bar{\theta}_{23}})\Re(\rho^Q \,e^{i\bar{\theta}_{23}})=0\,.\label{zq7}
\eeqa
\Eqs{z57}{zq7} have two solutions,
\beqa 
(a)\qquad\Im(Z_7e^{-i\bar{\theta}_{23}})&=& \quad \Im[\rho^Q e^{i\bar{\theta}_{23}}]\,=\,0\,,\label{hermit}\\
(b)\qquad\Re(Z_7e^{-i\bar{\theta}_{23}})&=& \quad \Re[\rho^Q e^{i\bar{\theta}_{23}}]\,=\,0\,. \label{antihermit}
\eeqa
Note that \eq{covcons} correctly defines the CP transformation for $Z_6 = 0$ provided that one replace $\theta_{23}$ by $\bar\theta_{23}$.  The transformation with the positive sign in \eq{covcons} corresponds to solution $(a)$ [\eq{hermit}] and the one with the negative sign to solution $(b)$ [\eq{antihermit}].

If neither \eq{hermit} nor \eq{antihermit} holds, then CP is violated, and the remaining two neutral scalar fields will have indefinite CP quantum numbers.  If the conditions of \eq{conditionsz} hold, all neutral scalars will be CP eigenstates, with CP quantum numbers displayed in Tables \ref{tabcpi}--\ref{tabcpii}.  Comparing the results of the tables with \eqthree{threemass}{threemass2}{threemass3}, one finds that in all three cases,
\beqa
m_{A^0}^2&=& Y_2 +\half[Z_3 + Z_4 - \Re(Z_5 \,e^{-2i\bar\theta_{23}})]~v^2\,,\,\,\rm{for}~\Im(Z_7\,e^{-i\bar\theta_{23}}) =\Im(\rho^Q\,e^{i\bar\theta_{23}}) =0\,,\nonumber \\
m_{A^0}^2&=& Y_2 +\half[Z_3 + Z_4 + \Re(Z_5 \,e^{-2i\bar\theta_{23}})]~v^2\,,\,\,\rm{for}~\Re(Z_7\,e^{-i\bar\theta_{23}}) =
 \Re(\rho^Q\,e^{i\bar\theta_{23}}) =0\,.\nonumber\\
&&\phantom{line}\label{oddmasses}
\eeqa
\beqa
m_{H^0}^2&=& Y_2 +\half[Z_3 + Z_4 + \Re(Z_5 \,e^{-2i\bar\theta_{23}})]~v^2\,,\,\,\rm{for}~\Im(Z_7\,e^{-i\bar\theta_{23}}) =\Im(\rho^Q\,e^{i\bar\theta_{23}}) =0\,,\nonumber \\
m_{H^0}^2&=& Y_2 +\half[Z_3 + Z_4 - \Re(Z_5 \,e^{-2i\bar\theta_{23}})]~v^2\,,\,\,\rm{for}~\Re(Z_7\,e^{-i\bar\theta_{23}}) =
 \Re(\rho^Q\,e^{i\bar\theta_{23}}) =0\,.\nonumber\\
&&\phantom{line}\label{evenmass}\eeqa
One can condense these results by defining the symbol $\varepsilon_{57}$:
\beq\label{fivesev}
\Re(Z_5^* Z_7^2)=\varepsilon_{57}|Z_5||Z_7|^2\,,\qquad
\varepsilon_{57}=\pm 1\,.
\eeq
The quantity $\varepsilon_{57}$ is independent of basis.  Note that
\beqa \Re(Z_5^*\thetdoubpl)& =& \frac{\Re(Z_5^* |Z_7|^2 \thetdoubpl)}{|Z_7|^2}\,,\nonumber\\
&=& \frac{\Re(Z_5^*Z_7^2)\Re[(Z_7^*)^2 \thetdoubpl]}{|Z_7|^2}\,.\eeqa
Then applying the definition of $\epsilon_{57}$ in \eq{fivesev},
\beqa \Re(Z_7 \,e^{-i\theta_{23}}) = 0  &\Rightarrow   &\varepsilon_{57}|Z_5| = - \Re(Z_5 \,e^{-2i\theta_{23}})\,, \nonumber  \\
\Im(Z_7 \,e^{-i\theta_{23}}) = 0& \Rightarrow & \varepsilon_{57}|Z_5| = + \Re(Z_5 \,e^{-2i\theta_{23}})\,. \eeqa  Thus, equations~(\ref{oddmasses}) can be expressed as a single equation, so the mass of the CP-odd field is given by
\beq
m_{A^0}^2=Y_2 +\half[Z_3 + Z_4 - \varepsilon_{57} |Z_5|]~v^2\,,\label{condense}\eeq
and the second CP-even field has mass
\beq
m_{H^0}^2= Y_2 +\half[Z_3 + Z_4 + \varepsilon_{57} |Z_5|]~v^2\,.\label{condense2}\eeq
If $Z_7 = 0$ but $\rho^Q \neq 0$, one can derive the analog of $\epsilon_{57}$ as follows:
From the conditions of CP symmetry in \eq{conditionsz}, 
\beq \Tr_f \Im[Z_5(\rho^Q)^2]=0\,,\,\quad\Rightarrow\quad\Im\{Z_5\Tr_f[(\rho^Q)^2]\}=0\,.\label{rhotr}\eeq
Any $2\times 2$ matrix $A$ satisfies its characteristic equation, $A^2 - A \Tr A + \det A = 0$.  Taking the trace of this equation yields the identity $\Tr(A^2)-(\Tr A)^2 + 2\det A = 0$  Thus, the condition in \eq{rhotr} is equivalent to
\beq \Im[Z_5(\Tr_f \rho^Q)^2-2 Z_5 \det (\rho^Q)]=0\,,\,\quad\Rightarrow\quad\Im\{Z_5\Tr_f[(\rho^Q)^2]\}=0\,.\label{trace}\eeq
Note that $\rho^Q \thet$ is purely real or imaginary [see \eqs{hermit}{antihermit}], so \beq\Im[Z_5 \det (\rho^Q)]=\zfiver\Im[\det (\rho^Q\thetminus)] =0\,.\eeq  Then \eq{trace} becomes
\beq \Im[Z_5(\Tr_f\rho^Q)^2]=0\,.\label{trace2}\eeq  Now we can define
\beq\label{fiveq}
\Re[Z_5 (\Tr_f\rho^{Q})^2]\equiv\varepsilon_{5Q}|Z_5||(\Tr\rho^Q)^2|\,,\qquad
\varepsilon_{5Q}=\pm 1\,.
\eeq
Using a similar calculation as in the case with $Z_7\neq 0$, one finds
\beqa \Re(\rho^Q \,e^{-i\theta_{23}}) = 0  &\Rightarrow   &\varepsilon_{5Q}|Z_5| = - \Re(Z_5 \,e^{-2i\theta_{23}})\,, \nonumber  \\
\Im(\rho^Q \,e^{-i\theta_{23}}) = 0& \Rightarrow & \varepsilon_{5Q}|Z_5| = + \Re(Z_5 \,e^{-2i\theta_{23}})\,. \eeqa
Thus, for $Z_6 = Z_7 = 0$, one replaces $\varepsilon_{57}$ by $\varepsilon_{5Q}$ in \eqs{condense}{condense2}.

To summarize, we find that for $Z_6 = 0$ there is one neutral Higgs field that is always CP-even, with mass squared equal to $Z_1 v^2$. Unless $Z_5^*Z_7^2$, $Z_7 \rho^Q$, and $Z_5 (\rho^Q)^2$ are all real-valued, the remaining neutral fields are mixtures of CP eigenstates, even though the mass matrix has a block diagonal form.  In the CP conserving case, the two solutions represented by \eq{hermit} and \eq{antihermit} correspond to different possibilities for the CP quantum numbers of those 2 remaining fields. An overview of the three cases and the associated CP quantum numbers for the CP-conserving case are given in Tables \ref{tabcpi}--\ref{tabcpiii}.  
\begin{table}[ht!]
\centering
\caption{Values of $q_{k\ell}$  in the
CP-conserving limit with $Z_6 = 0$.  Case ($i$): $s_{13} = c_{12}=\Im(Z_5 \,e^{-2i\theta_{23}})=0$. The CP quantum numbers are shown for $\Im(Z_7 \,e^{-i\theta_{23}})=\Im(\rho^Q e^{i\bar{\theta}_{23}})= 0$ (upper sign), and $\Re(Z_7 \,e^{-i\theta_{23}})=\Re(\rho^Q e^{i\bar{\theta}_{23}})= 0$ (lower sign). }
\vskip 0.08in
\label{tabcpi}
\begin{tabular}{|c||c|c|c|}\hline
$\phaa k\phaa $ &\phaa $q_{k1}\phaa $ & \phaa $\phm q_{k2}$ \phaa &\phaa CP \phaa  \\ \hline
$1$ & $0$ & $ 1$&$ \pm 1 $\\
$2$ & $-1$ & $0$&$ +1$ \\
$3$ & $0$ & $i$ & $ \mp 1 $\\
$4$ & $i$ & $ 0$& $-1$ \\ \hline 
\end{tabular}
\end{table}
\begin{table}[ht!]
\centering
\caption{Values of $q_{k\ell}$  in the
CP-conserving limit with $Z_6 = 0$. Case ($ii$): $s_{13} = s_{12}=\Im(Z_5 \,e^{-2i\theta_{23}})=0$. The CP quantum numbers are shown for $\Im(Z_7 \,e^{-i\theta_{23}})=\Im(\rho^Q e^{i\bar{\theta}_{23}})= 0$ (upper sign), and $\Re(Z_7 \,e^{-i\theta_{23}})=\Re(\rho^Q e^{i\bar{\theta}_{23}})= 0$ (lower sign). }
\vskip 0.08in
\label{tabcpii}
\begin{tabular}{|c||c|c|c|}\hline
$\phaa k\phaa $ &\phaa $q_{k1}\phaa $ & \phaa $\phm q_{k2}$ \phaa &\phaa CP \phaa  \\ \hline
$1$ & $1$ & $0$&$ +1 $\\
$2$ & $0$ & $1$&$ \pm 1$ \\
$3$ & $0$ & $i$ & $ \mp 1$\\
$4$ & $i$ & $ 0$& $-1$ \\ \hline 
\end{tabular}
\end{table}
\begin{table}[ht!]
\centering
\caption{Values of $q_{k\ell}$ in the
CP-conserving limit with $Z_6 = 0$. Case ($iii$): $c_{13}=\Im(Z_5 \,e^{2i(\theta_{12}-\theta_{23})})$. The CP quantum numbers are shown for $\Im(Z_7 \,e^{-i\theta_{23}})=\Im(\rho^Q e^{i\bar{\theta}_{23}})= 0$ (upper sign), and $\Re(Z_7 \,e^{-i\theta_{23}})=\Re(\rho^Q e^{i\bar{\theta}_{23}})= 0$ (lower sign). }
\vskip 0.08in
\label{tabcpiii}
\begin{tabular}{|c||c|c|c|}\hline
$\phaa k\phaa $ &\phaa $q_{k1}\phaa $ & \phaa $\phm q_{k2}$ \phaa &CP \phaa \\ \hline
$1$ & $0$ & $ ie^{i\theta_{12}}$&$\mp 1$\\
$2$ & $0$ & $ e^{i\theta_{12}}$&$ \pm 1$ \\
$3$ & $-1$ & $0$ & $ +1$\\
$4$ & $i$ & $ 0$& $-1$\\ \hline 
\end{tabular}
\end{table}
The preceding results are valid for the CP conserving case with $Z_6 =0$ as long either $\rho^Q$ or $Z_7$ is non-vanishing.  If $Z_6=Z_7=\rho^Q=0$, the model has some extra features which will be described in section \ref{sec:herquetcpcons}.
\subsection{$Z_6 = 0$ with Degenerate Neutral Scalars}\label{zsixzerodeg}
In the previous discussion we assumed that none of the of neutral scalar masses were degenerate.  If we relax that requirement, three additional ways of satisfying \eqthree{c23c12a}{c23c12b}{tan12} appear:
\beqa \label{zsixdeg}\begin{array}{ccc}
\Im(Z_5 \,e^{-2i\theta_{23}})=s_{13}= 0,&Z_1v^2&=Y_2+\half[Z_3 + Z_4+\zfiver]v^2\,, \qquad\rm{~Case}~(\it{iv})\,,\nonumber\\
\Im(Z_5 \,e^{-2i\theta_{23}})=c_{12}= 0,&Z_1v^2&=Y_2+\half[Z_3 + Z_4-\zfiver]v^2\,, \qquad\rm{~Case}~(\it{v})\,,\nonumber\\
\Im(Z_5 \,e^{-2i\theta_{23}})=s_{12}= 0,&Z_1v^2&=Y_2+\half[Z_3 + Z_4-\zfiver]v^2\,. \qquad\rm{~Case}~(\it{vi})\,.\end{array}
\eeqa
These cases are presented in Tables \ref{tabd4}--\ref{tabd6}.

For case ($iv$), diagonalizing the mass matrix yields
\beqa
m_1^2 &=& s_{12}^2 A^2 + c_{12}^2 Z_1 v^2 +v^2 s_{12}^2 \zfiver=Z_1 v^2\,, \nonumber \\
m_2^2 &=& c_{12}^2 A^2 + s_{12}^2 Z_1 v^2 +v^2 c_{12}^2 \zfiver=Z_1 v^2\,,\nonumber\\
m_3^2 &=& A^2=Y_2 +\half v^2 [Z_3+Z_4-\zfiver] \,,\label{deg4}
\eeqa
where $A^2$ has been replaced by $Z_1 v^2 -v^2 \zfiver$.
The analogous calculation for case ($v$) gives
\beqa
m_1^2 &=&  A^2+v^2\zfiver=Y_2 +\half v^2 [Z_3+Z_4+\zfiver]\,, \nonumber \\
m_2^2 &=& s_{13}^2 A^2 +v^2 c_{13}^2 Z_1 =Z_1 v^2   \,,\nonumber\\
m_3^2 &=& c_{13}^2 A^2 +v^2 s_{13}^2 Z_1 = Z_1 v^2    \,.\label{deg5}
\eeqa
Similarly, for case ($vi$) one finds
\beqa
m_1^2 &=&  Z_1 v^2\,, \nonumber \\
m_2^2 &=& v^2[Z_1 +\zfiver]=Y_2 +\half v^2 [Z_3+Z_4+\zfiver]\,,\nonumber\\
m_3^2 &=&  Z_1 v^2 \,.\label{deg6}
\eeqa
\begin{table}[ht!]
\centering
\caption{Values of $q_{k\ell}$ with $Z_6 = 0$.  Case ($iv$): $s_{13} =\Im(Z_5 \,e^{-2i\theta_{23}}) = Z_1 - A^2/v^2 - \zfiver = 0$. }
\vskip 0.08in
\label{tabd4}
\begin{tabular}{|c||c|c|}\hline
$\phaa k\phaa $ &\phaa $q_{k1}\phaa $ & \phaa $\phm q_{k2}$ \phaa   \\ \hline
$1$ & $c_{12}$ & $-s_{12}$\\
$2$ & $s_{12}$ & $c_{12}$ \\
$3$ & $0$ & $i$ \\
$4$ & $i$ & $ 0$ \\ \hline 
\end{tabular}
\end{table}
\begin{table}[ht!]
\centering\caption{Values of $q_{k\ell}$ with $Z_6 = 0$.  Case ($v$): $c_{12} =\Im(Z_5 \,e^{-2i\theta_{23}}) = Z_1 - A^2/v^2 = 0$. }
\vskip 0.08in
\label{tabd5}
\begin{tabular}{|c||c|c|}\hline
$\phaa k\phaa $ &\phaa $q_{k1}\phaa $ & \phaa $\phm q_{k2}$ \phaa   \\ \hline
$1$ & $0$ & $1$\\
$2$ & $-c_{13}$ & $i s_{13}$ \\
$3$ & $s_{13}$ & $i c_{13}$ \\
$4$ & $i$ & $ 0$ \\ \hline 
\end{tabular}
\end{table}
\begin{table}[ht!]
\centering\caption{Values of $q_{k\ell}$ with $Z_6 = 0$.  Case ($vi$): $s_{12} =\Im(Z_5 \,e^{-2i\theta_{23}}) = Z_1 - A^2/v^2 = 0$. }
\vskip 0.08in
\label{tabd6}
\begin{tabular}{|c||c|c|}\hline
$\phaa k\phaa $ &\phaa $q_{k1}\phaa $ & \phaa $\phm q_{k2}$ \phaa   \\ \hline
$1$ & $c_{13}$ & $-is_{13}$\\
$2$ & $0$ & $1$ \\
$3$ & $s_{13}$ & $i c_{13}$ \\
$4$ & $i$ & $ 0$ \\ \hline 
\end{tabular}
\end{table}
One observes that the degenerate fields have mass squared equal to $Z_1 v^2$ in all cases.
If $\Im[Z_5^*Z_7^2]=0 $, $\Im[Z_5 (\rho^Q)^2]=0$, and $\Im[Z_7\rho^Q]=0$, the scalar sector will be CP-conserving.  In case ($iv$), the interaction $\Re(q_{k2}Z_7 \thetminus )h_k H^+ H^-$ indicates that the non-degenerate field, $h_3$, is CP-odd if $\zseveni = 0$ and CP-even if $\zsevenr = 0$.  One also deduces that the combination $s_{12} h_1 +c_{12} h_2$ is CP-even if $\zseveni = 0$ and CP-odd if $\zsevenr = 0$.  [One obtains analogous results using the $\bar Q Q h_k$ interaction.] The orthogonal combination is always CP-even.  These results are summarized in Table \ref{tabdcp4}.  A similar analysis for cases ($v$) and ($vi$) gives the results shown in Tables  \ref{tabdcp5} and \ref{tabdcp6}, respectively.
\begin{table}[ht!]
\centering
\caption{The CP quantum numbers of the neutral scalars with $Z_6 = 0$.  Case ($iv$): $s_{13} =\Im(Z_5 \,e^{-2i\theta_{23}}) = Z_1 - A^2/v^2 - \zfiver = 0$. }
\vskip 0.08in
\label{tabdcp4}
\begin{tabular}{|c||c|c|}\hline
\phaa CP Eigenstates\phaa  &\phaa $\zsevenr  =\phaa $ & \phaa $\zseveni =$ \phaa   \\
  &\phaa $\Re(\rho^Q e^{i\theta_{23}}) = 0$ \phaa  & \phaa $\Im(\rho^Q e^{i\theta_{23}}) = 0$ \phaa \\
\hline
$c_{12}h_1 +s_{12}h_2$ & $+1$ & $+1$\\
$-s_{12}h_1 +c_{12}h_2$ & $-1$ & $+1$ \\
$h_3$ & $+1$ & $-1$ \\ \hline 
\end{tabular}
\end{table}
\begin{table}[ht!]
\centering\caption{The CP quantum numbers of the neutral scalars with $Z_6 = 0$.  Case ($v$): $c_{12} =\Im(Z_5 \,e^{-2i\theta_{23}}) = Z_1 - A^2/v^2 = 0$. }
\vskip 0.08in
\label{tabdcp5}
\begin{tabular}{|c||c|c|}\hline
\phaa CP Eigenstates\phaa  &\phaa $\zsevenr =\phaa $ & \phaa $\zseveni = $ \phaa   \\
  &\phaa $\Re(\rho^Q e^{i\theta_{23}}) = 0$ \phaa  & \phaa $\Im(\rho^Q e^{i\theta_{23}}) = 0$ \phaa \\
\hline
$h_1$ & $-1$ & $+1$\\
$c_{13}h_2 - s_{13}h_3$ & $+1$ & $+1$ \\
$s_{13}h_1 +c_{13}h_3$ & $+1$ & $-1$ \\ \hline 
\end{tabular}
\end{table}
\begin{table}[ht!]
\centering\caption{The CP quantum numbers of the neutral scalars with $Z_6 = 0$.  Case ($vi$): $s_{12} =\Im(Z_5 \,e^{-2i\theta_{23}}) = Z_1 - A^2/v^2 = 0$. }
\vskip 0.08in
\label{tabdcp6}
\begin{tabular}{|c||c|c|}\hline
\phaa CP Eigenstates\phaa  &\phaa $\zsevenr =  \phaa $ & \phaa $\zseveni = $\phaa   \\
  &\phaa $\Re(\rho^Q e^{i\theta_{23}}) = 0$ \phaa  & \phaa $\Im(\rho^Q e^{i\theta_{23}}) = 0$ \phaa \\
\hline
$c_{13}h_1 +s_{13}h_3$ & $+1$ & $+1$\\
$h_2$ & $-1$ & $+1$ \\
$-s_{13}h_1+c_{13}h_3$ & $+1$ & $-1$ \\ \hline 
\end{tabular}
\end{table}
Using Tables \ref{tabdcp4}--\ref{tabdcp6} and the equations for the masses in \eqthree{deg4}{deg5}{deg6}, one can write expressions for the mass of the CP-odd field that apply in all three cases, namely
\beq
m_{A^0}^2=Y_2 +\half[Z_3 + Z_4 - \varepsilon_{57} |Z_5|]~v^2\,,\label{oddmassesdeg}\eeq
and one of the CP-even field has mass
\beq
m_{H^0}^2= Y_2 +\half[Z_3 + Z_4 + \varepsilon_{57} |Z_5|]~v^2\,.\label{degeven}\eeq
as in \eqs{condense}{condense2}.
The mass of the other CP-even field is always one of the degenerate fields can be expressed in all three cases as follows:
\beq
m_{h^0}^2= Z_1 v^2\,.\label{degeven2}\eeq
(We have not imposed $m_{h^0}^2< m_{H^0}^2$).  Note that \eqthree{oddmassesdeg}{degeven}{degeven2} hold for all six cases ($i$)--($vi$).  The neutral scalar field with mass $m_{h^0}^2 = Z_1 v^2$ has exact Standard Model couplings.  As in the previous section, if $Z_7 = 0$, $\varepsilon_{57}$ can be replaced by $\varepsilon_{5Q}$.

\subsection{Special Case: The CP-Conserving Limit when $\boldsymbol{Z_6=0}$ and $\boldsymbol{Z_7=0}$}\label{sec:herquetcpcons}
In this section, we consider a 2HDM with $Y_3=Z_6=Z_7=0$.\footnote{Technically, since the potential minimum conditions require $Y_3 = -\half Z_6 v^2$, explicitly setting $Y_3 = 0$ is redundant.} For the moment, let us also assume that $\rho^Q = 0$.

This model is automatically CP conserving (since one can choose $Z_5$ to
be real and positive without loss of generality).  Normally, in a
CP-invariant 2HDM, starting from a real Higgs basis one can get to any possible generic real basis with an  O(2)
transformation.  However, in the model under study here, there exists a particular U(2)
transformation that is not an O(2) transformation, which has the effect of changing the sign of $Z_5$.  This corresponds to redefining the second Higgs field by multiplication by i, ie, 
\beq H_2 \rightarrow i H_2 \,.\eeq  
The relevant U(2) matrix is $diag(1,i)$.  In Appendix A of \cite{cpbasis}, the possible definitions of time reversal invariance are discussed.  In particular, it was argued that the definition of T is unique if the possible
transformations from the real Higgs basis to any real generic basis is O(2).  It was shown that T was not unique if the latter
transformation group was $O(2)\otimes\mathbb{D}$ where $\mathbb{D}$ is a nontrivial discrete
group.  Applying this to the model where $Y_3=Z_6=Z_7=0$, we identify
the relevant group as $O(2)\otimes\mathbb{Z}_2$, where $\mathbb{Z}_2$ is the discrete group
consisting of the identity and the transformation that changes the sign of $Z_5$.  Thus, we conclude that for the $Y_3=Z_6=Z_7=0$ model, there are two inequivalent definitions of T, or equivalently two definitions of CP (since the model is CPT invariant).  These two symmetries correspond to the two possible transformations defined in \eq{covcons}, where the plus sign applies to $\Im(Z_6e^{-i\bar{\theta}_{23}})=\Im(Z_7e^{-i\bar{\theta}_{23}})= \Im[\rho^Q e^{i\bar{\theta}_{23}}]=0$ and the negative sign to $\Re(Z_6e^{-i\bar{\theta}_{23}})=\Re(Z_7e^{-i\bar{\theta}_{23}})= \Re[\rho^Q e^{i\bar{\theta}_{23}}]=0$.

In this model, the Higgs/gauge boson interactions are insufficient to identify the CP-odd field.  We know from the results of the previous section that one of the neutral scalars has
couplings exactly identical to the Standard Model Higgs, with $m_{h^0}^2 = Z_1 v^2$.  We also know that the two remaining neutral fields have opposite CP.  However, since $\rho^Q = 0$ the interactions cannot distinguish
which one is CP-even and which one is CP-odd; the two inequivalent CP symmetries are both conserved.  

Now let us suppose $\rho^Q \neq 0$.  In section \ref{zsixzerocp}, we showed how to identify the CP-odd field based on the value of $\varepsilon_{5Q}$.  Thus, a non-zero value of $\rho^Q$ has the effect of picking out the definition of CP that is respected by the Yukawa interactions. In fact, a non-zero value of any one of $Z_6$, $Z_7$ or $\rho^Q$ identifies the respected CP symmetry; if any two are non-zero, CP violation arises unless the relative phases obey \eq{zrho}.
\section{The Custodial Limit of the 2HDM \label{custsec}}
The subject of custodial symmetry in the 2HDM doublet model has been addressed by Pomarol and Vega ~\cite{pomarol2}, in the context of two cases which they label ``case I'' and ``case II,'' \emph{not to be confused with the two cases of CP-conservation} defined in equations (\ref{sp0}) and (\ref{cp0}), to which they have no correlation.  We reproduce some of their work here in order to clarify the significance of their two cases.  Specifically, we show in this section that the only difference between their ``case I'' and ``case II'' lies in the conditions they impose on the vevs of the scalar fields, which is not a physically measurable distinction. 

First we will ignore the coupling of the Higgs doublets to fermions and just consider the scalar potential.  To replicate ``case I'' of \cite{pomarol2}, we construct two $2\times 2$ matrices whose columns are made up of the Higgs doublet fields in the generic basis:
\beq M_1 \equiv (\tilde{\Phi}_1\,, \Phi_1), \hspace{5mm}M_2 \equiv (\tilde{\Phi}_2\,,  \Phi_2), \label{matdef}\eeq
where $\tilde{\Phi}\equiv i \sigma_2 \Phi^*$.   
These matrices transform as  
\beq M_i \rightarrow L~M_i~R^\dagger\label{pvi}\eeq under global $SU(2)_L \times SU(2)_R $ transformations.  In order that a custodial $SU(2)_V$ symmetry be preserved after electroweak symmetry breaking, $<M_i>$ must be proportional to the identity matrix:
\beq\label{vevs} <M_i> = \frac{1}{\sqrt{2}}\left(\begin{array}{cc}v_i^*&0\\0&v_i\end{array}\right)\propto 1\quad\Rightarrow\quad v_i^* = v_i,.\eeq
Since $L = R$,  
\beq <M_i> \rightarrow L <M_i> R^\dagger =  \frac{v_i}{\sqrt{2}} L R^\dagger= <M_i>\,. \label{vevsym} \eeq
 
Note that
\beqa
\Tr[M_i^\dagger M_j] &=& \Tr\left[\left(\begin{array}{cc}\Phi_i^0&-\Phi_i^+\\\Phi_i^-&\Phi_i^{0*}\end{array}\right)
\left(\begin{array}{cc}\Phi_j^{0*}&\Phi_j^+\\-\Phi_j^-&\Phi_j^{0}\end{array}\right)\right]
=  \Phi_i^\dagger \Phi_j+ \Phi_j^\dagger \Phi_i. \eeqa
Thus, 
\beqa
\Tr[M_1^\dagger M_1] &=& 2 \Phi_1^\dagger \Phi_1 \nonumber\\
\Tr[M_2^\dagger M_2] &=& 2 \Phi_2^\dagger \Phi_2 \nonumber\\
\Tr[M_1^\dagger M_2] &=&  \Phi_2^\dagger \Phi_1 + h.c.\label{tr}\eeqa  
Using the expressions in \eq{tr} to construct a $SU(2)_L\times SU(2)_R$ symmetric scalar potential, we reproduce the following result of \cite{pomarol2}:
\beqa \mathcal{V} &=& \half m_{11}^2 \Tr[M_1^\dagger M_1] +\half m_{22}^2 \Tr[M_2^\dagger M_2]-m_{12}^2 \Tr[M_1^\dagger M_2]+ \frac{1}{4} \lambda_1 \left(\Tr[M_1^\dagger M_1]\right)^2 \nonumber\\
&& +\frac{1}{4} \lambda_2 \left(\Tr[M_2^\dagger M_2]\right)^2+ \quarter \lambda_3 \Tr[M_1^\dagger M_1] \Tr[M_2^\dagger M_2] +\half \lambda \left(\Tr[M_1^\dagger M_2]\right)^2 \nonumber\\
&&+\quarter \left( \lambda_6 \Tr[M_1^\dagger M_1] +\lambda_7 \Tr[M_2^\dagger M_2] \right)\Tr[M_1^\dagger M_2].\label{pompot}\eeqa
Comparing \eq{pompot} to the most general form of the scalar potential in \eq{pot}, we find that custodial symmetry imposes the following restrictions on the coefficients of the scalar potential:
\beq m_{12}^* = m_{12}\,,\,\,\lambda^*_{5,6,7} = \lambda_{5,6,7}\,,\,\,\lambda = \lambda_4 = \lambda_5\,.\eeq
Let us now convert from the generic basis to the Higgs basis. Plugging $\Phi_a= H_1 \widehat v_a+  H_2 \widehat v^*_{\bbar}\epsilon_{ba}$ into \eq{tr} yields
\beqa\label{traces}
\Tr[M_1^\dagger M_1] &=& 2 \left( |\vo|^2 H_1^\dagger H_1 + |\vt|^2 H_2^\dagger H_2 -\vo\vt H_2^\dagger H_1 - \vo^* \vt^* H_1^\dagger H_2 \right) \nonumber\\
\Tr[M_2^\dagger M_2] &=& 2 \left( |\vt|^2 H_1^\dagger H_1 + |\vo|^2 H_2^\dagger H_2 +\vo\vt H_2^\dagger H_1+ \vo^* \vt^* H_1^\dagger H_2 \right) \nonumber\\
\Tr[M_1^\dagger M_2] &=&   \vt^* \vo H_1^\dagger H_1 -\vo \vt^* H_2^\dagger H_2 +\left(\vo^2 H_2^\dagger H_1-\vt^{*^2} H_1^\dagger H_2  + h.c.\right).\eeqa  
Subsituting equations (\ref{traces}) into \eq{pompot} and grouping like terms yields the following:
\beqa \label{custpot}
\mathcal{V}&=& Y_1 H_1^\dagger H_1+ Y_2 H_2^\dagger H_2
+[Y_3 H_1^\dagger H_2+{\rm h.c.}]\nonumber \\[5pt]
&&\quad
+\half Z_1(H_1^\dagger H_1)^2 +\half Z_2(H_2^\dagger H_2)^2
+Z_3(H_1^\dagger H_1)(H_2^\dagger H_2)
+Z_4( H_1^\dagger H_2)(H_2^\dagger H_1) \nonumber \\[5pt]
&&\quad +\left\{\half Z_5 (H_1^\dagger H_2)^2
+\big[Z_6 (H_1^\dagger H_1)
+Z_7 (H_2^\dagger H_2)\big]
H_1^\dagger H_2+{\rm h.c.}\right\}\,,
\eeqa
  
where
\beqa
Y_1 &=& m_1^2 |\vo|^2 + m_2^2 |\vt|^2 -m_{12}^2\vcc \,,\nonumber\\
Y_2 &=& m_1^2 |\vt|^2 + m_2^2 |\vo|^2 +m_{12}^2  \vcc\,,\nonumber\\
Y_3 &=& m_{12}^2 (\vt^{*2} - \vo^{*2})  + \vt^* \vo^* (m_2^2 - m_1^2)\,, \nonumber\\
Z_1 &=& 2\left(\lambda_1 |\vo|^4 + \lambda_2 |\vt|^4 + \lambda_3 |\vo|^2 |\vt|^2\right)+\lambda\vcc^2 \nonumber \\&&+ (\lambda_6 |\vo|^2  + \lambda _7 |\vt|^2)\vcc\,,\nonumber\\
Z_2 &=& 2\left(\lambda_1 |\vt|^4 + \lambda_2 |\vo|^4 + \lambda_3 |\vo|^2 |\vt|^2\right)+\lambda\vcc^2 \nonumber \\&&- (\lambda_6 |\vt|^2  + \lambda _7 |\vo|^2)\vcc\,,\nonumber\\
Z_3&=&2(\lambda_1+\lambda_2)|\vo|^2|\vt|^2 + \lambda_3(|\vo|^4+|\vt|^4 )-\lambda\vcc^2 \nonumber \\
&&+ \half(\lambda_6-\lambda_7)\vcc\left(|\vt|^2-|\vo|^2\right)\,,\nonumber\\
Z_4&=&2(\lambda_1+\lambda_2-\lambda_3)|\vo|^2 |\vt|^2+\lambda|\vo-\vt|^2 |\vo+\vt|^2\nonumber \\
&&+\half(\lambda_6-\lambda_7)\left[\vt\vo\left(\vt^{*2}-\vo^{*2}\right)+c.c.\right]\,,\nonumber\\
Z_5  &=& 2(\lambda_1+\lambda_2-\lambda_3)(\vt^* \vo^*)^2 +\lambda(\vo^{*2}-\vt^{*2})^2 + (\lambda_6-\lambda_7)\vt^* \vo^*(\vt^{*2} - \vo^{*2})\,,\nonumber\\
Z_6 &=&\left[\lambda_2 |\vt|^2 - \lambda_1 |\vo|^2 + \lambda_3(|\vo|^2 - |\vt|^2) \right]\vo^* \vt^*+ \lambda\vcc(\vo^{*2}-\vt^{*2})\nonumber\\
&& +\half \left(\lambda_6|\vo|^2+\lambda_7|\vt|^2\right)(\vo^{*2}-\vt^{*2})+ \half \left(\lambda_7-\lambda_6\right)\vo^*\vt^*\vcc\,,\nonumber\\
Z_7&=&\left[\lambda_2 |\vo|^2 - \lambda_1 |\vt|^2 + \lambda_3(|\vt|^2 - |\vo|^2) \right]\vo^* \vt^*-\lambda\vcc(\vo^{*2}-\vt^{*2})\nonumber\\
&& +\half \left(\lambda_6|\vt|^2+\lambda_7|\vo|^2\right)(\vo^{*2}-\vt^{*2})- \half \left(\lambda_7-\lambda_6\right)\vo^*\vt^*\vcc\,.\label{zlist}
\eeqa

At this point, only $Y_1, Y_2, Z_1, Z_2, Z_3,$ and $Z_4$ are manifestly real.  Now, however, we can apply the custodial symmetry condition from \eq{vevsym}, and use the fact that all of the $m^2$ and $\lambda$ parameters are real.  Then we find
\beqa Y_3,\,Z_6,\,Z_7  \in \mathbb{R}\,,&\nonumber\\ 
Z_4\, = \,Z_5\,=& 2(\lambda_1+\lambda_2-\lambda_3)\vo^2\vt^2 + \lambda(\vo^2-\vt^2)^2+(\lambda_6-\lambda_7)\vt\vo(\vt^2-\vo^2) \in \mathbb{R}\,.\nonumber\\
\phantom{line}\label{real}\eeqa
Since we now have a Lagrangian written in terms of ``real basis" parameters, one can say that the condition for custodial invariance of the scalar potential is \beq Z_4 = Z_5 \mbox{ in the real basis.}\label{condcust} \eeq

To check that we do not expect any additional relations among the $Z$ parameters, we can compare the number of degrees of freedom in \eq{zlist} and \eq{pompot}.  In the most general CP-violating 2HDM, one starts with 6 real and 4 complex parameters in the scalar potential, plus $v$.  From these 15 degrees of freedom, 3 are removed by applying the scalar minimum conditions and one corresponds to an overall phase, which is not physically significant.  From these 11 physical degrees of freedom, the conditions of CP conservation [\eq{cpoddinv}] remove three, but there is no longer an overall phase to subtract, so there are 9 independent degrees of freedom. The condition of custodial symmetry removes an additional degree of freedom, leaving eight for the custodially-symmetric potential in \eqs{pompot}{zlist}.  

We will now replicate ``case II'' of Pomarol and Vega and show that it produces exactly the same condition on the parameters of the Lagrangian as \eq{condcust}.
The alternative to the matrices in \eq{matdef} is the following: 
\beq M_{21} \equiv (\tilde{\Phi}_2 \,,\Phi_1)\,, \label{matdeftwo}\eeq  
which transforms as  
\beq M_{21} \rightarrow L~M_{21}~R^\dagger\,,\eeq under $SU(2)_L \times SU(2)_R $. 
Preserving a custodial $SU(2)_V$ symmetry after EWSB requires that the vev of $M_{21}$ be proportional to the identity matrix:
\beq\label{vevs2} <M_{21}> = \frac{1}{\sqrt{2}}\left(\begin{array}{cc}v_2^*&0\\0&v_1\end{array}\right) \propto {\mathbb 1} \Rightarrow v_1^* = v_2\,,\eeq
since for $L=R$,
\beq <M_{21}> \rightarrow L <M_{21}> R^\dagger =  \frac{v_2}{\sqrt{2}} L R^\dagger= <M_{21}>\,. \eeq
From this matrix we construct the following:
\beqa
\Tr[M_{21}^\dagger M_{21}] &=& \Tr\left[\left(\begin{array}{cc}\Phi_2^0&-\Phi_2^+\\\Phi_1^-&\Phi_1^{0*}\end{array}\right)
\left(\begin{array}{cc}\Phi_2^{0*}&\Phi_1^+\\-\Phi_2^-&\Phi_1^{0}\end{array}\right)\right]
=  \Phi_1^\dagger \Phi_1+ \Phi_2^\dagger \Phi_2\,, \nonumber\\
\det[M_{21}^\dagger] &=& \det \left(\begin{array}{cc}\Phi_2^0&-\Phi_2^+\\\Phi_1^-&\Phi_1^{0*}\end{array}\right)=\Phi_1^\dagger \Phi_2\nonumber\,,\\
\det[M_{21}^\dagger M_{21}] &=&  \det\left[\left(\begin{array}{cc}\Phi_2^0&-\Phi_2^+\\\Phi_1^-&\Phi_1^{0*}\end{array}\right)
\left(\begin{array}{cc}\Phi_2^{0*}&\Phi_1^+\\-\Phi_2^-&\Phi_1^{0}\end{array}\right)\right]\nonumber\\
&=&(\Phi_1^\dagger \Phi_2) (\Phi_2^\dagger \Phi_1)\,.\label{tr2}\eeqa  

With the expressions in \eq{tr2}, one constructs the following $SU(2)_L\times SU(2)_R$ symmetric scalar potential, again replicating the results in \cite{pomarol2}:
\beqa \mathcal{V} &=& m^2 \Tr[M_{21}^\dagger M_{21}] - \left(m_{12}^2 \det[M_{21}]+h.c.\right)+ \lambda \Tr[M_{21}^\dagger M_{21}]^2 +  \lambda_4 \det[M_{21}^\dagger M_{21}]\nonumber\\
&&  +\half \left(\lambda_5 \det[M_{21}^\dagger]^2 + \lambda'  \det[M_{21}^\dagger]\Tr[M_{21}^\dagger M_{21}]+h.c. \right)\,.\label{pompot2}\eeqa
Hermiticity implies $m^2, \lambda, \lambda_4 \in \mathbb{R}$. As before, we convert to the Higgs basis: 
\beqa\label{traces2}
\Tr[M_{21}^\dagger M_{21}] &=&  H_1^\dagger H_1 +  H_2^\dagger H_2  \nonumber\\
\det[M_{21}^\dagger]  &=& \vo^* \vt \left(H_1^\dagger H_1 -H_2^\dagger H_2\right) -\vt^2 H_2^\dagger H_1+ \vo^{*2}  H_1^\dagger H_2  \nonumber\\
\det[M_{21}^\dagger M_{21}] &=&  \left[-\vt^2 \vo^2(H_2^\dagger H_1)^2+(|\vo|^2-|\vt|^2)\vo^*\vt^*\left(H_1^\dagger H_1 -H_2^\dagger H_2\right) H_1^\dagger H_2  + h.c.\right]\nonumber\\
&&+|\vt|^2 |\vo|^2 \left(H_1^\dagger H_1 -H_2^\dagger H_2\right)^2+\left(|\vo|^4+|\vt|^4\right) H_1^\dagger H_2 H_2^\dagger H_1 \,.\eeqa  
Subsituting equations (\ref{traces2}) into \eq{pompot2}, we again achieve \eq{custpot}, but now with different coefficients:
\beqa
Y_1 &=& m^2-m_{12}^2 \vo^* \vt - m_{12}^{*2} \vo \vt^*\,,\nonumber\\
Y_2 &=& m^2+m_{12}^2 \vo \vt^* + m_{12}^{*2} \vo^* \vt\,,\nonumber\\
Y_3 &=& -m_{12}^2 \vo^{*2}  + \vt^{*2}m_{12}^{*2}  \,, \nonumber\\
Z_1 &=& 2\lambda+ 2 \lambda_4|\vo|^2|\vt|^2+\left[\lambda_5(\vo^* \vt)^2 + \lambda' \vo^* \vt +c.c.\right]\,,\nonumber\\
Z_2 &=& 2\lambda+ 2 \lambda_4|\vo|^2|\vt|^2+\left[\lambda_5(\vo^* \vt)^2 - \lambda' \vo^* \vt +c.c.\right]\,,\nonumber\\
Z_3&=&2\lambda- 2 \lambda_4|\vo|^2|\vt|^2-\left[\lambda_5(\vo^* \vt)^2 +c.c.\right]\,,\nonumber\\
Z_4&=&  \lambda_4\left(|\vo|^4+|\vt|^4\right)-\left[\lambda_5(\vo^* \vt)^2 +c.c.\right]\,,\nonumber\\
Z_5&=&  -2\lambda_4 \vo^2 \vt^2 +\left[\lambda_5\vo^{*4} +\lambda_5^*\vt^{*4}\right]\,,\nonumber\\
Z_6 &=&\lambda_4(|\vo|^2-|\vt|^2)\vt^*\vo^* + \left(\lambda_5 \vo^{*3} \vt  - \lambda_5^* \vt^{*3} \vo+ \lambda'\vo^{*2}-\lambda'^*\vt^{*2}  \right)\,,  \nonumber\\
Z_7 &=&\lambda_4(|\vt|^2-|\vo|^2)\vt^*\vo^* + \left(-\lambda_5 \vo^{*3} \vt  + \lambda_5^* \vt^{*3} \vo+ \lambda'\vo^{*2} -\lambda'^*\vt^{*2}  \right)\,. \label{zlisttwo}\eeqa

As usual, $Y_1, Y_2, Z_1, Z_2, Z_3,$ and $Z_4$ are manifestly real.  Applying \eq{vevs2} and the fact that $m^2$, $\lambda_4$ and $\lambda$ are real, we find that in the custodial limit,
\beqa
iY_3,~iZ_6,~ iZ_7  \in \mathbb{R}\,,\nonumber\\ 
Z_4 = - Z_5  = 2\lambda_4|\vo|^4- \lambda_5 \vt^4 - \lambda_5^* \vo^4 \in \mathbb{R}\,.\eeqa
Again we expect no further relations, since we have reduced the number of independent degrees of freedom in \eq{zlisttwo} to 8, the same as in \eq{traces2}. Since the phase of $H_2$ is not physically meaningful, we are free to transform $H_2 \rightarrow i H_2$, which leads to $(Y_3,~Z_6,~Z_7) \rightarrow -i (Y_3,~Z_6,~Z_7)$ and $Z_5 \rightarrow - Z_5$, putting all of the parameters in the real basis. Furthermore, since the sign of $Z_5$ reverses, one can now make the statement that the condition for custodial invariance is ``$Z_4 = Z_5$ in the real basis,''\footnote{This statement is meaningful only if the sign of $Z_5$ in the real basis is a physical observable. If either $Z_6 \neq 0$ or $Z_7 \neq 0$, the operation $H_2 \rightarrow \pm i H_2$ transforms the couplings out of the real basis while changing the sign of $Z_5$, (as noted above).  Hence in this case the sign of $Z_5$ in the real basis is meaningful.  If $Z_6 = Z_7 = 0$, then $H_2 \rightarrow \pm i H_2$  changes the sign of $Z_5$ while preserving the real basis.  In this case the sign of $Z_5$ is not a meaningful and custodial symmetry implies that $Z_4 = |Z_5|$ in the real basis.} as in the previous case. This transformation has no effect on the vevs $<\Phi_i>$, as one can see from \eq{hbasis}.

Thus, we conclude that it is not possible to distinguish physically between these two ``cases'' of custodial symmetry--the only difference between them is the condition imposed on the vevs.  The relationship between the vevs depends on the choice of basis, and thus cannot be physically measurable.\footnote{In ref.~\cite{haberpomarol}, it was shown that for custodial symmetry in the quark-scalar sector of specialized versions of the 2HDM, the two ways to implement custodial symmetry can be distinguished based on the presence of the $A^0 G G$ effective interactions in the $E\gg m_W$ limit.  However, in their discussion, constraints on the Yukawa couplings effectively select a ``preferred'' basis. } 
\subsection{The Basis-Independent Condition for Custodial Symmetry in the Scalar Sector}
It is possible generalize the two implementations of custodial symmetry presented in the previous section by constructing a $SU(2)_L\times SU(2)_R$ invariant scalar potential using Higgs basis fields, which avoids having to impose conditions on the vevs. For the purposes of this section, we will take all scalar couplings to be in the real basis, so that
\beq Y_i \in \mathbb{R}, \quad Z_i \in \mathbb{R} \quad\quad \forall~i\,.\eeq
This requirement removes the freedom to redefine the phase of $H_2$ except by an overall sign (which would not change the results presented here).
Let us define\footnote{In the special case of $Z_6 = 0, c_{13}=0,$ and $\Im[Z_5 \,e^{2i(\theta_{12}- \theta_{23})}]=0$, \eq{matdefgen} applies with the substitution $\theta_{23} \rightarrow \theta_{23}-\theta_{12}$. In other mass-degenerate cases that we have treated here, in principle a different combination of mixing angles would play the role of $\theta_{23}$.} 
\beq \mathbb{M}_1 \equiv (\tilde{H}_1\,,H_1), \hspace{5mm}\mathbb{M}_2 \equiv (e^{-i\theta_{23}}\tilde{H_2}\,,e^{i\theta_{23}}H_2)\,, \label{matdefgen}\eeq
where the transformation under a global $SU(2)_L \times SU(2)_R $ is defined as usual:
\beq \mathbb{M}_i \rightarrow L~\mathbb{M}_i~R^\dagger\,.\label{htransfm}\eeq 
In section \ref{custsec} we explicitly required that $<\mathbb{M}_i>$ be proportional to the identity matrix so that the custodial symmetry is preserved after EWSB. Here, since we are writing the fields in the Higgs basis, we do not have impose any specific conditions on the vevs, since $<\mathbb{M}_i>$ are automatically proportional to the unit matrix.  Now the $SU(2)_L \times SU(2)_R $ invariant potential can be written as follows:
\beqa \mathcal{V} &=& \half Y_1 \Tr[\mathbb{M}_1^\dagger \mathbb{M}_1] +\half Y_2 \Tr[\mathbb{M}_2^\dagger \mathbb{M}_2]+ Y_3e^{-i\theta_{23}} \Tr[\mathbb{M}_1^\dagger \mathbb{M}_2]+ \frac{1}{8} Z_1 \left(\Tr[\mathbb{M}_1^\dagger \mathbb{M}_1]\right)^2 \nonumber\\
&& +\frac{1}{8} Z_2 \left(\Tr[\mathbb{M}_2^\dagger \mathbb{M}_2]\right)^2+ \quarter Z_3 \Tr[\mathbb{M}_1^\dagger \mathbb{M}_1] \Tr[\mathbb{M}_2^\dagger \mathbb{M}_2] +\half \lambda \left(\Tr[\mathbb{M}_1^\dagger \mathbb{M}_2]\right)^2 \nonumber\\
&&+\half \left( Z_6 e^{-i\theta_{23}}\Tr[\mathbb{M}_1^\dagger \mathbb{M}_1] +Z_7 e^{-i\theta_{23}}\Tr[\mathbb{M}_2^\dagger \mathbb{M}_2] \right)\Tr[\mathbb{M}_1^\dagger \mathbb{M}_2]\,,\label{cupo}\eeqa
where the coefficients have been adjusted to anticipate comparison with the standard form of the scalar potential.
\Eq{cupo} is equivalent to 
\beqa \nonumber
\mathcal{V}&=& Y_1 H_1^\dagger H_1+ Y_2 {H_2}^{\dagger} H_2
+[Y_3 H_1^\dagger {H_2}+{\rm h.c.}]\nonumber \\[5pt]
&&\quad+\half Z_1(H_1^\dagger H_1)^2 +\half Z_2({H_2}^\dagger {H_2})^2
+Z_3(H_1^\dagger H_1)({H_2}^\dagger {H_2})
+Z_4( H_1^\dagger {H_2})({H_2}^\dagger H_1) \nonumber \\[5pt]
&&\quad +\left\{\half Z_5  (H_1^\dagger {H_2})^2
+\big[Z_6 (H_1^\dagger H_1)+Z_7 ({H_2}^\dagger {H_2})\big]
H_1^\dagger {H_2}+{\rm h.c.}\right\}\,,\label{custpotgen}
\eeqa
with the condition \beq Z_4= Z_5 e^{-2i\theta_{23}}\,,\quad\zsixi=0\,,\quad\zseveni=0\,.\label{con}\eeq
Note that $\Im[Z_5 e^{-2i\theta_{23}}]=0$ since $Z_4$ is manifestly real.  \Eq{custpotgen} is consistent with the results derived previously in a specific basis.  One can rewrite \eq{con} so that $\theta_{23}$ does appear explicitly.  Custodial symmetry requires that $Z_5 Z_6^{*2}$, $Z_5 Z_7^{*2}$, and $Z_5 (\rho^Q)^{2}$ be real-valued, so \eq{con} is equivalent to
\beqa Z_4 &=& \pm\varepsilon_{56}|Z_5|\,\qquad \mbox{\text{if}}\,Z_6 \neq 0\,,\nonumber\\
Z_4 &=& \pm\varepsilon_{57}|Z_5|\,\qquad \mbox{\text{if}}\,Z_6 = 0\,,\,Z_7 \neq 0\,,\nonumber\\
Z_4 &=& \pm\varepsilon_{5Q}|Z_5|\,\qquad \mbox{\text{if}}\,Z_6 = Z_7 = 0\,,\,\rho^Q\neq 0\,,
\label{invcondit}\eeqa
where the positive sign corresponds to $\zsixi = \zseveni = \Im(\rho^Qe^{i\theta_{23}})=0$ and the negative sign to $\zsixr = \zsevenr = \Re(\rho^Qe^{i\theta_{23}})=0$. If $Z_6 = Z_7 = \rho^Q =0$, \eq{con} has two solutions, which in the real basis can be written
\beq Z_4 = \pm |Z_5|\,,\label{spec}\eeq reflecting the two possible definitions of CP symmetry which arise in this case. Since both CP symmetries are conserved, the condition in \eq{spec} is indeterminate.

\subsection{Degeneracy in the Custodial Limit}
Finally, one important consequence of custodial symmetry in the Higgs sector is that the charged Higgs boson is always degenerate with one of the neutral Higgs fields.  Since the custodial limit is CP-conserving, we can write the following expressions for the CP-odd mass from \eq{oddmasses}: 
\beqa m_{A^0}^2 &=& Y_2 + \half[Z_3 + Z_4 - \zfiver]v^2\nonumber\,,\\
&&\mbox{\text{for }}\, \zsixi = \zseveni = \Im(\rho^Qe^{i\theta_{23}})=0\,.\label{odddeg}\eeqa
Note that this is the only possible value of $m_{A^0}^2$ since the condition for custodial symmetry in \eq{con} eliminates the possibility that \eq{evenmass} is correct (as long as either $Z_6$ or $Z_7$ is non-zero).  Thus, the condition $Z_4 = \zfiver$ of \eq{invcondit} implies a degeneracy between the CP-odd Higgs and the charged Higgs, ie, 
\beq m_{A^0}^2 = \mhpm^2 =  Y_2 + \half Z_3 v^2\,. \label{de1}\eeq 
This observation agrees with that of \cite{pomarol2}.  However, in the case of $Z_6 = Z_7 = 0$, there is another possibility for the mass of the CP-odd field from \eq{evenmass},
\beq m_{A^0}^2 = Y_2 + \half[Z_3 + Z_4 + \zfiver]v^2\quad\mbox{\text{for }}\, \Re(\rho^Qe^{i\theta_{23}})=0\,.\eeq  This unique case arises because $\rho^Q \thet$ can be either imaginary or real, unlike $Z_{6,7} \thetminus$ which must be real for the scalar potential to be custodially symmetric.  (As we will see in \sect{hqcust}, imposing custodial symmetry on the Higgs-Quark sector does not require $\rho^Q \thet$ to be real.)
In this case it is the CP-even field $H^0$ that is degenerate with the charged Higgs boson:
\beq m_{H^0}^2 = \mhpm^2 =  Y_2 + \half Z_3v^2\,. \label{de2}\eeq 
These two possibilities for degeneracy arise from the two possible definitions of CP mentioned in section \ref{sec:herquetcpcons}.  This phenomenon of a degenerate charged Higgs boson and CP-even Higgs boson was described for the case $Z_6 = Z_7 = 0$ in  \cite{herquet}, in the context of a ``twisted" custodial symmetry.  In fact, this scenario can be analyzed without the authors' ``twisting" formalism, as we have seen here. 
Finally, we consider the case $Z_6 = Z_7 = \rho^Q = 0.$  With nothing to select out one of the two possible definitions of CP, it cannot be determined whether it is the CP-even or CP-odd neutral Higgs boson that is degenerate.  We emphasize, in contrast to the authors of \cite{herquet}, that these two possbilities in \eqs{de1}{de2} arise as a result of the two inequivalent definitions of the CP transformation, which exist (independently of the custodial limit) at the special point of parameter space where $Z_6 = Z_7 = \rho^Q = 0$.  This is the only case in which the mass-degeneracy $\mhpm = m_{H^0}$ corresponds to a custodial symmetry.
\subsection{Basis-Dependent Formulations of Custodial Symmetry in the Higgs-Quark Sector}
Let us now extend the two ways to impose $SU(2)_L \times SU(2)_R$ invariance discussed in the previous section to the Higgs-Quark sector.  As before, in the final analysis these two cases defined by Pomarol and Vega will be related by basis-transformations, and do not represent the most general way of implementing the symmetry. In Pomarol and Vega's ``case I,'' equation (\ref{ymodeliii}) becomes
\beq \label{custinvy}
-\mathscr{L}_{\rm Y}=\eta_1 \overline \qlcal~ M_1 ~\left(\begin{array}{c}\mathcal{U}_R\\D_R\end{array}\right)+\eta_2 \overline \qlcal~ M_2 ~\left(\begin{array}{c}\mathcal{U}_R\\D_R\end{array}\right)+{\rm h.c.}\,,\eeq
with $M_i$ defined in \eq{matdef}. This Lagrangian is manifestly invariant under  $SU(2)_L \times SU(2)_R$ given the transformations
\beqa {M}_i &\rightarrow& LM_i R^\dagger\,,\nonumber\\ 
\left(\begin{array}{c}\mathcal{U}_R\\D_R\end{array}\right) &\rightarrow& R \left(\begin{array}{c}\mathcal{U}_R\\D_R\end{array}\right),\nonumber\\
\overline \qlcal  &\rightarrow& \overline \qlcal L^\dagger\,.\label{transf}\eeqa
Comparing with \eq{ymodeliii}, one finds the conditions
\beq \label{caseonecon}
\eta^U_1= \eta^{D\dagger}_1\,,\qquad\qquad
\eta^U_2= \eta^{D\dagger}_2\,.
\eeq
With this constraint, the parameters $\kappa^Q$ and $\rho^Q$ defined in \eq{kapparho} become
\beqa \label{kappone}
\kappa^U&= v_1^*\eta^{U}_1+v_2^*\eta^{U}_2\,,\qquad\qquad
\kappa^{D\dagger}&= v_1\eta^{U}_1+v_2\eta^{U}_2\,,\nonumber\\
\rho^U&= - v_2 \eta^{U}_1+v_1\eta^{U}_2\,,\qquad\qquad
\rho^{D\dagger}&= - v_2^*\eta^{U}_1+v_1^*\eta^{U}_2\,.
\eeqa
Since the $v_i$ are real in this case [equation (\ref{vevsym})], we find $\kappa^U = \kappa^{D\dagger}$, or simply $M_U = M_D$.  Similarly, $\rho^{D\dagger}=\rho^U$.  The parameters $\rho^Q$ are not quite basis-independent, since they pick up a factor of det(U) under a U(2) transformation.  However, from the previous section, we know that we are in the real basis.  Thus, one state unambiguous conditions for custodial symmetry in the Yukawa sector:
\beq M_U = M_D,\,\,\, \rho^U = \rho^{D\dagger}\mbox{ in the real basis.}\label{unamb}\eeq

Let us now do the same for Pomarol and Vega's ``case II.''  We construct
\beq \label{custinvy2}
-\mathscr{L}_{\rm Y}=\eta_{21} \overline \qlcal~ {M}_{21} ~\left(\begin{array}{c}\mathcal{U}_R\\D_R\end{array}\right)+\eta_{12} \overline \qlcal~ {M}_{12} ~\left(\begin{array}{c}\mathcal{U}_R\\D_R\end{array}\right)+{\rm h.c.}\,,\eeq
with $M_{21}$ defined in \eq{matdeftwo}, and a similar matrix defined as
\beq M_{12} \equiv \tilde{\Phi}_1^* \Phi_2\,. \label{mattwo}\eeq 
This Lagrangian is invariant under  $SU(2)_L \times SU(2)_R$ given 
\beqa {M}_{12} &\rightarrow& L{M}_{12} R^\dagger\,,\nonumber\\
{M}_{21} &\rightarrow& L{M}_{21} R^\dagger\,. \eeqa

Comparing \eq{custinvy2} with \eq{ymodeliii}, one finds the conditions
\beq \label{casetwocon}
\eta^U_1= \eta^{D\dagger}_2\,,\qquad\qquad
\eta^U_2= \eta^{D\dagger}_1\,.
\eeq
Substituting equations (\ref{casetwocon}) and (\ref{vevs2}) into \eq{kapparho} yields
\beqa \label{kapptwo}
\kappa^U&= v_1^*\eta^{U}_1+v_2^*\eta^{U}_2\,,\qquad\qquad
\kappa^{D\dagger}&= v_2^*\eta^{U}_2+v_1^*\eta^{U}_1\,,\nonumber\\
\rho^U&= - v_2 \eta^{U}_1+v_1\eta^{U}_2\,,\qquad\qquad
\rho^{D\dagger}&= v_2\eta^{U}_1-v_1\eta^{U}_2\,.
\eeqa
Thus, one again finds $M_U = M_D$.  We also have $\rho^{D\dagger}=-\rho^U$.  From the previous section, we know that to express these quantities in the real basis, we need to make the transformation $H_2 \rightarrow i H_2$, which is equivalent to 
\beq   \rho^U \rightarrow i \rho^{U}   \,,\qquad\qquad \rho^{D\dagger} \rightarrow -i \rho^{D\dagger} .\eeq  Thus, the conditions for custodial invariance of the Yukawa sector are given by \eq{unamb}, as in the previous case.
\subsection{The Basis-Independent Custodially-Symmetric Higgs-Quark Lagrangian\label{hqcust}}
Now we will determine, in analogy to the previous section, the form of the  $SU(2)_L \times SU(2)_R$-invariant Higgs-Quark Lagrangian, assuming that the fields and scalar couplings are in the real basis.  Let us rewrite \eq{hbasisymodeliii} as follows:
\beq\label{rewrityuk}
-\mathscr{L}_{\rm Y}=\overline \qlcal (\tilde H_1\kappa^U+\tilde H_2\rho^U) \mathcal{U}_R
+\overline \qlcal (H_1\kappa^{D\,\dagger}+ H_2\rho^{D\,\dagger}) D_R +{\rm h.c.}\,.
\eeq
Then the desired lagrangian is of the form
\beq \label{custinvygen}
-\mathscr{L}_{\rm Y}= \kappa \overline \qlcal~ \mathbb{M}_1 ~\left(\begin{array}{c}\mathcal{U}_R\\D_R\end{array}\right)+ \rho \overline \qlcal~ \mathbb{M}_2 ~\left(\begin{array}{c}\mathcal{U}_R\\D_R\end{array}\right)+{\rm h.c.}\,,\eeq
with
$ \mathbb{M}_1 \equiv \tilde{H}_1~H_1 $ and $\mathbb{M}_2 \equiv (e^{-i\theta_{23}}\tilde{H_2}\,,e^{i\theta_{23}}H_2)$
as in \eq{matdefgen}.  
Comparing to \eq{rewrityuk}, one finds that the condition for invariance under $SU(2)_L \times SU(2)_R$ is 
\beq \kappa^U = \kappa^{D\dagger},   \quad\quad e^{-i\theta_{23}}\rho^U = e^{i\theta_{23}}\rho^{D\dagger}\,.\label{yukcondi}\eeq

Again, one can recover Pomarol and Vega's two cases by choosing to constrain the vevs of the Higgs fields. However, if there is no special symmetry imposed on the Yukawa sector (as we are assuming here), then there is nothing to select a ``preferred'' basis.  There are an infinite number of unphysical basis choices which can be made, but one can always write down unambiguous conditions either by going to real basis, or by using invariant combinations as we have done in this section.  

\chapter{Phenomenology of the 2HDM}
\section{The Oblique Parameters $S$, $T$ and $U$}
The parameters S, T, and U are independent UV-finite combinations of radiative corrections to gauge boson vacuum polarization diagrams (aka ``oblique'' corrections). $T$ is related to the parameter $\rho$ by $\rho - 1 = \alpha T$. They are calculated from the transverse part of the gauge boson two-point function:
\beqa
\label{obliqueSdef}
S &\equiv&-\frac{16\pi c_W}{s_W g^2}\frac{d}{dq^2}\Pi_{W^3 B}(q^2)\vert_{q^2=0}  \\
\label{obliqueTdef}\alpha T &\equiv& \frac{1}{m_W^2}\left[\Pi_{W^1W^1}(0)-\Pi_{W^3W^3}(0)\right]\,,\\
U&\equiv& \frac{16\pi}{g^2}\frac{d}{dq^2}\left[\Pi_{W^1W^1}(q^2)-\Pi_{W^3W^3}(q^2)\right]\vert_{q^2=0}\,.
\eeqa
Here, $\Pi_{ij}(q^2)$ is defined by
\beq \label{pi}i \Pi^{\mu \nu}_{ij} = i g^{\mu \nu} \Pi_{ij}(q^2) + (q^{\mu}q^\nu \mbox{ terms})\,. \eeq
Some care must be taken as this differs from the convention of ~\cite{appelquist,Peskin:1990zt,Peskin:1991sw}, and others, who pull out factors of $g^2,$ etc.    

The ``linear expansion approximation'' is often used in the literature, which extracts the dependence of the functions on $q^2$, as in \Refs{Peskin:1990zt}{Peskin:1991sw}:
\beq
\label{piA}\Pi_{ij}(q^2) = A_{ij}(0) + q^2 F_{ij}(q^2)\,. \eeq
For the case that the scale of the new physics is much greater than $m_Z$, one can take the momentum scale $q^2$ to be of order $m_Z^2$, and define the oblique parameters as follows~\cite{habertasi}:
\beqa
\label{Sdef}
\frac{g^2}{16\pi} S &\equiv& c_W^2\left[F_{ZZ}(m_Z^2) - F_{\gamma\gamma}(m_Z^2)+\left(\frac{2s_W^2-1}{s_W c_W}\right)F_{Z\gamma}(m_Z^2)\right]\,,\\
\label{Tdef}\alpha T &\equiv& \frac{A_{WW}(0)}{m_W^2}- \frac{A_{ZZ}(0)}{m_Z^2}\\
\frac{g^2}{16\pi}(S + U)&\equiv& F_{WW}(m_W^2)- F_{\gamma\gamma}(m_W^2)-\frac{c_W}{s_W}F_{Z\gamma}(m_W^2)\,.
\eeqa
The combination $S + U$ rather than $U$ has been taken for calculational simplicity. $S$, $T$ and $U$ are defined relative to the Standard Model, so that extending the Higgs sector has the effect of shifting the parameters away from zero.\footnote{Contributions to the oblique parameters from the Higgs-fermion Yukawa couplings will not be included here, since they arise from diagrams of order two-loops and higher.} Thus, in the calculations that follow, contributions from Standard Model processes have been subtracted out.  The Higgs mass used in this ``reference'' Standard Model ($m_\phi$) will be left arbitrary for the moment.

The result for $S$ is calculated in Appendix~\ref{app:stu} and found to be:
\beqa \label{seqn}\nonumber
 S&=& \frac{1}{\pi m_Z^2} \left[q_{k1}^2\mathcal{B}_{22}(m_Z^2;m_Z^2,m_k^2)-\mathcal{B}_{22}(m_Z^2;m_Z^2,m_\phi^2)- m_Z^2 q_{k1}^2\mathcal{B}_{0}(m_Z^2;m_Z^2,m_k^2) \right.\\\nonumber
& &+ m_Z^2\mathcal{B}_{0}(m_Z^2;m_Z^2,m_\phi^2)+q_{11}^2 \mathcal{B}_{22}(m_Z^2;m_2^2,m_3^2)+q_{21}^2\mathcal{B}_{22}(m_Z^2;m_1^2,m_3^2)\\
& &\left.+ q_{31}^2\mathcal{B}_{22}(m_Z^2;m_1^2,m_2^2) - \mathcal{B}_{22}(m_Z^2;\mhpm^2,\mhpm^2)\right]\,.
\eeqa
Here $\mhpm$ represents the mass of the charged Higgs $H^{\pm}$, as usual, and $m_k$ the masses of the neutral Higgs $h_k$ ($k = 1,2,3$). Repeated indices are summed over.  The notation $\mathcal{B}_{22}$ and $\mathcal{B}_{0}$ was introduced in ~\cite{habertasi}:
\beqa
\mathcal{B}_{22}(q^2;m_1^2,m_2^2) \equiv B_{22}(q^2;m_1^2,m_2^2)-B_{22}(0;m_1^2,m_2^2)\,,\\  
\mathcal{B}_{0}(q^2;m_1^2,m_2^2) \equiv B_{0}(q^2;m_1^2,m_2^2)-B_{0}(0;m_1^2,m_2^2)\,.
\eeqa
The functions $B_{22}$ and $B_{0}$ are defined in ref.~\cite{Passarino:1978jh} and come from the evaluation of two-point integrals.  They can be evaluated using the following formulae of ref.~\cite{habertasi}:
\beqa 
\label{eqnb}
\nonumber
B_{22}(q^2;m_1^2,m_2^2) &=& \frac{1}{4}(\Delta+1)[m_1^2+m_2^2-\frac{1}{3}q^2]-\half\int^1_0\,dx X \ln(X-i\epsilon)\,, \\\nonumber
B_{0}(q^2;m_1^2,m_2^2) &=& \Delta-\int^1_0\,dx \ln(X-i\epsilon)\,, \\
A_0(m^2) &=& m^2(\Delta+1-\ln m^2)\,,
\eeqa
where $X \equiv m_1^2 x + m_2^2(1-x) -q^2x(1-x)$ and $\Delta \equiv \frac{2}{4-d}+\ln(4\pi)-\gamma$, in $d$ space-time dimensions.

The calculation of $T$ and $S+U$ is also undertaken in the Appendix.  The result is (summing over repeated indices)
\beqa 
T &=& \frac{1}{16\pi m_W^2s_W^2} \left\{|q_{k2}|^2 F(\mhpm^2,m_k^2)-q_{21}^2 F(m_1^2,m_3^2)-q_{11}^2 F(m_2^2,m_3^2)\right.\nonumber\\
& &-q_{31}^2 F(m_1^2,m_2^2)+q_{k1}^2[F(m_W^2,m_k^2)-F(m_Z^2,m_k^2)]+4m_W^2 B_0(0;m_W^2,m_\phi^2)\nonumber\\
& &-4m_Z^2 B_0(0;m_Z^2,m_\phi^2)-4 q_{k1}^2[m_W^2 B_0(0;m_W^2,m_k^2)-m_Z^2 B_0(0;m_Z^2,m_k^2)]\nonumber\\
& &\left.+F(m_Z^2,m_\phi^2) -F(m_W^2,m_\phi^2)\right\}, \nonumber\\
S+U &=&\frac{1}{\pi m_W^2} \biggl[- q_{k1}^2 m_W^2 \mathcal{B}_{0}(m_W^2;m_W^2,m_k^2)+ m_W^2 \mathcal{B}_{0}(m_W^2;m_W^2,m_\phi^2)\biggr. \nonumber\\
& &- \mathcal{B}_{22}(m_W^2;m_W^2,m_\phi^2). + q_{k1}^2\mathcal{B}_{22}(m_W^2;m_W^2,m_k^2)\nonumber\\
& &\biggl.+|q_{k2}|^2 \mathcal{B}_{22}(m_W^2;\mhpm^2,m_k^2)-2 \mathcal{B}_{22}(m_W^2;\mhpm^2,\mhpm^2)\biggr]\,,\label{tsu} \eeqa
where 
\beq F(m_1^2,m_2^2) \equiv \half (m_1^2+m_2^2)-\frac{m_1^2m_2^2}{m_1^2-m_2^2}\ln\left(\frac{m_1^2}{m_2^2}\right)\,. \eeq
One can check that \beq   F(m^2,m^2) = 0\,.\eeq  This calculation $T$ has been exhibited in a basis-dependent formalism in \cite{grimus}{Grimus:2008nb}, and the result is consistent with \eq{tsu}.

\subsection{The Parameter T and the Custodial Limit}
The result for $T$ is exhibited in a different form in~\cite{pomarol2}, to emphasize which terms in $T$ arise from the breaking of the custodial symmetry and which arise from the non-custodially-invariant terms in the scalar potential.  In the custodial limit, $g' \rightarrow 0$ and $m_Z = m_W$. The terms resulting from the gauging of hypercharge, which vanish in this limit, may be rewritten so that they are proportional to $g'^2$:
\beqa \label{pomarolformT}
\alpha T&=&\frac{g'^2}{64 \pi^2}\sum_{k=1}^{3}\frac{q_{k1}^2}{m_W^2-m_Z^2}\left\{F(m_k^2,m_Z^2)-F(m_k^2,m_W^2) + F(m_\phi^2,m_W^2) - F(m_\phi^2,m_Z^2)\right.\nonumber\\
& &\quad\qquad+4[m_Z^2 B_0(0;m_Z^2,m_\phi^2)-m_W^2 B_0(0;m_W^2,m_\phi^2)]\nonumber\\
& &\qquad\quad\left.+4[m_W^2 B_0(0;m_W^2,m_k^2)-m_Z^2 B_0(0;m_Z^2,m_k^2)]\right\}\nonumber\\
& &+\frac{g^2}{64\pi^2m_W^2}\left[\sum_{k=1}^{3}|q_{k2}|^2F(m_k^2,\mhpm^2)-q_{11}^2 F(m_2^2,m_3^2)-q_{21}^2 F(m_1^2,m_3^2)\right.\nonumber\\
&&\left.-q_{31}^2 F(m_2^2,m_1^2)\right]\,,
\eeqa
where $\alpha \equiv \frac{e^2}{4 \pi}$.
Let us now focus on the terms proportional to $g^2$ in \eq{pomarolformT}.  One can verify that 
\beqa \sum_{k=1}^{3}|q_{k2}|^2F(m_k^2,\mhpm^2)-q_{11}^2 F(m_2^2,m_3^2)-q_{21}^2 F(m_1^2,m_3^2)-q_{31}^2 F(m_2^2,m_1^2) = 0\,,\nonumber\\
\hspace{3 in}\mbox{\text{[custodial limit]}}\label{partial}
\eeqa due to the degeneracy between one of the neutral fields and the charged Higgs boson.  For example, in CP Case I [see Table \ref{tab2}], the term displayed in \eq{partial} is proportional to \beq s_{12}^2 F(m_1^2,\mhpm^2) + c_{12}^2 F(m_2^2,\mhpm^2) + F(m_3^2,\mhpm^2) - c_{12}^2 F(m_2^2,m_3^2) - s_{12}^2 F(m_1^2,m_3^2)\,.\eeq Due to the degeneracy between the charged Higgs boson and the CP-odd field $m_3$, $F(m_3^2,\mhpm^2) = 0$ and the rest of the terms cancel. One may check the other cases of CP symmetry in a similar manner.  In the special case of $Z_6 = Z_7 = 0$, the charged Higgs may be degenerate with a CP-even field rather than the CP-odd field.   One can verify for the six cases of CP symmetry ($i$)--($vi$) that \eq{partial} vanishes when $\mhpm^2 = m_{H^0}^2$.  [For example, in case ($i$), this term is proportional to $F(m_{H^0}^2,\mhpm^2) + F(m_{A^0}^2,\mhpm^2) - F(m_{H^0}^2,m_{A^0}^2  ) = 0$.]  Thus, the only non-zero contribution to $\alpha T$ arises from the breaking of custodial symmetry [the first part of \eq{pomarolformT}].

\subsection{$S$, $T$, and $U$ in the CP-Conserving Limit}
The expressions for $S$, $T$, and $U$ in the CP-conserving limit can be calculated using Tables ~\ref{tab2}-~\ref{tab4}. Note that Case II has been divided into two subcases based on the mass ordering of the neutral scalars, Case IIa ($m_{h^0} < m_{A^0}$) and Case IIb ($m_{h^0} > m_{A^0}$). 
Plugging in the values of the $q_{k\ell}$ parameters from the tables into equations (\ref{seqn}) and (\ref{tsu}), one reproduces the results of ref.~\cite{habertasi} (the reference mass $m_\phi$ has been fixed to be $m_{h^0}$) :
\beqa \nonumber
S&=& \frac{1}{\pi m_Z^2} \biggl\{\sin^2(\beta-\alpha)\mathcal{B}_{22}(m_Z^2;m_{H^0}^2,m_{A^0}^2)+\cos^2(\beta-\alpha)\left[\mathcal{B}_{22}(m_Z^2;m_{h^0}^2,m_{A^0}^2)\right.\biggr.\\
& & + \mathcal{B}_{22}(m_Z^2;m_Z^2,m_{H^0}^2)- \mathcal{B}_{22}(m_Z^2;m_Z^2,m_{h^0}^2)- m_Z^2\mathcal{B}_{0}(m_Z^2;m_Z^2,m_{H^0}^2)\nonumber\\
& &\biggl.\left. +m_Z^2\mathcal{B}_{0}(m_Z^2;m_Z^2,m_{h^0}^2)\right]- \mathcal{B}_{22}(m_Z^2;\mhpm^2,\mhpm^2)\biggr\},\nonumber\\
T &=& \frac{1}{16\pi s_W^2 m_W^2} \biggl\{F(\mhpm^2,m_{A^0}^2)+\sin^2(\beta-\alpha)[F(\mhpm^2,m_{H^0}^2)-F(m_{A^0}^2,m_{H^0}^2)]\nonumber \biggr.\\
& &+\cos^2(\beta-\alpha)[F(\mhpm^2,m_{h^0}^2)-F(m_{A^0}^2,m_{h^0}^2)+F(m_W^2,m_{H^0}^2)-F(m_W^2,m_{h^0}^2)\nonumber\\
& &-F(m_Z^2,m_{H^0}^2)+F(m_Z^2,m_{h^0}^2)+4m_Z^2 B_0(0;m_Z^2,m_{H^0}^2)
\nonumber\\& &\biggl.-4m_Z^2 B_0(0;m_Z^2,m_{h^0}^2) -4m_W^2 B_0(0;m_W^2,m_{H^0}^2)+4m_W^2 B_0(0;m_W^2,m_{h^0}^2)] \biggr\},\nonumber\\
S+ U &=& \frac{1}{\pi m_W^2} \biggl\{\mathcal{B}_{22}(m_Z^2;\mhpm^2,m_{A^0}^2)-2\mathcal{B}_{22}(m_Z^2;\mhpm^2,\mhpm^2)\biggr.\nonumber\\
&&+\sin^2(\beta-\alpha)\mathcal{B}_{22}(m_W^2;\mhpm^2,m_{H^0}^2)+\cos^2(\beta-\alpha)[\mathcal{B}_{22}(m_W^2;m_{h^0}^2,\mhpm^2)\nonumber \\
& &+\mathcal{B}_{22}(m_W^2;m_W^2,m_{H^0}^2)-\mathcal{B}_{22}(m_W^2;m_W^2,m_{h^0}^2)\nonumber\\
& &\biggl.+ m_W^2\mathcal{B}_{0}(m_W^2;m_W^2,m_{h^0}^2)-m_W^2\mathcal{B}_{0}(m_W^2;m_W^2,m_{H^0}^2)]\biggr\}
\eeqa
The angle $\beta-\alpha$ is defined in \eq{cases12}. Note that these expressions hold for both cases of CP conservation defined in section \ref{indepsec}.

\section{Numerical Analysis}

The experimental determinations of $S$, $T$ and $U$ are as follows ~\cite{Amsler:2008zzb}:
\beqa \label{stu}
S &=& -0.10 \pm .10, \nonumber\\
T &=& -0.08 \pm .11, \nonumber\\
U &=& 0.15 \pm .11, \eeqa
for the value of the Standard Model Higgs mass $m_\phi = 117$ GeV.  The values in \eq{stu} have the Standard Model contributions subtracted out, so that they reflect the deviation from the SM prediction.

Fixing $U=0$, as required by some models, changes the experimental limits on $S$ and $T$ slightly due to correlations between the parameters.  The relevant constraints in the $U=0$ scenario for $m_\phi = 117$ GeV are shown in \Ref{Amsler:2008zzb} to be consistent with no deviation from the Standard Model: 
\beq\label{STdec}
S = -0.04 \pm .09,~~T = 0.02 \pm .09\eeq
For $m_\phi = 300$ GeV, the corresponding values are
\beq\label{STdecalt}
S = -0.07 \pm .09,~~T =  0.09 \pm .09\eeq
These limits indicate that new physics contributions to the oblique parameters are tightly constrained.  Their significance for the parameters of the 2HDM is the subject of this section.

The parameters of the 2HDM which are constrained by $S$, $T$, and $U$ can be taken to be $Z_1$, $Z_3$, $Z_3 +Z_4$, $Z_5\,e^{-2i\theta_{23}}$, $Z_6\, e^{-i\theta_{23}}$ and $Y_2$, since these 6 quantities determine the physical Higgs masses [see eqs.~(\ref{mtilmatrix}) and~(\ref{hplus})] and the invariant functions $q_{k\ell}$ [see Appendix \ref{app:three}].  At this point, it will be assumed that the lightest neutral Higgs mass ($m_1$) will be interpreted as the reference mass $m_\phi$. 

The procedure used here to study the effect of the 2HDM on the oblique parameters was to choose random values of the six parameters in the space allowed by the unitarity bounds calculated in Appendix~\ref{app:deriv}, subject to the additional requirement that $m_1$ falls within $15$ GeV of $117$ GeV, since $m_\phi$ was taken to be $117$ GeV in the PDG analysis. Then the Higgs masses and $q_{k\ell}$ are calculated numerically and inserted into \eq{seqn} and \eq{tsu} to obtain $S$, $T$, and $U$ for each point in parameter space.

\begin{figure}[h!]
  \begin{center}
$\begin{array}{c@{\hspace{.21in}}c}
\epsfxsize=3.1in
\epsffile{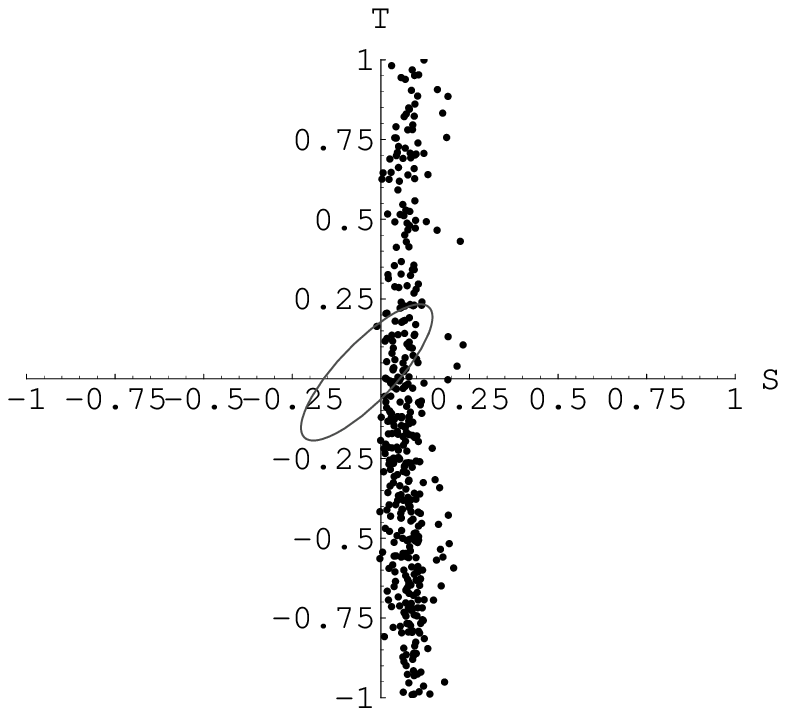} \\ \mbox{\bf (a)} \\
\epsfxsize=3.1in
\epsffile{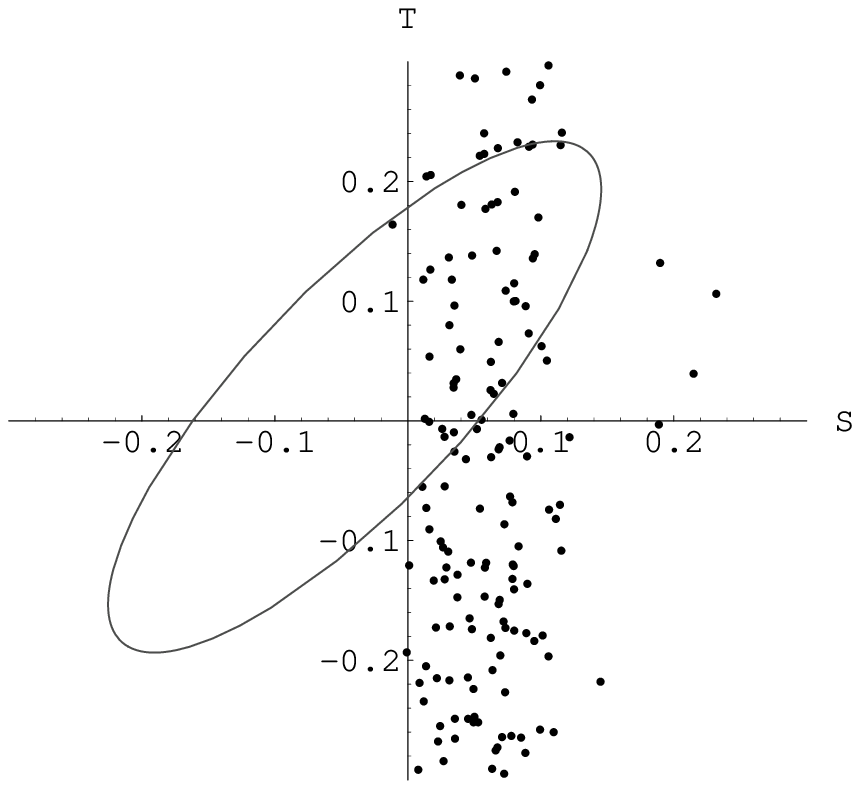}  \\ \mbox{\bf (b)}
\end{array}$

\end{center}
\caption{Scatterplots for $T$ as a function of $S$, with $m_1 = 117 \pm 15 $ GeV.  The ellipse, representing the $2~\sigma$ contour, is adapted from ~\cite{Amsler:2008zzb}.  The second plot shows a close-up of the allowed region.}
\label{scatter}
\end{figure}

It was found that the 2HDM consistently produces values of $U$ within $.02$ of zero.  This fact has important ramifications for comparison with experiment, since the limits in \eq{STdec} must be used to constrain $S$ and $T$, rather than the generic limits in \eq{stu}.  Scanning the parameter space and comparing with the allowed ``ellipse" in $S-T$ space produces the results shown in Fig.~\ref{scatter}.  
From the scatterplot it is evident that the values of $S$ produced are greater than indicated by the central value of $-.04$, although the discrepancy is not statistically significant.  Since the values of $T$ cover a broad range, the experimental value of $T$ is easily accomodated.  
  
\begin{figure}[h!]
\begin{center}
$\begin{array}{c@{\hspace{.21in}}c}
\epsfxsize=3.2in
\epsffile{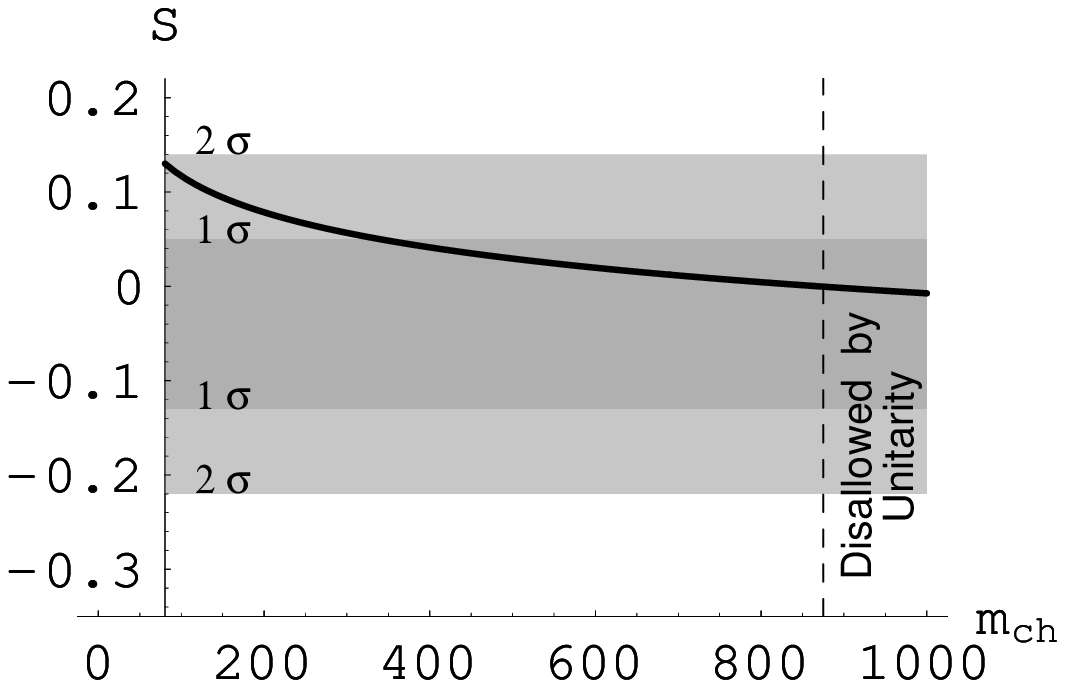} \\ \mbox{\bf (a)}\\
\epsfxsize=3.3in
\epsffile{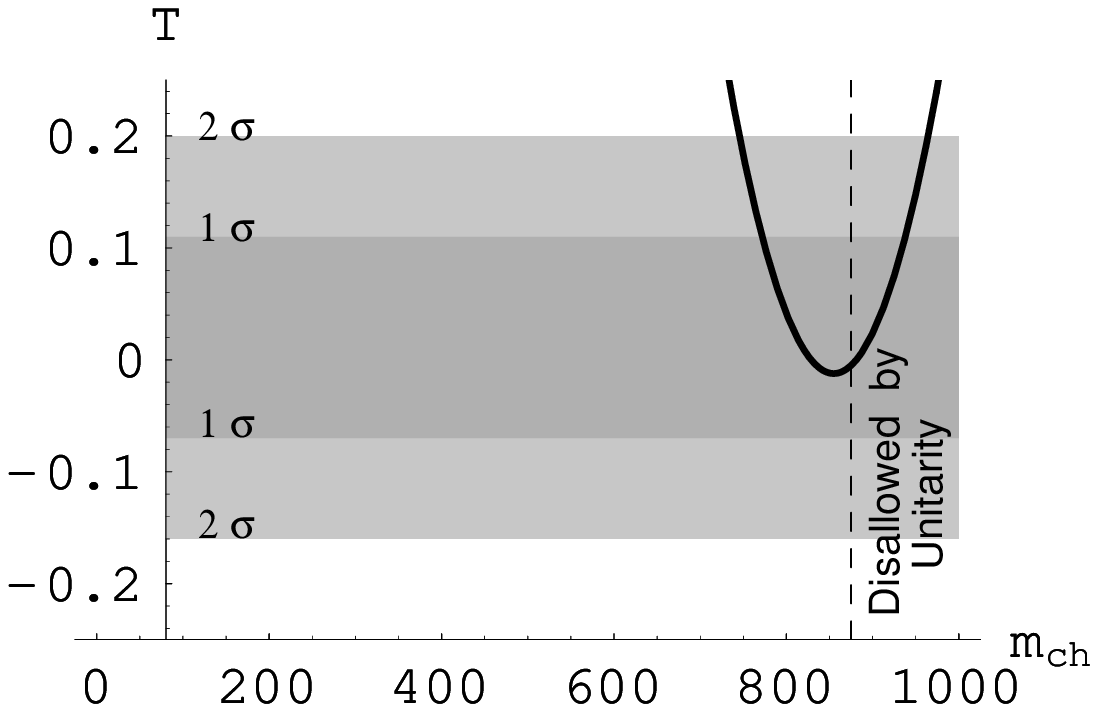} \\ 
\mbox{\bf (b)}
\end{array}$
\end{center}
\caption{S and T shown as a function of $\mhpm$ (solid line), with $Y_2$ fixed to be $m_W^2$.  The mass scale is in GeV. The unitarity bound (dashed line) applies to $Z_3$; all other parameters were picked so as to undersaturate unitarity.  The $1 \sigma$ and $2 \sigma$ contours (shaded regions) are obtained from \eq{STdec}.}
\label{SandT}
\end{figure}

Meanwhile, the interplay between the constraints from perturbative unitarity and the experimental limits on the oblique parameters is illustrated in Figs.~\ref{SandT}(a) and (b), which display $S$ and $T$ as a function of the charged Higgs mass $\mhpm$.  In these plots, we have fixed $Y_2 = m_W^2$ so that $\mhpm$ is limited by the unitarity bound on $Z_3$.  (Recall that $\mhpm^2 = Y_2 +\half Z_3 v^2$.)  The remaining parameters, shown below, are chosen such that $m_1 = 117$ GeV, to facilitate comparison with \eq{STdec}:
\beqa 
Z_1 =& 0.31\,,~~~~~~~~~~~~~Z_3+Z_4 =& 24,\nonumber \\
\Re(Z_6\,e^{-i\theta_{23}}) =& 0.1\,,~~~~~~~~\Im(Z_6\,e^{-i\theta_{23}}) =& -1.0,\nonumber \\ 
\Re(Z_5\,e^{-2i\theta_{23}}) =& -1.0\,,~~~~~\Im(Z_5\,e^{-2i\theta_{23}}) =& 1.0.\label{choices}\eeqa
Fixing the quantity $Z_3 + Z_4$, rather than $Z_4$ alone, ensures that $m_1$ (and thus $m_\phi$) does not vary as $Z_3$ is increased.  This simplifies the task of using the experimental limits in \eq{STdec}.  The sharp minimum in Fig. \ref{SandT}(b) results from the fact that corrections to $T$ are minimized when the difference between the charged Higgs and heavy neutral Higgs masses is small.  This effect will be exhibited in more detail in section \ref{sec:spl}.

The values for the $Z_i$ in \eq{choices} and our choice of $Y_2 = (80.4~\gev)^2$ generate the following masses for the Higgs particles:
\beqa m_1 &=& 117 \mbox{ GeV}\,, \nonumber \\
 m_2 &=& 831 \mbox{ GeV}\,, \nonumber \\
 m_3 &=& 882 \mbox{ GeV}\,.
\eeqa
The charged Higgs mass can be at most $875$ GeV for $Y_2 = m_W^2$, based on the unitarity limit $Z_3 \leq 8 \pi$.  

This numerical analysis shows that the current experimental limits on the oblique parameters are consistent with contributions to new physics from a second Higgs doublet over a large region of the 2HDM parameter space. However, experimental bounds on the $S$, $T$, and $U$ parameters do impose some non-trivial constraints. Our results also suggest that the 2HDM favors a slightly positive value of $S$ and a value of $U$ within $.02$ of zero, which is consistent with present data within the statistical error.  The 2HDM produces a broad range of values for $T$, which overlap completely with the experimental 2~$\sigma$ bounds. Finally, we have shown that taken together, the bounds on $S$, $T$, and $U$ and the unitarity limits act synergistically to restrict much the 2HDM parameter space.

\subsection{Splitting Between Neutral and Charged Higgs Masses\label{sec:spl}}
\begin{figure}[h!]
\begin{center}
$\begin{array}{c@{\hspace{.22in}}c}
\epsfxsize=3.2in
\epsffile{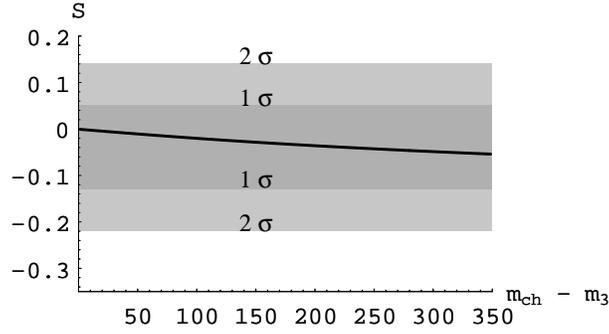} \\\mbox{\bf (a)} \\
\epsfxsize=3.2in
\epsffile{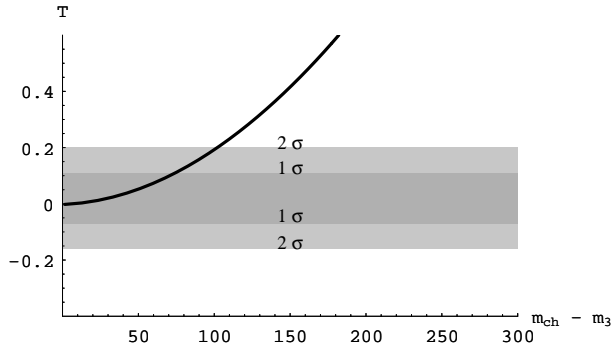} \\ 
 \mbox{\bf (b)}
\end{array}$
\end{center}
\caption{S and T shown as a function of the mass splitting between the charged Higgs and the heaviest neutral Higgs.  The mass scale is in GeV. The central values (dashed lines) and the $1 \sigma$ and $2 \sigma$ contours (shaded regions) are obtained from \eq{STdec}. Splitting of more than $100$ GeV is disfavored by the constraints from $T$.}
\label{SandTsplit}
\end{figure}

One can consider the case where the neutral masses $m_1$, $m_2$, and $m_3$ are light, and the charged Higgs mass heavier ($\mhpm > m_3$).  For a given choice of $Z_3$, the size of the splitting $\Delta m \equiv \mhpm - m_3$ is maximized when the other $Z_i$ are small.  As we will see in this section, the measured value of $T$ puts a limit on $\Delta m$.  The following values for the $Z_i$ have been chosen to maximize $\Delta m$, while preserving our requirement that $m_1 \approx 117 $ GeV at $Y_2 = m_W^2$:  
\beqa \label{splitZ}
Z_1 &= 0.25\,,~~~~ \Re(Z_6\,e^{-i\theta_{23}}) &= 0.01,\nonumber \\
Z_3+Z_4 &= .26\,,~~~~ \Im(Z_6\,e^{-i\theta_{23}}) &= 0.01,\nonumber \\ 
\Re(Z_5\,e^{-2i\theta_{23}}) &= 0.01\,,~~~~\Im(Z_5\,e^{-2i\theta_{23}}) &= 0.01.
\eeqa
With the parameters fixed as in \eq{splitZ}, the size of the splitting (controlled by $Z_3$) can easily become large; for example,  $\Delta m  = 400$ GeV at $Z_3 = 9$.  However, one can now impose constraints from the oblique parameters.  Using the values of the $Z_i$ in \eq{splitZ}, $S$ and $T$ as functions of $\Delta m$ are shown in Figs.~\ref{SandTsplit}(a) and (b), respectively.

In these plots, as in the previous section, the experimental limits are taken from \eq{STdec}, since $U\approx 0$.  The graph of $T$ shows that the mass of the charged Higgs can at most be roughly $100$ GeV greater than $m_3$ and still be consistent with the experimental limits, which corresponds to the value $Z_3 = 1.5$.  
\section{Measuring $\boldsymbol{tan(\beta)}$ \label{tanbsec}} 
In sections \ref{sec:five} and \ref{sec:six}, we have written out the entire
interaction Lagrangian for the Higgs bosons of the 2HDM.  Yet, the
famous parameter $\tan\beta$, given by $\tan\beta\equiv v_2/v_1$
in a generic basis [see \eq{emvev}], does not appear in any
physical Higgs (or Goldstone) boson 
coupling.  This is rather surprising given the
large literature of 2HDM phenomenology in which the parameter
$\tan\beta$ is ubiquitous.  For example, 
numerous methods have been proposed for
measuring $\tan\beta$ at future 
colliders~\cite{ilctanb1,tanbprecision1,ilctanb2,ilctanb3,boosetal,plctanb1,plctanb2,lhctanb}.
In a generic basis, one can also define
the relative phase of the two vacuum expectation values,
$\xi={\rm arg}~(v_2 v_1^*)$.  However, neither $\tanb$ nor $\xi$
are basis-independent.  One can remove $\xi$ by rephasing one
of the two Higgs doublet fields, and both $\xi$ and
$\tan\beta$ can be removed
entirely by transforming to the Higgs basis.  Thus, in a
general 2HDM, $\tan\beta$ is an unphysical parameter with no 
significance \textit{a priori}.

The true significance of $\tanb$ emerges only in specialized
versions of the 2HDM, where
$\tan\beta$ is promoted to a
physical parameter.  As noted in section \ref{sec:six}, the general 2HDM generally
predicts FCNCs in conflict with experimental data.  One way to avoid
this phenomenological problem is to constrain the theoretical
structure of the 2HDM.  Such constraints often pick out a preferred
basis.  Relative to that basis, $\tan\beta$ is then a meaningful
parameter.

The most common 2HDM constraint is the requirement that
some of the Higgs-fermion Yukawa couplings vanish in a
``preferred'' basis.
This leads to the well known type-I and type-II 2HDMs~\cite{hallwise}
(henceforth
called 2HDM-I and 2HDM-II).  In the 2HDM-I, there exists a 
preferred basis where
$\eta^U_2=\eta^D_2=0$~\cite{type1,hallwise}.
In the 2HDM-II, there exists a preferred basis where
$\eta^U_1=\eta^D_2=0$~\cite{type2,hallwise}.
These conditions can be enforced by a suitable
symmetry.  For example, the MSSM possesses a type-II Higgs-fermion
interaction, in which case the supersymmetry guarantees
that $\eta^U_1=\eta^D_2=0$.  In non-supersymmetric models, appropriate
discrete symmetries can be found to enforce the type-I or type-II
Higgs-fermion couplings.\footnote{These discrete symmetries also
imply that some of the coefficients of the scalar potential must also
vanish in the same preferred basis~\cite{type1,type2,hallwise,lavoura2}.}

The conditions for type-I and type-II Higgs-fermion interactions given
above are basis-dependent.  But, there is also a basis-independent
criterion that was first given in \Ref{davidson}:\footnote{In this
paper, we have slightly modified our definition of the Yukawa
coupling.  What is called $\eta^D$ in \Ref{davidson} is called
$\eta^{D\,\dagger}$ here.}
\beqa
\epsilon_{\abar\bbar}\eta^D_a\eta^U_b=
\epsilon_{ab}\eta^{D\,\dagger}_{\abar}\eta^{U\,\dagger}_{\bbar}&=&0\,,
\qquad \hbox{\rm type-I}
\,,\label{typecond1} \\
\delta_{a\bbar}\,\eta^{D\,\dagger}_{\abar}\eta^U_b &=& 0
\,,\qquad \hbox{\rm type-II}\,.
\label{typecond2}
\eeqa

We can now prove that $\tan\beta$ is a physical parameter in the 2HDM-II
(a similar analysis holds for the 2HDM-I\footnote{\Eq{typecond1} involves pseudo-invariant quantities.
Nevertheless, setting these quantities to zero yields a U(2)-invariant
condition.}).
In the preferred basis where $\eta^U_1=\eta^D_2=0$, we shall denote:
$\widehat v=e^{i\eta}(\cos\beta\,,\,
\sin\beta\,e^{i\xi})$ and $\widehat w=e^{-i\eta}(-\sin\beta
e^{-i\xi}\,,\,\cos\beta)$.  Evaluating $\kappa^Q=\widehat v^*\cdot\eta^Q$
and $\rho^Q=\widehat w^*\cdot\eta^Q$ in the
preferred basis, and recalling that the $\kappa^Q$ are diagonal 
real matrices, it follows that:
\beq  \label{modeliitanbdefs}
\mathbold{\it I}e^{-i(\xi+2\eta)}\tanb=-\rho^{D\,\dagger}(\kappa^D)^{-1}
=(\rho^U)^{-1}\kappa^U\,,
\eeq
where $\mathbold{\it I}$ is the identity matrix in quark flavor space
and $\kappa^Q=\sqrt{2}M_Q/v$ [see \eqs{diagumass}{diagdmass}].
These two definitions are consistent if
$\kappa^D\kappa^U+\rho^{D\,\dagger}\rho^U=0$ is satisfied. 
But the latter is equivalent to the type-II condition [which can be verified
by inserting \eq{kapparhoinv} into \eq{typecond2}].

To understand the phase factor that appears in \eq{modeliitanbdefs},
we note that only unitary matrices of the form
$U={\rm diag}(e^{i\chi_1}\,,\,e^{i\chi_2})$ that span a
U(1)$\times$U(1) subgroup of the flavor-U(2) group preserve the
the type-II conditions $\eta^U_1=\eta^D_2=0$ in the preferred basis.
Under transformations of this type, $\eta\to \eta+\chi_1$ and
$\xi\to\xi+\chi_2-\chi_1$.   Using \eq{rhotrans}, it follows that
$\rho^Q\to e^{i(\chi_1+\chi_2)}\rho^Q$ .  Hence $\rho^Q e^{-i(\xi+2\eta)}$
is invariant with respect to such U(1)$\times$U(1) transformations.
We conclude that \eq{modeliitanbdefs} is covariant with respect to
transformations that preserve the type-II condition.

The conditions specified in \eq{modeliitanbdefs} are quite
restrictive.  In particular, they determine the matrices
$\rho^Q$:
\beq \label{phfacts}
\rho^D e^{-i(\xi+2\eta)}=\frac{-\sqrt{2}M_D \tan\beta}{v}\,,\qquad
\rho^U e^{-i(\xi+2\eta)}=\frac{\sqrt{2}M_U \cot\beta}{v}\,.
\eeq
Up to an overall phase, $\rho^U$ and $\rho^D$ are real diagonal matrices
with non-negative entries.  There is also some interesting
information in the phase factors of \eq{phfacts}.  Although  the $\rho^Q$ are
pseudo-invariants, we have noted below
\eq{hffu2} that $e^{i\theta_{23}}\rho^Q$ is U(2)-invariant.
This means that the phase factor $e^{-i(\theta_{23}+\xi+2\eta)}$ is a physical
parameter.  
Moreover, we can now define $\tan\beta$ as a physical parameter of the 2HDM-II
as follows:
\beq
\tan\beta=\frac{v}{3\sqrt{2}}\,\bigl|\Tr\left(\rho^D M_D^{-1}\right)\bigr|\,,
\eeq
where $0\leq\beta\leq\pi/2$.  This is a manifestly
basis-independent definition, so $\tan\beta$ is indeed physical.

In Higgs studies at future colliders, suppose one encounters phenomena
that appear consistent with a 2HDM.  It may not be readily apparent
that there is any particular structure in the Higgs-fermion
interactions.  In particular, it could be that \eq{modeliitanbdefs} is
simply false.  Here we present a model-independent version of the $\tanb$ parameter.  For simplicity, we assume that the Yukawa couplings of the Higgs bosons to the third
generation fermions
dominate, in which case we can ignore the effects of the first
two generations.\footnote{This is probably not a bad assumption,
since $\kappa^Q$ is proportional to the quark mass 
matrix $M_Q$.}  In a one-generation model, one can introduce
three $\tanb$-like parameters
\beq \label{tanblike}
\tanb_b\equiv \frac{v}{\sqrt{2}}\frac{|\rho^D|}{m_b}\,,\qquad
\tanb_t\equiv  \frac{\sqrt{2}}{v}\frac{m_t}{|\rho^U|}\,,\qquad
\tanb_\tau\equiv \frac{v}{\sqrt{2}}\frac{|\rho^E|}{m_\tau}\,,
\eeq
where $\tan\beta_\tau$ is analogous to $\tan\beta_d$ and depends on
the third generation Higgs-lepton interaction.  In a
type-II model, one indeed has $\tanb_b=\tanb_t=\tanb_\tau=\tanb$.
In the more general
(type-III) 2HDM, there is no reason for the three parameters
above to coincide.  However, these three parameters are indeed
U(2)-invariant quantities, and thus correspond to physical observables
that can be measured in the laboratory.
The interpretation of these parameters is straightforward.  In the
Higgs basis, up and down-type quarks interact with both Higgs doublets.
But, clearly there exists some basis (\textit{i.e.}, a rotation by an angle
$\beta_t$ from the Higgs basis) for which only one of the two
up-type quark Yukawa couplings is non-vanishing.
This defines the physical angle $\beta_t$.  The interpretation of the
other two angles is similar.

Since the phase of $e^{i\theta_{23}}\rho^Q$ is a physical parameter,
one can generalize \eq{tanblike} by defining
\beq \label{tanblike2}
e^{i(\theta_{23}-\chi_b)}\tanb_b\equiv \frac{v}{\sqrt{2}}
\frac{\rho^{D\,*}}{m_b}\,,\qquad
e^{i(\theta_{23}-\chi_t)}\tanb_t\equiv \frac{\sqrt{2}}{v}
\frac{m_t}{\rho^U}\,,
\eeq
and similarly for $\tan\beta_\tau$.  Thus, in addition to three
$\tan\beta$-like parameters, there are three independent
physical phases $\chi_b$, $\chi_t$ and $\chi_\tau$
that could in principle be deduced from experiment.  Of course,
in the 2HDM-II,
one must have $\beta_b=\beta_t=\beta_\tau$ and
$\chi_b=\chi_t=\chi_\tau$.

A similar analysis can be presented for the case of the 2HDM-I.  In
this case, one is led to define slightly different $\tan\beta$-like
physical parameters.  But, these would be related to those defined in
\eq{tanblike} in a simple way.  A particular choice could be motivated
if one has evidence that that either the type-I or type-II conditions
are approximately satisfied.

We conclude this section by illustrating the utility of this approach
in the case of the MSSM.  This example has already been presented in
\Ref{davidson} in the case of a CP-conserving Higgs sector.  We
briefly explain how that analysis is generalized in the case of a
CP-violating Higgs sector.
The MSSM Higgs sector is a CP-conserving type-II 2HDM in
the limit of exact supersymmetry.  However, when supersymmetry breaking
effects are taken into account, loop corrections to the Higgs
potential and the Higgs-fermion interactions can lead to both
CP-violating effects in the Higgs sector, and the (radiative) generation of
the Higgs-fermion Yukawa couplings that are absent in the type-II
limit.  In particular, in the approximation that
supersymmetric masses are significantly
larger than $m_Z$, the effective Lagrangian that describes the
coupling of the Higgs bosons to the third generation quarks is given
(in the notation of \cite{Carena:2002es}) by
\beq \label{leff}
-\call_{\rm eff}=
(h_b+\delta h_b)(\anti q_L\Phi_1) b_R
+ (h_t+\delta h_t)(\anti q_L \wtil\Phi_2) t_R 
+\Delta h_b\,(\anti q_L \Phi_2) b_R +\Delta h_t\,(\anti q_L \wtil\Phi_1) t_R
+{\rm h.c.}\,,
\eeq
where $\anti q_L\equiv (\anti u_L\,,\,\anti d_L)$.
Note that the terms proportional to $\Delta h_{b}$ and
$\Delta h_{t}$, which are absent in the tree-level MSSM, 
are generated at one-loop due to supersymmetry-breaking effects
Thus, we identify $\eta^{D}=((h_b+\delta h_b)^*\,,\,\Delta h_b^*)$
and $\eta^{U}=(\Delta h_t\,,\,h_t+\delta h_t)$.
The tree-level MSSM is CP-conserving, 
and $\xi=0$ in the supersymmetric basis.
At one-loop, CP-violating effects can shift $\xi$ away from
zero, and we shall denote this quantity by $\Delta\xi$.\footnote{In
practice, one would rephase the fields after computing the radiative
corrections.  But, since we are advocating basis-independent methods
in this paper, there is no need for us to do this.}
Evaluating $\kappa^Q=\widehat v^*\cdot\eta^Q$
and $\rho^Q=\widehat w^*\cdot\eta^Q$ as we did above \eq{modeliitanbdefs},
\beqa
e^{i\eta}\kappa^D &=& c_\beta (h_b+\delta h_b)^*+e^{-i\Delta\xi}s_\beta(\Delta h_b)^*\,,\nonumber\\
e^{-i\eta}\rho^D &=& -e^{i\Delta\xi} s_\beta (h_b+\delta h_b)^*+c_\beta(\Delta h_b)^*,\nonumber\\
e^{i\eta}\kappa^U &=& c_\beta \Delta h_t+e^{-i\Delta\xi}s_\beta(h_t+\delta h_t)\,,\nonumber\\
e^{-i\eta}\rho^U &=& -e^{i\Delta\xi} s_\beta \Delta h_t+c_\beta(h_t+\delta h_t)\,.
\eeqa
By definition, the $\kappa^Q$ are real and non-negative, and related
to the top and bottom quark masses via \eqs{diagumass}{diagdmass}.
Thus, the tree-level relations between $m_b$, $m_t$ and $h_b$, $h_t$
respectively are modified~\cite{db1,db2,db3,db4,db5}:\footnote{If 
one of the Higgs fields is rephased in order to
remove the phase $\Delta\xi$, then one simultaneously rephases
$\Delta h_{b,t}$ such that the quantities $\Delta h_{b,t} e^{i\Delta\xi}$ 
are invariant
with respect to the rephasing.  In particular,
$h_b$ and $h_t$ are not rephased, since these tree-level
quantities are always real and positive
and proportional to the tree-level values of $m_b$
and $m_t$, respectively.}
\beqa 
m_b=\frac{v\kappa^D}{\sqrt{2}}=
\frac{v \cb h_b}{\sqrt{2}}\left[1+\Re\left(\frac{\delta h_b}{h_b}+
\frac{\Delta h_b}{h_b}e^{i\Delta\xi}\tan\beta\right)\right]
\equiv \frac{v \cb h_b}{\sqrt{2}}
\left[1+\Re(\Delta_b)\right]\,,\label{modmassesb}\\[6pt]
m_t=\frac{v\kappa^U}{\sqrt{2}}=
\frac{v \sb h_t}{\sqrt{2}}\left[1+\Re\left(\frac{\delta h_t}{h_t}+
\frac{\Delta h_t}{h_t}e^{i\Delta\xi}\cot\beta\right)\right]
\equiv \frac{v \sb h_t}{\sqrt{2}}
\left[1+\Re(\Delta_t)\right]\,,\label{modmassest}
\eeqa
which define the complex quantities $\Delta_b$ and 
$\Delta_t$.\footnote{In deriving 
\eqs{modmassesb}{modmassest}, we computed $\kappa^Q=|\kappa^Q|$ by
expanding up to linear order
in the one-loop quantities $\Delta h_{b,t}$ and
$\delta h_{b,t}$.  Explicit
expressions for $\Delta_b$ and $\Delta_t$ in terms of supersymmetric masses and
parameters, and references to the original literature
can be found in \Ref{Carena:2002es}.}\Eq{tanblike2} then yields:
\beqa
\tan\beta_b&=&\left|\frac{-e^{-i\Delta\xi} s_\beta (h_b+\delta h_b)
+c_\beta\Delta h_b}
{c_\beta (h_b+\delta h_b)+e^{i\Delta\xi}s_\beta\Delta h_b}\right|\,,\qquad
\,\chi_b=\theta_{23}+\psi_b+\eta\,,\\[6pt]
\tan\beta_t&=&\left|\frac{\cb\Delta h_t
+e^{-i\Delta\xi} s_\beta (h_t+\delta h_t)}
{-\Delta h_t\sb e^{i\Delta\xi}+c_\beta(h_t+\delta h_t)}\right|\,,\qquad
\quad\chi_t=\theta_{23}+\psi_t+\eta\,,
\eeqa
where $\psi_{t,b}\equiv\arg(e^{-i\eta}\rho^{U,D})$.
Expanding the numerators and denominators 
above and dropping terms of 
quadratic order in the one-loop quantities, we end up with
\beqa 
\tan\beta_b&=&\frac{\tan\beta}{1+\Re\,\Delta_b}
\left[1+\frac{1}{\sb^2}\,\Re\left(\frac{\delta h_b}{h_b}
-\cb^2\Delta_b\right)\right]\,,\label{tanbb} \\[8pt]
\cot\beta_t&=&\frac{\cot\beta}{1+\Re\,\Delta_t}
\left[1+\Re\left(\Delta_t-\frac{1}{\cb\sb}\frac{\Delta h_t}{h_t}
e^{i\Delta\xi}\right)\right]\,.\label{tanbt} 
\eeqa

We have chosen to write $\tan\beta_b/\tan\beta$ in terms of $\Delta_b$
and $\delta h_b/h_b$, and $\cot\beta_t/\cot\beta$ in terms of $\Delta_t$
and $\Delta h_t/h_t$ in order to emphasize 
the large $\tanb$ behavior of the deviations of these quantities from one.
In particular,
keeping only the leading $\tan\beta$-enhanced corrections, 
\eqs{modmassesb}{modmassest} imply that\footnote{Because the one-loop
corrections $\delta h_b$, $\Delta h_b$, $\delta h_t$ and $\Delta h_t$
depend only on Yukawa and gauge couplings and the supersymmetric
particle masses, they
contain no hidden $\tan\beta$ enhancements 
or suppressions~\cite{Carena:2001uj}.} 
\beq
\Delta_b\simeq e^{i\Delta\xi}\,\frac{\Delta h_b}{h_b}\,\tanb\,,\qquad\qquad
\Delta_t\simeq  \frac{\delta h_t}{h_t}\,.
\eeq
That is, the complex quantity $\Delta_b$ is $\tanb$-enhanced.
In typical models at large $\tanb$, the quantity $|\Delta_b|$ can be
of order 0.1 or larger and of either sign.  
Thus, keeping only the one-loop corrections that are
$\tan\beta$-enhanced,\footnote{In \Ref{davidson} the one-loop
$\tan\beta$-enhanced correction to $\cot\beta_t$ was incorrectly omitted.}
\beq
\tan\beta_b\,\simeq\,\frac{\tanb}{1+\Re\,\Delta_b}
\,,\qquad\quad
\cot\beta_t\simeq \cot\beta\left[
1-\tan\beta\,\Re\left(\frac{\Delta h_t}{h_t}e^{i\Delta\xi}\right)\right]\,.
\eeq
Thus, we have expressed the basis-independent quantities $\tan\beta_b$
and $\tan\beta_t$ in
terms of parameters that appear in the natural basis of the MSSM Higgs
sector.  Indeed, we find that  $\tan\beta_b\neq\tan\beta_t$ as a consequence of
supersymmetry-breaking loop-effects.

		\chapter{Conclusion}			

The basis-independent formalism will allow multiple Higgs bosons to be interpreted in experiments before the underlying dynamics (MSSM, etc.) are understood.   Once the physical observables ($\rho^Q$, scalar masses, etc.) are measured, they can then be related to physical parameters such as the $\tanb$-like parameters to identify whether the scalars are consistent with a specific model, such as supersymmetric model.  
At this point, we have a complete theoretical description of the scalar, boson, and Yukawa sectors of the basis-independent 2HDM in both the CP-conserving and CP-violating cases.  We have defined a mass matrix for neutral Higgs particles in terms of basis-independent scalar couplings [$Y_2$, $Z_1$, $Z_3$ ,$Z_4$, $\zfive$, and $\zsix$.]  Since $Z_7$ does not appear in the mass matrix, it is possible to have CP violation in the scalar sector of the theory (arising from $Z_7$ terms) even in the absence of mixing betweeen CP-even and CP-odd Higgs eigenstates.  When the physical Higgs bosons are CP eigenstates (regardless of whether CP violation appears in the Lagrangian), there exist three ways to implement the mixing of the neutral Higgs particles, which we have called cases I, IIa, and IIb.  These cases, which can be physically distinguished based on the invariant $\zfive$, correspond to different orderings of the neutral Higgs fields in the mass matrix.  We have also described the mixing angles in the special cases of $Z_6 = 0$ and $Z_6 = Z_7 = 0$.   For $Z_6 = 0$, we find that barring degeneracy in the neutral scalars, the mass matrix breaks up into $1\times 1$ and $2\times 2$ blocks, as it does in the CP-conserving cases.  However, the neutral scalars may still have indefinite CP quantum numbers, since CP violation may arise from the interactions involving $Z_7$. For $Z_6 = 0$ there are three cases that can be distinguished, labeled cases ($i$),($ii$) and ($iii$) in this document, which can again be related to the different patterns of mass ordering.  For completeness, we also describe the situations in which $Z_6 = 0$ and the neutral scalars are doubly degenerate. In special case of $Z_6 = Z_7 = 0$, we find that there are two physically distinguishable cases corresponding to different definitions of a time reversal (T) transformation.  

The decoupling limit and the custodial limit have been developed in this basis-independent formalism, and the phenomenology of the oblique parameters and $\tanb$-like parameters have been defined in such a way as to be physically measurable without imposing symmetries or other assumptions on the scalar sector. The oblique parameters of the CP-violating 2HDM were analyzed numerically and found to be consistent with the experimental values within statistical error when plausible values of the scalar couplings were postulated (based on unitary limits).  However, since our analysis shows that the 2HDM favors specific ranges for $S$ and $U$, it may be possible to rule out the model in the future if experimental limits change.  (Our analysis suggests that the 2HDM can accommodate equally well both positive and negative values of $T$, so that parameter is unlikely to falsify the model.)  We also made use of the oblique parameters to constrain the mass difference between the neutral Higgs bosons and the charged Higgs boson. Other phenomenological impications of the 2HDM, such as FCNCs, remain open to further research.

\appendix
\chapter{The 2HDM scalar potential in a generic
basis} \label{app:one}

Let $\Phi_1$ and
$\Phi_2$ denote two complex hypercharge-one, SU(2)$\ls{\rm L}$ doublets 
of scalar fields.
The most general gauge-invariant scalar potential is given
by
\beqa  \label{pot}
\mathcal{V}&=& m_{11}^2\Phi_1^\dagger\Phi_1+m_{22}^2\Phi_2^\dagger\Phi_2
-[m_{12}^2\Phi_1^\dagger\Phi_2+{\rm h.c.}]\nonumber\\[6pt]
&&\quad +\half\lambda_1(\Phi_1^\dagger\Phi_1)^2
+\half\lambda_2(\Phi_2^\dagger\Phi_2)^2
+\lambda_3(\Phi_1^\dagger\Phi_1)(\Phi_2^\dagger\Phi_2)
+\lambda_4(\Phi_1^\dagger\Phi_2)(\Phi_2^\dagger\Phi_1)
\nonumber\\[6pt]
&&\quad +\left\{\half\lambda_5(\Phi_1^\dagger\Phi_2)^2
+\big[\lambda_6(\Phi_1^\dagger\Phi_1)
+\lambda_7(\Phi_2^\dagger\Phi_2)\big]
\Phi_1^\dagger\Phi_2+{\rm h.c.}\right\}\,,
\eeqa
where $m_{11}^2$, $m_{22}^2$, and $\lam_1,\cdots,\lam_4$ are real parameters.
In general, $m_{12}^2$, $\lambda_5$,
$\lambda_6$ and $\lambda_7$ are complex.  The form of \eq{pot} holds
for any generic choice of $\Phi_1$--$\Phi_2$ basis, whereas the coefficients
$m_{ij}^2$ and $\lambda_i$ are basis-dependent quantities.  Matching
\eq{pot} to the U(2)-covariant form of \eq{genericpot}, we identify:
\beq \label{ynum}
Y_{11}=m_{11}^2\,,\qquad\qquad
Y_{12}=Y_{21}^\ast=-m_{12}^2 \,,\qquad\qquad
Y_{22}=m_{22}^2\,,
\eeq
and
\beqa \label{znum}
&& Z_{1111}=\lam_1\,,\qquad\qquad \,\,\phantom{Z_{2222}=}
Z_{2222}=\lam_2\,,\nonumber\\
&& Z_{1122}=Z_{2211}=\lam_3\,,\qquad\qquad
Z_{1221}=Z_{2112}=\lam_4\,,\nonumber \\
&& Z_{1212}=\lam_5\,,\qquad\qquad \,\,\phantom{Z_{2222}=}
Z_{2121}=\lam_5^\ast\,,\nonumber\\
&& Z_{1112}=Z_{1211}=\lam_6\,,\qquad\qquad
Z_{1121}=Z_{2111}=\lam_6^\ast\,,\nonumber \\
&& Z_{2212}=Z_{1222}=\lam_7\,,\qquad\qquad
Z_{2221}=Z_{2122}=\lam_7^\ast\,.
\eeqa
Explicit formulae for the coefficients of the Higgs basis scalar potential
in terms of the corresponding coefficients of \eq{pot} in a generic
basis can be found in \Ref{davidson}.
\chapter{The neutral Higgs boson squared-mass matrix in a generic
basis} \label{app:two}
Starting from \eq{genericpot}, one can obtain the neutral Higgs
squared-mass matrix from the quadratic part of the scalar potential:
\beq \label{vpmp}
\mathcal{V}_{\rm mass}=\frac{1}{2}\left(\begin{array}{cc}
\Phi^0_a & \quad \Phi^{0\,\dagger}_{\bbar}\end{array}\right) \mathscr{M}^2
\left(\begin{array}{c} \Phi^{0\,\dagger}_{\cbar} \\[4pt]
\!\!\!\!\Phi^0_d\end{array}\right)\,.
\eeq
Thus, $\mathscr{M}^2$ is given by the following matrix of second
derivatives:
\beq \label{scriptm2}
\mathscr{M}^2={\left(\begin{array}{cc}  \displaystyle
\frac{\partial^2 \mathcal{V}}{\partial\Phi^0_a
\partial\Phi^{0\,\dagger}_{\cbar}} &
\qquad\displaystyle
\frac{\partial^2 \mathcal{V}}{\partial\Phi^0_a\partial\Phi^0_d} \\ \\
\displaystyle
\frac{\partial^2 \mathcal{V}}{\partial\Phi^{0\,\dagger}_{\bbar}
{\partial\Phi^{0\,\dagger}_{\cbar}}} &
\qquad\displaystyle
\frac{\partial^2 \mathcal{V}}
{\partial\Phi^{0\,\dagger}_{\bbar}\partial\Phi^0_d}
\end{array}\right)}_{\Phi^0_a=v_a}\,,
\eeq
where $v_a\equiv v\widehat v_a/\sqrt{2}$ and
$\widehat v_{\abar}^*\widehat v_a=1$.  With $\mathcal{V}$ given by
\eq{genericpot}, one finds:
\beq \label{genericm2}
\mathscr{M}^2=\left(\begin{array}{cc} (Y_{a\cbar})^*
+\half v^2\bigl[(Z_{a\cbar f\ebar}
+Z_{f\cbar a\ebar})\,\widehat v_e\widehat v^*_{\fbar}\bigr]^* &\qquad
\quarter v^2(Z_{e\abar f\dbar}+Z_{e\dbar f \abar})
\widehat v^*_{\ebar}\widehat v^*_{\fbar} \\ \\
\quarter v^2(Z_{b\ebar c\fbar}+Z_{c\ebar b \fbar})
\widehat v_e\widehat v_f & \qquad
Y_{b\dbar}+\half v^2(Z_{e\fbar b\dbar}+Z_{e\dbar b\fbar})
\widehat v^*_{\ebar}\widehat v_f \end{array}\right)\,.
\eeq
In deriving this result, we used the hermiticity properties of
$Y$ and $Z$ to rewrite the upper left hand block so that the indices
appear in the standard order for matrix multiplication in \eq{vpmp}.  
In addition, we employed:
\beq
\frac{\partial\Phi^0_e}{\partial\Phi^0_a}=\delta_{e\abar}\,,\qquad\qquad
\frac{\partial\Phi^{0\,\dagger}_{\fbar}}
{\partial\Phi^{0\,\dagger}_{\bbar}}=\delta_{b\fbar}\,.
\eeq

It is convenient to express the squared-mass matrix in terms of
(pseudo)-invariants.  To do this, we note that we can expand an
hermitian second-ranked tensor [which satisfies $A_{a\bbar}=(A_{b\abar})^*$]
in terms of the eigenvectors of
$V_{a\bbar}\equiv \widehat v_a \widehat v^*_{\bbar}$:
\beq \label{aab}
A_{a\bbar}=
\Tr(VA)V_{a\bbar}+\Tr(WA)W_{a\bbar}+\bigl[(\widehat v^*_{\cbar}
\widehat w_d A_{c\dbar})\widehat v_a \widehat w^*_{\bbar}+
(\widehat w^*_{\cbar}\widehat v_d
 A_{c\dbar})\widehat w_a \widehat v^*_{\bbar}\bigr]\,,
\eeq
where $W_{a\bbar}\equiv \widehat w_a\widehat
w^*_{\bbar}=\delta_{a\bbar}-V_{a\bbar}$.
Likewise, we can expand a second-ranked symmetric tensor
with two unbarred (or two barred indices), \textit{e.g.}, 
\beq
A_{ab}= (\widehat v^*_{\cbar}\widehat v^*_{\dbar}
A_{cd})\widehat v_a \widehat v_b +
(\widehat w^*_{\cbar}\widehat w^*_{\dbar}A_{cd})\widehat w_a \widehat w_b +
(\widehat v^*_{\cbar}\widehat w^*_{\dbar}A_{cd})
(\widehat v_a \widehat w_b + \widehat w_a \widehat v_b)\,.
\eeq
We can therefore rewrite the upper and lower right hand
$2\times 2$ blocks of the squared-mass matrix [\eq{genericm2}]
respectively as:
\beqa
[\mathscr{M}^2]_{\abar \dbar}&=&\half v^2\left[Z_1 v^*_{\abar} v^*_{\dbar}
+Z_5 w^*_{\abar} w^*_{\dbar} +Z_6(\widehat v^*_{\abar} \widehat w^*_{\dbar}
+ \widehat w^*_{\abar} \widehat v^*_{\dbar})\right]\,, \label{upright}\\[5pt]
[\mathscr{M}^2]_{b\dbar}
&=& (Y_1+Z_1 v^2)V_{b\dbar}
+[Y_2+\half (Z_3+Z_4)v^2]W_{b\dbar}\nonumber\\
&&\qquad+[(Y_3+Z_6 v^2)\widehat v_b
\widehat w^*_{\dbar} +(Y^*_3+Z^*_6 v^2)\widehat w_b
\widehat v^*_{\dbar} ]\,.\label{downright}
\eeqa
The upper and lower left hand blocks are
then given by the hermitian adjoints of the
lower and upper right hand blocks, respectively.
Note that \eq{downright} can be
simplified further by eliminating $Y_1$ and
$Y_3$ using the scalar potential minimum conditions [\eq{hbasismincond}].

Let us apply this result to the Higgs bases, where $\widehat
v=(1\,,\,0)$ and $\widehat w=(0\,,\,1)$.  After imposing the
scalar potential minimum conditions,
\beq \label{hbasism2}
\mathscr{M}^2=\frac{1}{2} v^2\left(\begin{array}{cccc} Z_1 &\quad
    Z_6^* &\quad Z_1 &\quad Z_6 \\
  Z_6  &\quad Z_3+Z_4+2Y_2/v^2 &\quad  Z_6  &\quad
 Z_5  \\
 Z_1 &\quad    Z_6^* &\quad Z_1 &\quad Z_6\\
 Z_6^* &\quad Z_5^* &\quad Z_6^*
 &\quad Z_3+Z_4+2Y_2/v^2 \end{array}\right)\,.
\eeq
The massless Goldstone boson eigenvector
\beq
G^0=\frac{-i}{\sqrt{2}}\left(\begin{array}{c} 1\\\ \!0\\ \!\!-1\\ \,\,0
\end{array}
\right)\,,
\eeq
can be determined by inspection [the normalization factor is chosen
for consistency with \eq{hbasisfields}].
Thus, we can perform a (unitary) similarity
transformation on $\mathscr{M}^2$ to remove the Goldstone boson from
the neutral Higgs squared-mass matrix.  Explicitly, with the unitary matrix
\beq
V=\frac{1}{\sqrt{2}}\left(\begin{array}{cccc} 1 &\quad 0 &\quad \,\,0
& \quad -i\\  0 &\quad 1 &\quad -i & \quad \,\,\,0\\
 1 &\quad 0 &\quad \,\,0 & \quad \!\phm i\\
0 &\quad 1 &\quad \!\phm i & \quad
\,\,\,0\end{array}\right)\,,
\eeq
it follows from \eq{hbasism2} that:
\beq
V^\dagger \mathscr{M}^2 V=\left(\begin{array}{cc}\mathcal{M} &\quad 0
    \\
0 &\quad 0 \end{array}\right)\,,
\eeq
where $\mathcal{M}$ is the $3\times 3$ neutral Higgs squared-mass
matrix in the $\varphi_1^0$--$\varphi_2^0$--$a^0$ basis
obtained in \eq{matrix33}.

We diagonalize $\mathcal{M}$ as described in \sect{sec:four}.
The corresponding diagonalization of $\mathscr{M}^2$ is given by:
\beq \label{rvmvr}
\mathcal{D} \mathscr{M}^2\mathcal{D}^\dagger\equiv
\left(\begin{array}{cc} R &\quad 0\\ 0 &\quad 1\end{array}\right)
V^\dagger \mathscr{M}^2 V
\left(\begin{array}{cc} R^T &\quad 0\\ 0 &\quad 1\end{array}\right)
= \left(\begin{array}{cc}\mathcal{M}_D &\quad 0\\
0 &\quad 0 \end{array}\right)\,,
\eeq
where $\mathcal{M}_D= {\rm diag}~(m_1^2\,,\,m_2^2\,,\,m_3^2)$
and $m_k$ is the mass of the neutral Higgs mass-eigenstate $h_k$.
The diagonalizing matrix $\mathcal{D}$ is given by:
\beq
\mathcal{D}\equiv
\left(\begin{array}{cc} R &\quad 0\\ 0 &\quad 1\end{array}\right) V^\dagger
=\frac{1}{\sqrt{2}}\left(\begin{array}{cccc} d_{11} &\quad d_{12}
&\quad d^*_{11} &\quad d^*_{12} \\
 d_{21} &\quad d_{22}
&\quad d^*_{21} &\quad d^*_{22} \\ d_{31} &\quad d_{32}
&\quad d^*_{31} &\quad d^*_{32} \\ d_{41} &\quad d_{42}
&\quad d^*_{41} &\quad d^*_{42}\end{array}\right)\,,
\eeq
where
\beqa
\!\!\!\!\! d_{11}&=&c_{13}c_{12}\,,\quad\qquad
d_{21}=c_{13}s_{12}\,,\quad\qquad
d_{31}=s_{13}\,,\qquad\,\,\,\,
d_{41}=i\,,\nonumber \\
\!\!\!\!\! d_{12}&=&-s_{123}e^{-i\theta_{23}}\,,\quad\!
d_{22}=c_{123}e^{-i\theta_{23}}\,,\quad\!
d_{32}= ic_{13}e^{-i\theta_{23}}\,,\,\,
d_{42}=0\,,
\eeqa
with the $c_{ij}$ and $s_{ij}$ defined in \eq{rmatrix} and
\beq
c_{123}\equiv c_{12}-is_{12}s_{13}\,,\qquad\quad s_{123}\equiv
s_{12}+ic_{12}s_{13}\,.
\eeq
Note that $\mathcal{D}$ is a unitary matrix and $\det\mathcal{D}=1$.
Unitarity implies that:
\beqa
&&\qquad\qquad\qquad\quad
\Re\left(d_{k1} d_{\ell 1}^* + d_{k2}d_{\ell 2}^* \right)=\delta_{k\ell}\,,
\label{dunitarity1} \\[6pt]
&&\half\sum_{k=1}^4\,|d_{k1}|^2=\half\sum_{k=1}^4\,|d_{k2}|^2=1\,,\qquad\qquad
\sum_{k=1}^4\,d_{k2}^{\,2}=\sum_{k=1}^4\,d_{k1}d_{k2}=0\,. \label{dunitarity2}
\eeqa
Noting that $d_{41}=i$ and $d_{42}=0$ [and using \eq{dtoq}], 
these equations reduce to
\eqs{unitarity1}{unitarity2} given in \sect{sec:four}.  In addition, 
${\rm det}~\mathcal{D}=-i\det RW =1$,
where $RW$ is given in \eq{RW}.  This yields an additional constraint
on the $d_{k\ell}$ [\textit{c.f.} \eq{detrw}].

The matrix $\mathcal{D}$ converts the neutral
Higgs basis fields into the neutral Higgs mass-eigenstates:
\beq \label{conversion}
\left(\begin{array}{c} h_1\\ h_2 \\ h_3 \\ G^0\end{array}\right)
=\mathcal{D}
\left(\begin{array}{c} \overline{H}\lsup{0\,\dagger}_1\\ H^{0\,\dagger}_2 \\
\overline{H}_1\lsup{0} \\ H^0_2\end{array}\right)\,,
\eeq
where $\overline{H}_1\lsup{0}\equiv H_1^0-v/\sqrt{2}$.

The mass-eigenstate fields do not depend on the choice of basis.
Using the fact that $H_1$ is invariant and $H_2$ is pseudo-invariant
with respect to flavor-U(2) transformations,
\eq{conversion} implies that the $d_{k1}$ are invariants whereas the
$d_{k2}$ are pseudo-invariants with the same transformation law as $H_2$
[\eq{h2pseudo}].  One can also check this directly from \eq{rvmvr},
using the fact that the physical Higgs masses must be basis-independent.
These results then imply that $\theta_{12}$ and $\theta_{13}$ are
invariant whereas $e^{i\theta_{23}}$ is a pseudo-invariant, \textit{i.e.},
$e^{i\theta_{23}}\to (\det U)^{-1} e^{i\theta_{23}}$
under an arbitrary flavor-U(2) transformation $U$.

Finally, using the results of this appendix, we can eliminate the
Higgs basis fields entirely and obtain the diagonalizing matrix that
converts the neutral Higgs fields in the generic basis into the
neutral Higgs mass-eigenstates:
\beq \label{gentomass}
\left(\begin{array}{c} h_1\\ h_2 \\ h_3 \\ G^0\end{array}\right)
=\mathcal{D}\left(\begin{array}{c}
\overline\Phi_{\bbar}\lsup{0\,\dagger}\widehat U\lsup{\dagger}_{b\abar} \\
\widehat U_{a\bbar}\overline\Phi_b\lsup{0}
\end{array}\right)\,,
\eeq
where $\overline\Phi_a\lsup{0}\equiv \Phi_a^0-v\widehat v_a/\sqrt{2}$ and
$\widehat U$ is the matrix
that converts the generic basis fields into the Higgs basis fields
[see \eq{ugenhiggs}].
\Eq{gentomass} then yields:
\beq
h_k=\frac{1}{\sqrt{2}}\left[\overline\Phi_{\abar}\lsup{0\,\dagger}
(d_{k1}\widehat v_a+d_{k2}\widehat w_a)
+(d^*_{k1}\widehat v^*_{\abar}+d^*_{k2}\widehat w^*_{\abar})
\overline\Phi_a\lsup{0}\right]\,,
\eeq
where $h_4\equiv G^0$.
Note that the U(2)-invariance of the $h_k$ imply
that the $d_{k1}$ are invariants and the $d_{k2}$ are pseudo-invariants
that transform oppositely to $\widehat w$ as $d_{k2}\to (\det
U)d_{k2}$ in agreement with the previous results above.  Indeed, it
is useful to define:
\beq \label{dtoq}
d_{k1}\equiv q_{k1}\,,\qquad {\rm and}\qquad
d_{k2}\equiv q_{k2} e^{-i\theta_{23}}\,,
\eeq
where \textit{all} the $q_{k\ell}$ are U(2)-invariant [see \eq{qtrans}].
In particular, $\widehat w_a e^{-i\theta_{23}}$ is a proper vector
with respect to flavor-U(2) transformations.
Hence,
\beq \label{apphk}
h_k=\frac{1}{\sqrt{2}}\left[\overline\Phi_{\abar}\lsup{0\,\dagger}
(q_{k1} \widehat v_a+q_{k2}\widehat w_a e^{-i\theta_{23}})
+(q^*_{k1}\widehat v^*_{\abar}+q^*_{k2}\widehat w^*_{\abar}e^{i\theta_{23}})
\overline\Phi_a\lsup{0}\right]\,,
\eeq
provides an invariant expression for the neutral
Higgs mass-eigenstates.
\chapter{Explicit formulae for the neutral Higgs masses and mixing angles}
\label{app:three}
To obtain expressions for the neutral Higgs masses and mixing angles,
we insert \eq{master} into \eq{genericpot}, and expand out the
resulting expression, keeping only terms that are linear and quadratic
in the fields.  Using \eqs{invariants}{pseudoinvariants}, one can
express the resulting expression in terms of the invariants
($Y_1$, $Y_2$ and $Z_{1,2,3,4}$) and pseudo-invariants ($Y_3$, $Z_{5,6,7}$).
The terms linear in the fields vanish if the potential
minimum conditions [\eq{hbasismincond}] are satisfied.  We then
eliminate $Y_1$ and $Y_3$ from the expressions of the quadratic terms.
The result is:
\beqa \label{V2}
\mathcal{V}_2\!\!\!&=&\!\!\! H^+H^-(Y_2+\frac{v^2 Z_3}{2})+\frac{v^2}{2} h_j h_k \biggl\{
Z_1 \Re(q_{j1})\Re(q_{k1})
+[\frac{Z_3+Z_4}{2}+Y_2/v^2]\Re(q_{j2}q^*_{k2}) \nonumber \\
&&\!\!\!\!\!\quad +\half\Re(Z_5 q_{j2}q_{k2}\,e^{-2i\theta_{23}})
+\Re(q_{j1})\Re(Z_6 q_{k2}\, e^{-i\theta_{23}})
+\Re(q_{k1})\Re(Z_6 q_{j2}\, e^{-i\theta_{23}})
\biggr\}\nonumber \\[6pt]
&=&
m_{H^\pm}^2 H^+H^- +\half \sum_k m_k^2 (h_k)^2 +\half v^2\sum_{j\neq k}
C_{jk}h_j h_k\,.\eeqa
In \eq{V2}, there is an implicit sum over $j,k=1,\ldots,4$ (with $h_4\equiv G^0$), where the $C_{kj}$ are given by\footnote{For convenience we provide linear combinations of $C_{23}$ and $C_{13}$, but the explicit forms can be obtained by a trivial calculation.}
\beqa
C_{23} c_{12}-C_{13} s_{12} &=& s_{13}\,\Re(Z_6\,e^{-i\theta_{23}})-
\half c_{13}\,\Im(Z_5\,e^{-2i\theta_{23}}) \label{c23c12a}\,,\\
C_{23} s_{12}+C_{13} c_{12} &=& \half(Z_1 - A^2/v^2) \sin 2\theta_{13}
-\cos 2\theta_{13}\,\Im(Z_6\,e^{-i\theta_{23}})\,,\label{c23c12b}
\eeqa
\beqa C_{12} &=& c_{12}s_{12}\left[c_{12}^2(Z_1 - A) -\Re(Z_5\,e^{-2i\theta_{23}}) +2 s_{12}c_{12}\Im(Z_6\,e^{-i\theta_{23}}) \right]\nonumber\\
&& (c_{12}^2 - s_{12}^2)\left[\half s_{13}\Im(Z_5\,e^{-2i\theta_{23}}) +c_{13} \Re(Z_6\,e^{-i\theta_{23}})\right]\,.\label{th12eq}\eeqa

However, since $q_{41}=i$ and $q_{42}=0$, it is clear that there are
no terms in \eq{V2} involving $G^0$.  Hence, we may restrict the sum
to run over $j,k=1,2,3$.  The charged Higgs mass obtained above
confirms the result quoted in \eq{hplus}.  The neutral Higgs boson
masses are given by:
\beq \label{hmassesinv}
m_k^2 = |q_{k2}|^2 A^2+ v^2\left[q_{k1}^2 Z_1
+\Re(q_{k2})\,\Re(q_{k2}Z_5\,e^{-2i\theta_{23}})
+2q_{k1}\Re(q_{k2}Z_6\, e^{-i\theta_{23}})\right]\,,
\eeq
where $A^2$ is defined in \eq{madef}.  It is often convenient to
assume that $m_1\leq m_2\leq m_3$.

Note that the right-hand side of \eq{hmassesinv} is manifestly
U(2)-invariant.  Moreover, by using \eqs{unitarity1}{unitarity2}, 
one finds that
the sum of the three neutral Higgs boson squared-masses is given by
\beq \label{tracesum}
{\rm Tr}~\mathcal{M}=\sum_k\,m_k^2=2Y_2+(Z_1+Z_3+Z_4)v^2\,,
\eeq
as expected. A more explicit form for the neutral Higgs squared-masses
than the one obtained in \eq{hmassesinv}
would require the solution of the cubic characteristic equation
[\eq{charpoly}].  Although an analytic solution can be found, it is too
complicated to be of much use (a numerical evaluation is more practical).

For the neutral scalar states $h_i$ to correspond to physical mass-eigenstates, the coefficients $C_{jk}$ of \eqthree{c23c12a}{c23c12b}{th12eq} must vanish.  Since $C_{jk}$ is
symmetric under the interchange of its indices, the conditions $C_{jk}=0$
yield three independent equations that determine the two
mixing angles $\theta_{12}$ and $\theta_{13}$ and 
an invariant combination of $\theta_{23}$ and the phase of $Z_6$ (or
$Z_5$).  These three invariant angles are defined
modulo $\pi$ once a definite convention is established for the signs
of neutral Higgs mass-eigenstate fields (as discussed at the end
of \sect{sec:four}).  Unique solutions for the invariant angles within
this domain are obtained after a mass ordering for the three neutral 
Higgs bosons is specified (except at certain singular points of the 2HDM
parameter space as noted in footnote \ref{fnmass}). 

To determine explicit formulae for the invariant angles, we shall initially
assume that $Z_6\equiv |Z_6|e^{i\theta_6}\neq 0$ and define the
invariant angles $\phi$ and $\theta_{56}$:
\beq\label{app:invang}
\begin{array}{c}
\phm\phi\equiv\theta_6-\theta_{23}\,, \\[6pt]
\theta_{56}\equiv \theta_5-\theta_6\,,\end{array}
\quad\qquad {\rm where} \qquad \begin{cases}
\,\theta_6\equiv\arg Z_6\,, &\\
\,\theta_{5}\equiv\half\arg Z_5\,. & \end{cases}
\eeq
The factor of $1/2$ in the definition of $\theta_5$ has been inserted for
convenience.  As discussed in \sect{sec:four}, we can fix the
conventions for the overall signs of the $h_k$ fields by restricting 
the domain of $\theta_{12}$, $\theta_{13}$ and $\phi$ to the region:
\beq \label{app:domains}
-\pi/2\leq\theta_{12}\,,\,\theta_{13}<\pi/2\,,\qquad\quad
0\leq \phi<\pi\,.
\eeq

Setting $C_{13}=C_{23}=0$ in \eqs{c23c12a}{c23c12b} yields:
\beqa
\tan\theta_{13}&=&\frac{\Im(Z_5\,e^{-2i\theta_{23}})}
{2\,\Re(Z_6\,e^{-i\theta_{23}})}
\,, \label{tan13}\\[8pt]
\tan 2\theta_{13}&=&\frac{2\,\Im(Z_6\,e^{-i\theta_{23}})}
{Z_1-A^2/v^2}\,. \label{tan213}
\eeqa
Using the well known identity
$\tan 2\theta_{13}=2\tan\theta_{13}/(1-\tan^2\theta_{13})$,
one can use \eqs{tan13}{tan213} to eliminate $\theta_{13}$ and 
obtain an equation for $\phi$.%
\footnote{Recall that the
quantity $A^2$ [\eq{madef}] depends on $\phi$ via
$\Re(Z_5\,e^{-2i\theta_{23}})=|Z_5|\cos 2(\theta_{56}+\phi)$.}
The resulting equation for $\phi$ has more than one solution.
Plugging a given solution for $\phi$ back into \eq{tan13} yields
a corresponding solution for $\theta_{13}$. 
Note that if $(\theta_{13}\,,\,\phi)$ is a solution to
\eqs{tan13}{tan213}, then so is $(-\theta_{13}\,,\,\phi\pm\pi)$,
in agreement with \eqs{signflip13}{signflip23}.  
By restricting to the domain of $\theta_{13}$ and $\phi$ specified
by \eq{app:domains}, only one of these two solutions survives.
However, multiple solutions to \eqs{tan13}{tan213} still exist within the
allowed domain, which
correspond to different choices for the mass
ordering of the three neutral Higgs fields.  
By imposing a particular mass ordering, a unique
solution is selected [see \eqs{s13}{c13s12}].
  
Finally, having obtained $\phi$ and
$\tan\theta_{13}$, we use $C_{12}=0$ in \eq{th12eq} to compute $\theta_{12}$, with the result
\beq \label{tan12}
\tan 2\theta_{12}=\frac{s_{13}\,\Im(Z_5\,e^{-2i\theta_{23}})+2c_{13}\,
\Re(Z_6\,e^{-i\theta_{23}})}{c_{13}^2\left(A^2/v^2-Z_1\right)
+\Re(Z_5\,e^{-2i\theta_{23}})-2s_{13} c_{13}\,
\Im(Z_6\,e^{-i\theta_{23}})}\,.
\eeq
We can simplify the above result by using \eqs{tan13}{tan213}
to solve for $\Im(Z_5\,e^{-2i\theta_{23}})$ and
$\Im(Z_6\,e^{-i\theta_{23}})$ and eliminate these factors from \eq{tan12}.
The end result is:
\beq \label{tan12f}
\tan 2\theta_{12}=\frac{2\cos 2\theta_{13}\,\Re(Z_6\,e^{-i\theta_{23}})}
{c_{13}\left[c_{13}^2(A^2/v^2-Z_1)+\cos 2\theta_{13}\,
\Re(Z_5\,e^{-2i\theta_{23}})\right]}\,.
\eeq
Note that if $\theta_{12}$ is a solution to \eq{tan12f}, then
$\theta_{12}\pm\pi/2$ is also a solution.  That is, \eq{tan12f}
yields two solutions for $\theta_{12}$ 
in the allowed domain [\eq{app:domains}], which
correspond to the two possible mass orderings of $h_1$ and $h_2$
as shown below \eq{m2m1}.

The neutral Higgs boson masses were given in \eq{hmassesinv}.
With the help of \eqthree{tan13}{tan213}{tan12f}, one can 
can express these masses in terms of 
$Z_1$, $Z_6$ and the invariant angles:
\beqa
m_1^2 &=&
\left[Z_1-\frac{s_{12}}{c_{12}c_{13}}\Re(Z_6\,e^{-i\theta_{23}})
+\frac{s_{13}}{c_{13}}\Im(Z_6\,e^{-i\theta_{23}})\right]v^2\,,
\label{m12}\\[5pt]
m_2^2 &=&
\left[Z_1+\frac{c_{12}}{s_{12}c_{13}}\Re(Z_6\,e^{-i\theta_{23}})
+\frac{s_{13}}{c_{13}}\Im(Z_6\,e^{-i\theta_{23}})\right]v^2\,,
\label{m22}\\[5pt]
m_3^2 &=&
\left[Z_1-\frac{c_{13}}{s_{13}}\Im(Z_6\,e^{-i\theta_{23}})\right]v^2\,.
\label{m32}
\eeqa

For the subsequent analysis, it is useful to invert \eqst{m12}{m32}
and solve for $Z_1$, $\Re(Z_6\,e^{-i\theta_{23}})$ and 
$\Im(Z_6\,e^{-i\theta_{23}})$:
\beq
Z_1 v^2=m_1^2 c_{12}^2 c_{13}^2+m_2^2 s_{12}^2 c_{13}^2 + m_3^2
s_{13}^2\,,\label{z1v} 
\eeq
\beqa
\Re(Z_6\,e^{-i\theta_{23}})\,v^2 &=& c_{13}s_{12}c_{12}(m_2^2-m_1^2)\,,
\label{z6rv} \\[5pt]
\Im(Z_6\,e^{-i\theta_{23}})\,v^2 &=& s_{13}c_{13}(c_{12}^2 m_1^2+s_{12}^2
m_2^2-m_3^2) \,. \label{z6iv}
\eeqa
In addition, \eqs{tan213}{tan12f} can be used to express 
$\Re(Z_5\,e^{-i\theta_{23}})$ in terms of~$Z_6$:
\beq \label{z56}
\Re(Z_5\,e^{-2i\theta_{23}})=\frac{c_{13}}{s_{13}}
\Im(Z_6\,e^{-i\theta_{23}})
+\frac{c_{12}^2-s_{12}^2}{c_{13}s_{12}c_{12}}\,
\Re(Z_6\,e^{-i\theta_{23}})\,.
\eeq
Inserting \eqs{z6rv}{z6iv} into \eqs{tan13}{z56} then yields
expressions for $\Im(Z_5\,e^{-i\theta_{23}})$
and $\Re(Z_5\,e^{-i\theta_{23}})$ in terms of the invariant angles and
the neutral Higgs masses.
The above results can be used to derive an expression for
\beqa \label{imz56} \hspace{-.2 in}
\Im(Z_5^* Z_6^2)&=& 2\,\Re(Z_5 e^{-2i\theta_{23}})\,
\Re(Z_6\,e^{-i\theta_{23}})\,\Im(Z_6\,e^{-i\theta_{23}}) \nonumber \\
&& \qquad -\,\Im(Z_5 e^{-2i\theta_{23}})\left\{
[\Re(Z_6\,e^{-i\theta_{23}})]^2-[\Im(Z_6\,e^{-i\theta_{23}})]^2\right\}\,.
\eeqa
Using \eq{tan13} and \eqst{z6rv}{z56}, 
one can simplify the right hand side of \eq{imz56} to obtain:
\beq \label{imz56f}
\Im(Z_5^* Z_6^2)\,v^6=2s_{13}c_{13}^2 s_{12}c_{12}\,(m_2^2-m_1^2)(m_3^2-m_1^2)
(m_3^2-m_2^2)\,.
\eeq
\Eq{imz56f} was first derived in \Ref{cpx}; it is 
equivalent to a result initially obtained in \Ref{pomarol}.
In particular, if any two of the 
neutral Higgs masses are degenerate, then $\Im(Z_5^* Z_6^2)=0$,
in which case one can always find a basis in which 
the pseudo-invariants $Z_5$ and $Z_6$ are simultaneously real. 
The neutral scalar squared-mass matrix [\eq{matrix33}] then 
breaks up into a block diagonal form consisting of a
$2\times 2$ block and a $1\times 1$ block.  The 
diagonalization of the $2\times 2$ block has a simple analytic form,
and the neutral scalar mixing
can be treated more simply by introducing one invariant
mixing angle instead of the three needed in the general case.
Note that  $\Im(Z_5^* Z_6^2)=0$ is a necessary (although not
sufficient) requirement for a CP-conserving Higgs sector, as discussed
in Chapter \ref{cpcoch}.  For the remainder of this Appendix, 
we shall assume that the
neutral Higgs boson masses are non-degenerate.

In order to facilitate the discussion of the CP-conserving limit and
the decoupling limit of the 2HDM (which are treated in  Chapter \ref{cpcoch} and \App{app:four}, respectively),
it is useful to derive a number of additional relations for the
invariant angles.  First, we employ \eqst{m12}{m32} to
eliminate $\theta_{12}$ and $\phi$ and obtain a single equation
for $\theta_{13}$:
\beq \label{s13}
s_{13}^2=\frac{(Z_1 v^2-m_1^2)(Z_1 v^2-m_2^2)+|Z_6|^2 v^4}
{(m_3^2-m_1^2)(m_3^2-m_2^2)}\,.
\eeq
\Eq{s13} determines $c_{13}$ (in the convention where 
$c_{13}\geq 0$).
The sign of $s_{13}$ is determined
from \eq{m32}, which can be rewritten as: 
\beq \label{phieq}
\sin\phi=\frac{(Z_1v^2-m_3^2)\tan\theta_{13}}{|Z_6|v^2}\,.
\eeq
Since $\sin\phi\geq 0$ in the angular domain specified by \eq{app:domains}, it
follows that the sign of $s_{13}$ is equal to the sign of 
the quantity $Z_1 v^2-m_3^2$.  In particular, if $m^2_3$ is the largest 
eigenvalue of $\widetilde{\mathcal{M}}$ [\eq{matrix33}], then it must
be greater than the largest diagonal element of
$\widetilde{\mathcal{M}}$.  That is, $Z_1 v^2-m_3^2< 0$ if 
$m_3> m_{1,2}$, in which case $s_{13}\leq 0$.

However,
\eq{phieq} does not fix the sign of $\cos\phi$.  To determine this sign, we
can use \eq{tan13} to eliminate $\theta_{13}$ from
\eq{phieq}.  Consequently, one obtains a single equation for $\phi$:
\beq \label{tan2phi}
\tan 2\phi=\frac{\Im(Z_5^* Z_6^2)}{\Re(Z_5^* Z_6^2)+
\displaystyle{\frac{|Z_6|^4 v^2}{m_3^2-Z_1 v^2}}}\,.
\eeq
Given $\sin\phi\geq 0$ and $\tan 2\phi$ in the region
$0\leq\phi<\pi$, one can uniquely determine the value of $\phi$ 
(and hence the sign of $\cos\phi$).
Thus, for a fixed ordering of the neutral Higgs masses, 
\eqst{s13}{tan2phi} provide a unique solution for $(\theta_{13},\phi)$
in the domain $-\pi/2\leq\theta_{13}<\pi/2$ and $0\leq\phi<\pi$.

Next, we note that \eq{z6rv} can be rewritten as:
\beq \label{m2m1}
\sin 2\theta_{12}=\frac{2\,|Z_6|\,v^2\cos\phi}
{c_{13}(m_2^2-m_1^2)}\,.
\eeq
As advertised below \eq{tan12f}, the mass ordering of $m_1$ and $m_2$ 
fixes the sign of $\sin 2\theta_{12}$.
In particular, in the angular domain 
of \eq{app:domains}, $m_2>m_1$ implies that $s_{12}\cos\phi\geq 0$.
The sign of $s_{12}$ is then fixed after using \eq{imz56f} to infer that
$\sin 2\theta_{56}\cos\phi\geq 0$ for $m_3>m_2>m_1$.

An alternative expression for $\theta_{12}$ can be obtained by
combining \eqs{z1v}{s13}: 
which yields:
\beq \label{c13s12}
c_{13}^2 s_{12}^2=\frac{(Z_1 v^2-m_1^2)(m_3^2-Z_1 v^2)-|Z_6|^2 v^4}
{(m_2^2-m_1^2)(m_3^2-m_2^2)}\,.
\eeq 
Note the similarity of the expressions given by \eqs{s13}{c13s12};
both these results play an important role in determining the 
conditions that govern the decoupling limit.
 
A simpler form for $\tan^2\theta_{13}$ can also be obtained by combining 
\eqs{tan213}{phieq}:
\beq \label{t213}
\tan^2\theta_{13}=\frac{m_3^2-A^2}{m_3^2-Z_1 v^2}\,.
\eeq
Finally, one can derive an
expression for $m_2^2-m_3^2$, after eliminating $\Im(Z_6
e^{-i\theta_{23}})$ in favor of $\Re(Z_5 e^{-2i\theta_{23}})$ using \eq{z56}:
\beq \label{m2m3}
m_2^2-m_3^2=\frac{v^2}{c_{13}^2}\left[\Re(Z_5 e^{-2i\theta_{23}})
+\frac{c_{13}(s_{12}^2-c_{12}^2 s_{13}^2)}{s_{12}c_{12}}
\Re(Z_6 e^{-i\theta_{23}})\right]\,.
\eeq
The expressions for the differences of squared-masses
[\eqs{m2m1}{m2m3}] 
take on rather simple forms in the CP-conserving limit.

In this discussion we have assumed that $Z_6 \neq 0$.  The $Z_6 = 0$ case is treated separately in section \ref{zsixzero}.

\chapter{The Decoupling Limit of the 2HDM}
\label{app:four}
One can consider the case in which all but one Higgs have masses at some high scale $\Lambda$.  The effective low energy theory is a one-Higgs-doublet
model that corresponds to the Higgs sector of the Standard Model; the heavier scalar particles are therefore ``decoupled" from the low energy theory~\cite{ghdecoupling,habernir}.  The light neutral (SM-like) Higgs has a mass of order the electroweak scale: $m_1 \sim v$.  This decoupling limit corresponds to $Y_2\gg v^2$ and $|Z_i|\lsim\mathcal{O}(1)~\forall~i$.  In the first section of this chapter, I present the basis-independent description of the 2HDM that was derived in~\cite{haberoneil}.  Section \ref{opesec} describes the impact of CP-violation on the effective dimension-6 operators in the decoupling limit of the 2HDM.   

\section{The Decoupling Condition in the Basis-Independent Formalism}

We shall order the neutral scalar masses according to $m_1< m_{2,3}$
and define the invariant Higgs mixing angles accordingly.
Thus, we expect one light CP-even Higgs boson, $h_1$, with couplings
identical (up to small corrections) to those of the 
Standard Model (SM) Higgs boson.
Using the fact that $m^2_1$, $|Z_i| v^2 \ll m^2_2$, $m^2_3$, $\mhpm^2$ in the
decoupling limit, \eqs{m22}{m32} yield:
\beq \label{decoupconds}
|s_{12}|\lsim\mathcal{O}\left(\frac{v^2}{m_2^2}\right)\ll 1\,,
\qquad \qquad |s_{13}|\lsim\mathcal{O}\left(\frac{v^2}{m_3^2}\right)\ll 1\,,
\eeq
and \eq{tan2phi} imples that $\tan 2\phi+\tan 2\theta_{56}\ll 1$,
where $\theta_{56}\equiv \half\arg Z_5-\arg Z_6$.  This latter
inequality is equivalent to:
\beq \label{z5decoup}
\Im(Z_5\,e^{-2i\theta_{23}})\lsim\mathcal{O}
\left(\frac{v^2}{m_3^2}\right)\ll 1\,.
\eeq
Note that \eq{z5decoup} is also satisfied if $\theta_{23}\to\theta_{23}+\pi/2$.
These two respective solutions (modulo~$\pi$) correspond to the two
possible mass orderings of $h_2$ and $h_3$.

One can explicitly verify the assumed mass hierarchy of the  
Higgs bosons in the decoupling limit.  
Using \eqs{m12}{decoupconds}, it follows 
that $m_1^2=Z_1 v^2$, with corrections 
$\lsim\mathcal{O}(v^4/m^2_{2,3})$.
\Eq{t213} yields $m_3^2=A^2$, with corrections 
$\lsim\mathcal{O}(v^2)$, and \eq{m2m3} yields
$m_3^2-m_2^2\lsim\mathcal{O}(v^2)$.  Finally, \eqs{hplus}{madef} 
imply that $\mhpm^2-m_3^2\lsim\mathcal{O}(v^2)$.  That is,
$m_1\ll m_2\simeq m_3\simeq \mhpm$.

The values of the $q_{k\ell}$ in the exact decoupling 
limit, where \beq s_{12}=s_{13}=\Im(Z_5\,e^{-2i\theta_{23}})=0\,,\eeq
are tabulated in Table~\ref{tab5}.
\begin{table}[ht!]
\centering
\caption{The U(2)-invariant quantities $q_{k\ell}$ in the exact
decoupling limit.}\vskip 0.15in
\label{tab5}
\begin{tabular}{|c||c|c|}\hline
$\phaa k\phaa $ &\phaa $q_{k1}\phaa $ & \phaa $q_{k2} \phaa $ \\ \hline
$1$ & $1$ & $0$ \\
$2$ & $0$ & $1$ \\
$3$ & $0$ & $i$ \\
$4$ & $i$ & $0$ \\ \hline
\end{tabular}
\end{table}

\noindent
It is a simple exercise to insert the values of the
$q_{k\ell}$  in the exact decoupling limit
into the Higgs couplings of \sects{sec:five}{sec:six}.
The couplings of $h_1\equiv h$ are then given by:
\beqa
&&\hspace{-0.2in} \mathscr{L}_{\rm h}= \half(\partial_\mu h)^2-\half
Z_1v^2 h^2
-\half vZ_1 h^3 -\eighth vZ_1 h^4
+\left(gm_W W_\mu^+W^{\mu\,-}+\frac{g}{2c_W}
m_Z Z_\mu Z^\mu\right) h \nonumber \\
&& \qquad
+\left[\frac{g^2}{4}  W_\mu^+W^{\mu\,-}
+\frac{g^2}{8c_W^2}Z_\mu Z^\mu\right] h^2
+\biggl\{\left(\frac{eg}{2} A^\mu W_\mu^+ -\frac{g^2s_W^2}{2c_W}Z^\mu
  W_\mu^+\right)G^-h+{\rm h.c.}\biggr\} \nonumber \\
&& \qquad
-\half ig\left[W_\mu^+G^-\ddel\lsup{\,\mu} h+{\rm h.c.}\right]
+\frac{g}{2c_W}  Z^\mu G^0\ddel_\mu h
+\frac{1}{v}\overline D M_D D h+\frac{1}{v}\overline U M_U U h\,.
\eeqa
This is precisely the SM Higgs Lagrangian.  Even in the
most general CP-violating 2HDM, the interactions of the $h$ in the
decoupling limit are CP-conserving and diagonal in quark flavor space.
CP-violating and flavor non-diagonal effects in the Higgs interactions
are suppressed by factors of $\mathcal{O}(v^2/m^2_{2,3})$.
In contrast to the SM-like Higgs boson $h$, 
the interactions of the heavy neutral Higgs bosons ($h_2$
and $h_3$) and the charged Higgs bosons ($\hpm$) 
exhibit both CP-violating and quark flavor non-diagonal couplings
(proportional to the $\rho^Q$) in the decoupling limit.
In particular, whereas \eq{z5decoup} implies that
$\sin 2(\theta_5-\theta_{23})\ll 1$, the CP-violating invariant
quantities
$\sin(\theta_6-\theta_{23})$ and $\sin(\theta_7-\theta_{23})$ 
[\textit{c.f.} \eq{cpang}] need
not be small in the most general 2HDM.

One can understand the origin of 
the decoupling conditions [\eqs{decoupconds}{z5decoup}] as follows.
First, using \eq{hmassesinv}, we see that we can decouple $h_2$ and $h_3$
(and $H^\pm$) by
taking $A^2\gg v^2$ while sending $q_{12}\to 0$.  Thus, in the
convention in which the mass ordering of
the three neutral Higgs states is $m_1\leq m_2\leq m_3$, it follows
that the exact decoupling limit is formally achieved when
$A^2\to\infty$ and
$|q_{12}|^2=s_{12}^2+c_{12}^2 s_{13}^2=0$, which implies that
$s_{12}=s_{13}=0$.  Inserting these results into
\eq{rtil} yields $\widetilde R=I$, where $I$ is the
$3\times 3$ identity matrix.  Consequently, 
$\wtil{\mathcal{M}}$
[see \eqst{mtilmatrix}{diagtil}] must be diagonal up to
corrections of $\mathcal{O}(v^2/A^2)$.  However, because 
\eq{mtilmatrix} is dominated in the decoupling limit
by its $22$ and $33$ elements (which are approximately degenerate), 
it follows that the $23$ element must vanish exactly in leading order.
Thus, in the exact decoupling limit, 
$\Im(Z_5\,e^{-2i\theta_{23}})=0$.  Note that this latter
constraint is consistent with \eq{tan13}, as $\theta_{13}=0$ in
the decoupling limit.

For further details and a more comprehensive treatment of the decoupling
limit, see \Ref{ghdecoupling}.
\section{Dimension 6 Operators in the Decoupling Limit\label{opesec}}
One can consider the case in which all but one Higgs have masses at some high scale $\Lambda$.  The light neutral (SM-like) Higgs has a mass of order the electroweak scale: $m_1 \sim v$. In this limit, 
\beqa
s_{12} & \sim  s_{13}  \sim & \mathcal{O}\left(\frac{v^2}{\LL}\right), \nonumber\\ 
c_{12} & \sim  c_{13}  \sim & 1, \label{sandc}
\eeqa
as discussed in~\cite{haberoneil}.
In this scenario one can imagine that one has integrated out the higher mass fields, leaving the Standard Model as the effective low-energy theory.  In a linear model of electroweak symmetry breaking, deviations from the Standard Model can be realized in two dimension 6 operators of the effective Lagrangian:
\[ \mathcal{L}_{eff} = \mathcal{L}_{SM}+ \frac{1}{v^2}(S~ \mathcal{O}_S+ T~ \mathcal{O}_T), \]
where $S$ and $T$ are of order $\mathcal{O}(\frac{v^2}{\LL})$.\footnote{If the new physics involves additional vectors, such as gauge bosons, two additional parameters ($Y$ and $W$) associated with dimension 6 operators are added to the effective Lagrangian---see ref. ~\cite{barbieri}.}  The parameter $U$, which corresponds to a dimension 8 operator and is of order  $\mathcal{O}(\frac{v^4}{\Lambda^4})$, will be negligible here ~\cite{grinstein}. Thus, in a decoupling scenario, contributions to $S$ and $T$ from an extended Higgs sector are constrained by requiring that they not spoil the rough agreement with the Standard Model demonstrated in equations (\ref{STdec}) -- (\ref{STdecalt}).  
Using a notation similar to that of Grojean et al.~\cite{grojean}, one can define the operators as:
\beqa \mathcal{O}_S = \frac{\alpha}{4 s_W c_W}\mathcal{O}_{WB},&  \mathcal{O}_T = - 2 \alpha \mathcal{O}_h, \eeqa
where 
\beqa \label{op}\mathcal{O}_{WB} = \Phi^\dagger \sigma^a \Phi W^a_{\mu \nu}B^{\mu \nu},& \mathcal{O}_h = |\Phi^\dagger D_{\mu} \Phi |^2.  \eeqa
The covariant derivative is given by $D_\mu = \delta_\mu -\frac{i g}{2}\sigma^a W^a_\mu - ig' Y B_\mu$, and the field strengths by $W^a_{\mu \nu}= \delta_\mu W^a_\nu - \delta_\nu W^a_\mu + g \epsilon_{abc}W^b_\mu W^c_\nu$ and $B_{\mu \nu}= \delta_\mu B_\nu - \delta_\nu B_\mu$. $\mathcal{O}_{WB}$ and $\mathcal{O}_h$ are related to two well-known CP-conserving operators  ($O_{WB}$ and $O^{(3)}_\varphi$) from the work of Buchm\"{u}ller and Wyler~\cite{buchmuller}.

Before calculating $S$ and $T$ in this limit, it is useful to derive relations between the scalar masses. This is done in Appendix ~\ref{app:two}, with the results
\beqa \label{relations}
m_3^2 &\equiv& \LL, \nonumber\\
m_2^2 &=& \LL + \Re(\zfive) v^2, \nonumber\\
\mhpm^2 &=& \LL + \half[\Re(\zfive) - Z_4]v^2. \eeqa
These relations allow $S$ to be expanded in the decoupling limit in powers of $\frac{v^2}{\LL}$.

Now one can calculate $S$ in the decoupling limit.  It is convenient to set the reference point $m_\phi$ equal to $m_1$.  Thus, one can write the general expression for $S$ as follows:
\beqa 
S&=& \frac{1}{\pi m_Z^2} \biggl\{q_{31}^2\left[\mathcal{B}_{22}(m_Z^2;m_Z^2,m_3^2)+\mathcal{B}_{22}(m_Z^2;m_1^2,m_2^2) - m_Z^2 \mathcal{B}_{0}(m_Z^2;m_Z^2,m_3^2) \right]\biggr. \nonumber\\
& &+q_{21}^2\left[\mathcal{B}_{22}(m_Z^2;m_Z^2,m_2^2)+\mathcal{B}_{22}(m_Z^2;m_1^2,m_3^2) - m_Z^2 \mathcal{B}_{0}(m_Z^2;m_Z^2,m_2^2) \right]\nonumber\\
& &-(q_{21}^2+q_{31}^2)\left[\mathcal{B}_{22}(m_Z^2;m_Z^2,m_1^2)- m_Z^2\mathcal{B}_{0}(m_Z^2;m_Z^2,m_1^2)\right]\nonumber\\
& &\biggl.+q_{11}^2 \mathcal{B}_{22}(m_Z^2;m_2^2,m_3^2)- \mathcal{B}_{22}(m_Z^2;\mhpm^2,\mhpm^2)\biggr\},\eeqa
where the identity $\sum_{k=1}^{3}q_{k1}^2 = 1$ has been used.
In the decoupling limit, $q_{31}\sim q_{21} \sim \mathcal{O}\left(\frac{v^2}{\LL}\right)$, and $q_{11} \sim 1$; so $S$ becomes
\beqa \label{sdec}
S&=& \frac{1}{\pi m_Z^2} \biggl\{\mathcal{B}_{22}(m_Z^2;m_2^2,m_3^2)- \mathcal{B}_{22}(m_Z^2;\mhpm^2,\mhpm^2)+\mathcal{O}\left(\frac{v^4}{\Lambda^4}\right) \left[\mathcal{B}_{22}(m_Z^2;m_Z^2,m_3^2) \biggr.   \right.\nonumber\\
& &+\mathcal{B}_{22}(m_Z^2;m_1^2,m_2^2) - m_Z^2 \mathcal{B}_{0}(m_Z^2;m_Z^2,m_3^2)+\mathcal{B}_{22}(m_Z^2;m_Z^2,m_2^2) +\mathcal{B}_{22}(m_Z^2;m_1^2,m_3^2)  \nonumber\\
& &\biggl.\left.- m_Z^2 \mathcal{B}_{0}(m_Z^2;m_Z^2,m_2^2)+\mathcal{B}_{22}(m_Z^2;m_Z^2,m_1^2)- m_Z^2\mathcal{B}_{0}(m_Z^2;m_Z^2,m_1^2)\right]\biggr\}\,.\eeqa
The expression above can be evaluated with equations~(\ref{relations}) and the following rules, where the order parameter $y$ is defined by $y \equiv \frac{v^2}{\LL}$, and $a$ is order $1$:
\beqa
\mathcal{B}_{22}(m_Z^2;\LL + a v^2,\LL) &=&  -\frac{m_Z^2}{12} \left[\Delta - \ln\LL - \frac{a}{2}y + \frac{1}{10}\frac{m_Z^2}{v^2}y + \mathcal{O}(y^2)\right]\,,\nonumber \\
\mathcal{B}_{22}(m_Z^2;\LL + a v^2,\LL + a v^2) &=& -\frac{m_Z^2}{12} \left[\Delta - \ln\LL - a y + \frac{1}{10}\frac{m_Z^2}{v^2}y + \mathcal{O}(y^2)\right]\,, \nonumber \\
\mathcal{B}_{22}(m_Z^2;\LL + a v^2,v^2) &=&  -\frac{m_Z^2}{12} \left[\Delta +\frac{5}{6}- \ln\LL + \mathcal{O}(y)\right]\,, \nonumber \\
m_Z^2\mathcal{B}_{0}(m_Z^2;m_Z^2,\LL + a v^2) &=& \frac{v^2}{2}y + \mathcal{O}(y^2)\,.
\eeqa

Thus one finds that $S= \frac{1}{\pi m_Z^2} \left[ \frac{m_Z^2}{12}(\frac{m_3^2+m_2^2}{2 v^2}-\frac{\mhpm^2}{v^2})y  + m_Z^2 \mathcal{O}\left(y^2 \ln y\right)  \right]$.
Since $y^2 \ln y \ll y$ for $y \ll 1$, the result may be written
\beq  \label{decS}S \approx \frac{1}{24 \pi}\frac{m_3^2+m_2^2-2 \mhpm^2}{m_3^2}.\eeq
Hence, the limits on $S$ given in equations (\ref{STdec}) -- (\ref{STdecalt}) constrain the splitting between the neutral heavy Higgs massses and the charged Higgs mass to be small.
This is equivalent to constraining differences of the scalar couplings, as one can see from using \eq{relations} to rewrite \eq{decS}:
\beq  S \approx \frac{1}{24 \pi}\frac{Z_4 v^2}{m_3^2}. \eeq

One can evaluate $T$ in a similar manner, using the expansion
\beq F(\LL + a v^2, \LL+ b v^2) \approx \LL y^2 \frac{(a-b)^2}{6}.\label{expF} \eeq
The expression for $T$ in \eq{tsu} can be written 
\beqa T&=& \frac{1}{16 \pi s_W^2 m_W^2}\biggl\{|q_{22}|^2 F(\mhpm^2,m_2^2)+ |q_{32}|^2 F(\mhpm^2,m_3^2)- q_{11}^2 F(m_2^2,m_3^2)  \biggr. \nonumber \\
& &+(q_{31}^2+q_{21}^2)\left[F(\mhpm^2,m_1^2)-F(m_W^2,m_1^2)+F(m_Z^2,m_1^2)+4m_W^2 B_0(0;m_W^2,m_1^2)\right.\nonumber\\
& &\left.- 4 m_Z^2 B_0(0;m_Z^2,m_1^2)\right] + q_{31}^2 \left[F(m_W^2,m_3^2)-F(m_1^2,m_2^2)-F(m_Z^2,m_3^2)\right.\nonumber\\
& &\left.-4m_W^2 B_0(0;m_W^2,m_3^2)+ 4 m_Z^2 B_0(0;m_Z^2,m_3^2)\right]\nonumber\\
& & + q_{21}^2 \left[F(m_W^2,m_2^2)-F(m_1^2,m_3^2)-F(m_Z^2,m_2^2)\right.\nonumber\\
& &\biggl.\left.-4m_W^2 B_0(0;m_W^2,m_2^2)+ 4 m_Z^2 B_0(0;m_Z^2,m_2^2)\right] \biggr\}. 
\eeqa
Noting that $|q_{22}|^2 \sim |q_{32}|^2 \sim q_{11}^2 \sim 1$, $T$ can be expanded in the decoupling limit as
\beq
T = \frac{1}{16\pi s_W^2 m_W^2}\bigl[F(\mhpm^2,m_2^2)+  F(\mhpm^2,m_3^2)-F(m_2^2,m_3^2) + q_{31}^2\mathcal{O}(v^2\ln y) + q_{21}^2\mathcal{O}(v^2\ln y)\bigr]. \eeq
As in the previous calculation, the answer contains terms of $\mathcal{O}(y)$ and terms of $\mathcal{O}(y^2 \ln y)$.  Neglecting the latter, and using \eq{expF}, one obtains the following:
\beq T \label{decT} \approx \frac{(\mhpm^2 - m_3^2)(\mhpm^2 - m_2^2)}{48\pi s_W^2 m_W^2 m_3^2},\eeq
or 
\beq T \approx \frac{\left[Z_4  - \Re(\zfive)\right]\left[Z_4 + \Re(\zfive)\right]v^4}{192\pi s_W^2 m_W^2 m_3^2}.\eeq

As a check on this calculation, one can use eqs.~(\ref{decS}) and~(\ref{decT}) to calculate the effect of the MSSM Higgs sector on $S$ and $T$ in the decoupling limit.  In the supersymmetric limit CP is conserved, so $m_2$ becomes $m_{H^0}$ and $m_3$ becomes $m_{A^0}$.  The MSSM mass relations can be approximated in the decoupling limit by the following:
\beqa \mhpm^2 &=& m_{A^0}^2 + m_W^2, \nonumber\\
m_{H^0}^2 &=& m_{A^0}^2 + m_Z^2 \sin^22\beta + \mathcal{O}(v^2/m_{A^0}^2). \label{mssm}\eeqa 
Substituting \eq{mssm} into ~(\ref{decS}) and~(\ref{decT}) gives
\beqa S(MSSM-Higgs) & \approx &\frac{m_Z^2(\sin^22\beta - 2 c_W^2)}{24\pi m_{A^0}^2},\nonumber\\
 T(MSSM-Higgs) &\approx& \frac{m_Z^2(c_W^2 - \sin^22\beta)}{48\pi s_W^2  m_{A^0}^2},\eeqa
which agree with the results in \cite{habertasi}.  

\section{The Lack of CP-Violating Effects in Dimension 6 Operators}
One notes that despite allowing for the physical Higgs fields to be mixings of CP-eigenstates, no new phenomena related to CP-violation have emerged in this calculation.  In fact, although there exist CP-violating operators of dimension 6, the following discussion will show that they do not contribute to $S$ or $T$. 

The CP-violating dimension 6 operators, as listed in ref. ~\cite{buchmuller}, are:
\beqa \label{first}\mathcal{O}_{\Phi \tilde{G}}&=&(\Phi^\dagger \Phi)\tilde{G}^{A \mu\nu}G^{A }_{\mu\nu}, \\
\label{second}\mathcal{O}_{\Phi \tilde{W}}&=&(\Phi^\dagger \Phi)\tilde{W}^{a\mu\nu}W^{a}_{ \mu\nu}, \\
\label{third}\mathcal{O}_{\Phi \tilde{B}}&=&(\Phi^\dagger \Phi)\tilde{B}^{\mu\nu}B_{\mu\nu}, \\
\label{last}\mathcal{O}_{\tilde{W}B}&=&(\Phi^\dagger \sigma^a \Phi)\tilde{W}^{a\mu\nu}B_{\mu\nu}, \eeqa where $\tilde{V}^{\mu\nu}\equiv \frac{1}{2} \epsilon^{\mu\nu\rho\sigma}V_{\rho \sigma}$.

The operators in equations ~(\ref{first})~--~(\ref{third}) cannot contribute to $T$ [see \eq{obliqueTdef}]; there is no CP-violating equivalent of $\mathcal{O}_h$. However, one might ask whether the operator $\mathcal{O}_{\tilde{W}B}$, which is the CP-violating analogue of $\mathcal{O}_{WB}$, contributes to $S$.  Starting with eq.~(\ref{last}), one obtains the following:
\beqa \mathcal{O}_{\tilde{W}B} &=& \frac{1}{2} (\Phi^\dagger \sigma^a \Phi) \epsilon^{\mu\nu\rho\sigma}W^{a}_{\rho \sigma}B_{\mu \nu} \nonumber \\
&=& \frac{1}{2} (\Phi^\dagger \sigma^a \Phi) \epsilon^{\mu\nu\rho\sigma}(\delta_\rho W^{a}_{\sigma}- \delta_\sigma W^{a}_{\rho}+ g \epsilon_{a b c}W^b_\mu W^c_\nu )(\delta_\mu B_\nu-\delta_\nu B_\mu).\eeqa
After some manipulation of indices and replacing $\Phi$ with $\frac{1}{\sqrt{2}}\left(\begin{smallmatrix}0\\v\end{smallmatrix}\right)$, one finds that the contribution to $i \Pi^{\mu \nu}_{W^3 B} (q^2)$ would be $- i v^2 \epsilon^{\sigma \nu \mu \rho}q_\rho q_\sigma$.  As this has no part proportional to $g^{\mu \nu}q^2$, one concludes that the operator $\mathcal{O}_{\tilde{W}B}$  makes no contribution to $S$ [see equations ~(\ref{obliqueSdef}) and ~(\ref{pi})]. 
Other CP-violating operators (of higher dimension) are suggested in ~\cite{longhitano} and ~\cite{appelquist}.  However, in the decoupling scenario, operators above dimension 6 are suppressed by powers of $\frac{v^2}{\LL}$ ~\cite{grinstein}, and thus will not produce measurable effects on oblique corrections.

\chapter{Derivation of Basis-Independent Conditions for CP-invariance of Scalar Potential} \label{app:cptrans}
In this section we derive  the conditions for an explicitly CP-conserving scalar potential.  Let us start by applying a CP-transformation of the scalar potential in \eq{genericpot}, \ie~ $\,\mathcal{CP} ~\mathcal{V} ~\mathcal{CP}^{-1}$. Beginning with the quadratic ($Y_{a\bbar}$) term and using the definition in \eq{covcons}, we note that
\beq 
 \mathcal{CP}~ \Phi_\abar^\dagger\Phi_b ~\mathcal{CP}^{-1} =  (\widehat{v}_\abar^\ast \widehat{v}_\cbar^\ast \pm \thetdoubpl \widehat{w}_\abar^\ast \widehat{w}_\cbar^\ast)   (\widehat{v}_b \widehat{v}_d\pm\thetdoub \widehat{w}_b \widehat{w}_d)  \Phi_c^T  \Phi_\dbar^\ast \,. \eeq
Using $\Phi_c^T  \Phi_\dbar^\ast  =   \Phi_\dbar^\dagger \Phi_c$, the first term in \eq{genericpot} becomes
\beqa  \mathcal{CP}~\mathcal{V}_{quad} ~\mathcal{CP}^{-1} &=& Y_{a\bbar}(\widehat{v}_\abar^\ast \widehat{v}_\cbar^\ast  \widehat{v}_b \widehat{v}_d + \widehat{w}_\abar^\ast \widehat{w}_\cbar^\ast \widehat{w}_b \widehat{w}_d \pm\thetdoubpl \widehat{w}_\abar^\ast \widehat{w}_\cbar^\ast\widehat{v}_b \widehat{v}_d\nonumber\\
&&\quad\pm\thetdoub \widehat{v}_\abar^\ast \widehat{v}_\cbar^\ast \widehat{w}_b \widehat{w}_d)  \Phi_\dbar^\dagger \Phi_c \nonumber\\
&=& (Y_1 + Y_2 \pm \thetdoubpl Y_3^*  \widehat{v}_d \widehat{w}_\cbar^\ast \pm \thetdoub Y_3  \widehat{w}_d \widehat{v}_\cbar^\ast  \Phi_\dbar^\dagger \Phi_c\,.
\eeqa
Applying \eq{hbasis}, and using the orthogonality relation $\widehat{v}_a \widehat{w}_\abar^\ast=0$, this becomes
\beqa  \mathcal{CP}~\mathcal{V}_{quad} ~\mathcal{CP}^{-1}&=& Y_1 + Y_2 \pm \thetdoub Y_3 \widehat{v}_c \widehat{v}_\cbar^\ast \widehat{w}_d \widehat{w}_\dbar^\ast H_2^\dagger H_1  \pm \thetdoubpl Y_3^* \widehat{v}_c \widehat{v}_\cbar^\ast \widehat{w}_d \widehat{w}_\dbar^\ast H_1^\dagger H_2\nonumber\\
&=& Y_1 + Y_2 \pm (\thetdoub Y_3 H_2^\dagger H_1  + h.c.)\,.\eeqa
Now requiring the Lagrangian to be invariant under CP,  ie $ \mathcal{CP}~\mathcal{V}_{quad} ~\mathcal{CP}^{-1}= \mathcal{V}_{quad}$,  yields the following condition:
\beq Y_1 + Y_2 \pm (\thetdoub Y_3 H_2^\dagger H_1  + h.c. )= Y_1 + Y_2 + (Y_3 H_1^\dagger H_2  + h.c.)\,,\eeq
or
\beq  \thetminus Y_3 = \pm Y_3^* \thet\,.\eeq
The analogous calculation for the quartic term in \eq{genericpot} yields similar conditions for $\zsix$ and $Z_7 e^{-i \theta_{23}}$:
  \beq  \thetminus Z_{6,7} = \pm Z_{6,7}^* \thet\,.\label{zsixtran}\eeq
Meanwhile, for the $Z_5$ part of the potential, the relevant term after doing a CP-transformation appears as
\beq 
 \mathcal{CP}~\mathcal{V} ~\mathcal{CP}^{-1}~\ni \half \left[e^{- 4 i \theta_{23}}Z_5 (H_2^\dagger H_1)^2+h.c.\right]\,,\eeq
which leads to the condition \beq Z_5^* \thetdoubpl = Z_5 \thetdoub\,.\label{zfivetran}\eeq
One can now calculate, for example,
 \beqa  Z_5^*Z_6^2 &=& (Z_5 e^{-4 i \theta_{23}})(Z_6^{*2}e^{4 i \theta_{23}})= Z_5Z_6^{*2},\nonumber\\
Z_6Z_7^* &=& (\pm Z_6^* e^{2 i \theta_{23}})(\mp Z_7e^{-2 i \theta_{23}})= -Z_6^*Z_7,\nonumber\\
Z_5^*(Z_6^2+Z_7^2) &=& (Z_5 e^{-4 i \theta_{23}})(Z_6^{*2}e^{4 i \theta_{23}}+Z_7^{*2}e^{4 i \theta_{23}})= Z_5(Z_6^{*2}+Z_7^{*2}),\label{zcond}
\eeqa
using \eq{zsixtran} and \eq{zfivetran}. 
Thus, we replicate the CP conservation conditions of \eq{cpoddinv}.
\section{Calculation of $S$, $T$ and $U$}\label{app:stu}
The one-loop corrections to the gauge boson propagators contain $3$- and $4$-point interactions between gauge bosons and the Higgs bosons of the 2HDM, the form of which can be read off from eqs.~(\ref{VVH})~--~(\ref{VVHH}).  The resulting Feynman rules in t'Hooft-Feynman gauge are shown in Table~\ref{tabFeyn}. To simplify the Feynman rules, we have made use of \eqs{unitarity1}{epsid}.
The 2HDM contributions to $S$ are shown in Tables~\ref{tabS} and \ref{tabS}; Contributions to $T$ and $S+U$ are displayed in Tables~\ref{tabT1}, \ref{tabT2} and~\ref{tabU}, respectively. The reference Standard Model contributions, which are subtracted out from the 2HDM contributions, are shown in Table~\ref{tabSM}. The integrals are evaluated as in ref.~\cite{langacker}:
\beqa
\int \frac{d^4k}{(2\pi)^4} \frac{k^\mu k^\nu}{(k^2-m_1^2)((k+q)^2-m_2^2)} = \frac{i}{16\pi^2}g^{\mu\nu}B_{22}(q^2;m_1^2,m_2^2), \\
\int \frac{d^4k}{(2\pi)^4} \frac{1}{(k^2-m_1^2)((k+q)^2-m_2^2)} = \frac{i}{16\pi^2}B_0(q^2;m_1^2,m_2^2),\\
\int \frac{d^4k}{(2\pi)^4} \frac{1}{(k^2-m^2)} = \frac{i}{16\pi^2}A_0(m^2).
\eeqa
The contributions to $S$ from the diagrams in Table~\ref{tabS} and Table~\ref{tabSM} are compiled according to \eq{piA} and \eq{Sdef}, with the following result:
\beqa \nonumber
\frac{g^2}{16\pi c_W^2} S &\equiv& F_{ZZ}^{2H}(m_Z^2) - F_{\gamma\gamma}^{2H}(m_Z^2)-\frac{c_{2W}}{s_W c_W}F_{Z\gamma}^{2H}(m_Z^2) \nonumber\\
& & - F_{ZZ}^{SM}(m_Z^2) + F_{\gamma\gamma}^{SM}(m_Z^2)+\frac{c_{2W}}{s_W c_W}F_{Z\gamma}^{SM}(m_Z^2)\nonumber\\\nonumber
&=& \frac{g^2}{16 \pi^2 m_Z^2} \left[q_{k1}^2\mathcal{B}_{22}(m_Z^2;m_Z^2,m_k^2)-\mathcal{B}_{22}(m_Z^2;m_Z^2,m_\phi^2) \right.\nonumber\\
& &- m_Z^2 q_{k1}^2\mathcal{B}_{0}(m_Z^2;m_Z^2,m_k^2)+ m_Z^2\mathcal{B}_{0}(m_Z^2;m_Z^2,m_\phi^2)\nonumber\\
& & +q_{11}^2 \mathcal{B}_{22}(m_Z^2;m_2^2,m_3^2)+q_{21}^2\mathcal{B}_{22}(m_Z^2;m_1^2,m_3^2)\nonumber\\
& &\left.+ q_{31}^2\mathcal{B}_{22}(m_Z^2;m_1^2,m_2^2) - \mathcal{B}_{22}(m_Z^2;\mhpm^2,\mhpm^2)\right].
\eeqa
The parameter $T$ can be calculated in a similar manner, using the following relations provided in ref.~\cite{habertasi}:
\beq
\label{BtoF}
4 B_{22}(0;m_1^2,m_2^2) = F(m_1^2,m_2^2)+A_0(m_1^2)+A_0(m_2^2)\,,
\eeq
\beq
B_{0}(0;m_1^2,m_2^2) = \frac{A_0(m_1^2)-A_0(m_2^2)}{m_1^2-m_2^2}\,,
\eeq
with $F(m_1^2,m_2^2) \equiv \half (m_1^2+m_2^2)-\frac{m_1^2m_2^2}{m_1^2-m_2^2}\ln\left(\frac{m_1^2}{m_2^2}\right)$ as before.   
Adding the contributions to $T$ from all the diagrams shown in Tables~\ref{tabT1}, \ref{tabT2} and~\ref{tabSM} yields
\beqa 
\alpha T&\equiv&\frac{A^{2H}_{WW}(0)}{m_W^2}- \frac{A^{2H}_{ZZ}(0)}{m_Z^2}-\left[\frac{ A^{SM}_{WW}(0)}{m_W^2}- \frac{A^{SM}_{ZZ}(0)}{m_Z^2}\right]\nonumber\\
&=&\frac{g^2}{16\pi^2m_W^2}\biggl\{|q_{k2}|^2 B_{22}(0;\mhpm^2,m_k^2)-q_{21}^2 B_{22}(0;m_1^2,m_3^2)-q_{11}^2 B_{22}(0;m_2^2,m_3^2)\biggr.\nonumber\\
& &\left.-q_{31}^2 B_{22}(0;m_1^2,m_2^2)
-\frac{1}{2}A_0(\mhpm^2) +q_{k1}^2[B_{22}(0;m_W^2,m_k^2)-B_{22}(0;m_Z^2,m_k^2)]\right.\nonumber\\
& &-B_{22}(0;m_W^2,m_\phi^2)-q_{k1}^2[m_W^2 B_0(0;m_W^2,m_k^2)-m_Z^2 B_0(0;m_Z^2,m_k^2)]\nonumber\\
& & \biggl.+B_{22}(0;m_Z^2,m_\phi^2)+m_W^2 B_0(0;m_W^2,m_\phi^2)-m_Z^2 B_0(0;m_Z^2,m_\phi^2)\biggr\}.
\eeqa
This can be simplified using \eq{BtoF} and $\alpha =\frac{g^2 s_W^2}{4\pi}$:
\beqa \label{generalformT}
\nonumber T=& &\frac{1}{16\pi m_W^2s_W^2} \bigl\{|q_{k2}|^2 F(\mhpm^2,m_k^2)-q_{21}^2 F(m_1^2,m_3^2)-q_{11}^2 F(m_2^2,m_3^2)-q_{31}^2 F(m_1^2,m_2^2)\bigr.\\\nonumber& &
-F(m_W^2,m_\phi^2)+q_{k1}^2[F(m_W^2,m_k^2)-F(m_Z^2,m_k^2)]+F(m_Z^2,m_\phi^2)\\\nonumber
& & -4 q_{k1}^2[m_W^2 B_0(0;m_W^2,m_k^2)-m_Z^2 B_0(0;m_Z^2,m_k^2)]\\
& &\bigl.
+4m_W^2 B_0(0;m_W^2,m_\phi^2)-4m_Z^2 B_0(0;m_Z^2,m_\phi^2)\bigr\}.
\eeqa
Lastly, adding all of the contributions to $S+U$ in Tables~\ref{tabU} and~\ref{tabSM} gives the following:
\beqa
\frac{g^2}{16\pi} (S+U) &=& F_{WW}(m_W^2)- F_{\gamma\gamma}(m_W^2)-\frac{c_W}{s_W}F_{Z\gamma}(m_W^2)\nonumber\\
&=&\frac{g^2}{16 \pi^2 m_W^2} \biggl[- q_{k1}^2 m_W^2 \mathcal{B}_{0}(m_W^2;m_W^2,m_k^2)+ m_W^2 \mathcal{B}_{0}(m_W^2;m_W^2,m_\phi^2)\biggr. \nonumber\\
& &- \mathcal{B}_{22}(m_W^2;m_W^2,m_\phi^2)+ q_{k1}^2\mathcal{B}_{22}(m_W^2;m_W^2,m_k^2)\nonumber\\
& &\biggl. +|q_{k2}|^2 \mathcal{B}_{22}(m_W^2;\mhpm^2,m_k^2)-2 \mathcal{B}_{22}(m_W^2;\mhpm^2,\mhpm^2)\biggr], 
\eeqa
or
\beqa S+U &=&\frac{1}{\pi m_W^2} \biggl[- q_{k1}^2 m_W^2 \mathcal{B}_{0}(m_W^2;m_W^2,m_k^2)+ m_W^2 \mathcal{B}_{0}(m_W^2;m_W^2,m_\phi^2)\biggr. \nonumber\\
& &- \mathcal{B}_{22}(m_W^2;m_W^2,m_\phi^2)+ q_{k1}^2\mathcal{B}_{22}(m_W^2;m_W^2,m_k^2)\nonumber\\
& &\biggl. +|q_{k2}|^2 \mathcal{B}_{22}(m_W^2;\mhpm^2,m_k^2)-2 \mathcal{B}_{22}(m_W^2;\mhpm^2,\mhpm^2)\biggr]. 
\eeqa

\begin{table}[ht!]
\centering
\caption{Feynman rules used in the calculation of the oblique parameters.}
\label{tabFeyn}
\begin{tabular}{|c|c|}\hline
\feynrulefour{W_+}{h_i}{h_i}{\frac{ig^2}{2}\Re(|q_{i1}|^2+|q_{i2}|^2)}{= \frac{ig^2}{2} g^{\mu\nu} }& \feynrulefour{Z}{h_i}{h_i}{\frac{ig^2}{2c_W^2}\Re(|q_{i1}|^2+|q_{i2}|^2)}{= \frac{ig^2}{2c_W^2} g^{\mu\nu} }\\
\feynrulefour{W_+}{H^+}{H^+}{\frac{ig^2}{2}}{}& \feynrulefour{Z}{H^+}{H^+}{\frac{ig^2}{2c_W^2}c_{2W}^2}{}\\ 
\feynruleVV{W_+}{W_+}{h_i}{i q_{i1}m_W g }  & \feynruleVV{Z}{Z}{h_i}{i q_{i1}m_Z g } \\
\feynrule{W_+}{H^-}{h_i}{-i q_{i2} \frac{g}{2}}&\feynruletwo{Z}{h_1,G^0}{h_3,h_2}{\Im(q_{32}q_{12}^*)\frac{g}{2 c_W}}{q_{21}\frac{g}{2 c_W}}\\
\feynrule{W_+}{G^-}{h_i}{-i q_{i1} \frac{g}{2}}&\feynruletwo{Z}{h_2,h_1}{h_3,G^0}{\Im(q_{22}q_{32}^*)\frac{g}{2 c_W}}{-q_{11}\frac{g}{2 c_W}}\\
\feynrule{W_+}{G^-}{\phi}{-i \frac{g}{2}}&\feynruletwo{Z}{h_1,G^0}{h_2,h_3}{\Im(q_{12}q_{22}^*)\frac{g}{2 c_W}}{q_{31}\frac{g}{2 c_W}}\\
\feynrule{\gamma}{H^+,G^+}{H^+,G^+}{i g s_W}&\feynrule{Z}{H^+}{H^+}{(c_W^2-s_W^2)\frac{ig}{2 c_W}}\\
\feynruleVV{W_+}{W_+}{\phi}{i m_W g}& \feynruleVV{Z}{Z}{\phi}{i m_Z g } \\
& \feynrule{Z}{\phi}{G^0}{- \frac{g}{2c_W}}\\\hline
\end{tabular}
\end{table}

\begin{table}[ht!]
\centering
\caption{Diagrams representing the 2HDM contributions to $S$, part 1.}
\label{tabS}
\begin{tabular}{|c|}\hline
Contributions to $\Pi_{ZZ}^{2H}(m_Z^2)$ \T \Bot\\ \hline
\Gaugeprop{Z}{Z}{Z}{\hi}{=-\frac{g^2M_Z^2}{16 \pi^2 c_W^2}q_{i1}^2B_0(m_Z^2;m_Z^2,m_i^2)}\\
\Loopgraph{Z}{Z}{G^0}{\hi}{=\gc q_{i1}^2B_{22}(m_Z^2;m_Z^2,m_i^2)}{}\\
\Loopgraph{Z}{Z}{h_3}{h_1}{=\frac{g^2}{16 \pi^2 c_W^2}q_{21}^2B_{22}(m_Z^2;m_1^2,m_3^2)}{}\\
\Loopgraph{Z}{Z}{h_3}{h_2}{=\frac{g^2}{16 \pi^2 c_W^2}q_{11}^2B_{22}(m_Z^2;m_2^2,m_3^2)}{}\\
\Loopgraph{Z}{Z}{h_1}{h_2}{=\frac{g^2}{16 \pi^2 c_W^2}q_{31}^2B_{22}(m_Z^2;m_1^2,m_2^2)}{}\\
\Loopgraph{Z}{Z}{H^+}{H^+}{=\gc c_{2W}^2 B_{22}(m_Z^2;\mhpm^2,\mhpm^2)}{}\\ \hline
\end{tabular}
\end{table}

\begin{table}[ht!]
\centering
\caption{Diagrams representing the 2HDM contributions to $S$, part 2.}
\label{tabS2}
\begin{tabular}{|c|}\hline
Contributions to $\Pi_{\gamma\gamma}^{2H}(m_Z^2)$ and $\Pi_{Z\gamma}^{2H}(m_Z^2)$\T \Bot\\ \hline
 \Loopgraph{\gamma}{\gamma}{H^+}{H^+}{=4\gtwo s_W^2 B_{22}(m_Z^2;\mhpm^2,\mhpm^2)}{}\\ 
\Loopgraph{Z}{\gamma}{H^+}{H^+}{= \frac{2 g^2}{16\pi^2c_W} s_W  c_{2W} B_{22}(m_Z^2;\mhpm^2,\mhpm^2)}{} \\ \hline
\end{tabular}
\end{table}

\begin{table}[ht!]
\addtolength{\tabcolsep}{5pt}
\centering
\caption{Diagrams representing the 2HDM contributions to $T$, part 1.}
\label{tabT1}
\begin{tabular}{|c|}\hline
\qquad \quad \qquad \quad Contributions to $A_{WW}^{2H}(0)$ \T \Bot \qquad \quad \qquad \quad\\ \hline
\gaugeprop{W^+}{W^+}{W^+}{\hi}{=-\frac{g^2m_W^2}{16 \pi^2}q_{i1}^2B_0(0;m_W^2,m_i^2)}\\
\loopgraph{W^+}{W^+}{G^+}{\hi}{=\gtwo q_{i1}^2 B_{22}(0;m_W^2,m_i^2)}{}\\
\loopgraph{W^+}{W^+}{h_1}{H^+}{=\gtwo |q_{12}|^2B_{22}(0;\mhpm^2,m_1^2)}{}\\
\loopgraph{W^+}{W^+}{h_2}{H^+}{=\gtwo |q_{22}|^2B_{22}(0;\mhpm^2,m_2^2)}{}\\
\loopgraph{W^+}{W^+}{h_3}{H^+}{=\gtwo |q_{32}|^2B_{22}(0;\mhpm^2,m_3^2)}{}\\
\fourpoint{W^+}{W^+}{\hi}{=-\frac{1}{2}\gtwo A_0(m_i^2)}\\
\fourpoint{W^+}{W^+}{H^+}{=-\frac{1}{2}\gtwo A_0(\mhpm^2)}\\ \hline
\end{tabular}
\end{table}

\begin{table}[ht!]
\addtolength{\tabcolsep}{20pt}
\centering
\caption{Diagrams representing the 2HDM contributions to $T$, part 2.}
\label{tabT2}
\begin{tabular}{|c|}\hline
\qquad \quad \qquad \quad Contributions to $A_{ZZ}^{2H}(0)$\qquad \quad \qquad \quad\\ \hline
\gaugeprop{Z}{Z}{Z}{\hi}{=-\frac{g^2m_Z^2}{16 \pi^2 c_W^2}q_{i1}^2B_0(0;m_Z^2,m_i^2)}\\
\loopgraph{Z}{Z}{G^0}{\hi}{=\gc q_{i1}^2B_{22}(0;m_Z^2,m_i^2)}{}\\
 \loopgraph{Z}{Z}{h_3}{h_1}{=\frac{g^2}{16 \pi^2 c_W^2}q_{21}^2B_{22}(0;m_1^2,m_3^2)}{}\\
\loopgraph{Z}{Z}{h_3}{h_2}{=\frac{g^2}{16 \pi^2 c_W^2}q_{11}^2B_{22}(0;m_2^2,m_3^2)}{}\\
\loopgraph{Z}{Z}{h_1}{h_2}{=\frac{g^2}{16 \pi^2 c_W^2}q_{31}^2B_{22}(0;m_1^2,m_2^2)}{}\\
\fourpoint{Z}{Z}{\hi}{=-\frac{1}{2}\gc A_0(m_i^2)}\\
\fourpoint{Z}{Z}{H^+}{=-\frac{1}{2}\gc c_{2W}^2 A_0(\mhpm^2)}\\
\loopgraph{Z}{Z}{H^+}{H^+}{=\gc c_{2W}^2 B_{22}(0;\mhpm^2,\mhpm^2)}{=\frac{1}{2}\gc c_{2W}^2 A_0(\mhpm^2)}\\ \hline
\end{tabular}
\end{table}

\begin{table}[ht!]
\addtolength{\tabcolsep}{20pt}
\centering
\caption{Diagrams representing the 2HDM contributions to $S+U$.}
\label{tabU}
\begin{tabular}{|c|}\hline
Contributions to $\Pi_{WW}^{2H}(m_W^2)$ \T \Bot\\ \hline
\gaugeprop{W^+}{W^+}{W^+}{\hi}{=-\frac{g^2m_W^2}{16 \pi^2}q_{i1}^2B_0(m_W^2;m_W^2,m_i^2)}\\
\loopgraph{W^+}{W^+}{G^+}{\hi}{=\gtwo q_{i1}^2 B_{22}(m_W^2;m_W^2,m_i^2)}{} \\
\loopgraph{W^+}{W^+}{H^+}{\hi}{=\gtwo |q_{i2}|^2B_{22}(m_W^2;\mhpm^2,m_i^2)}{}\\ \hline
\qquad \quad Contributions to $\Pi_{\gamma\gamma}^{2H}(m_W^2)$ and $\Pi_{Z\gamma}^{2H}(m_W^2)$ \qquad\quad\\ \hline
\loopgraph{\gamma}{\gamma}{H^+}{H^+}{=4\gtwo s_W^2 B_{22}(m_W^2;\mhpm^2,\mhpm^2)}{}\\
\loopgraph{Z}{\gamma}{H^+}{H^+}{= 2\frac{g^2s_W c_{2W} }{16\pi^2c_W} B_{22}(m_W^2;\mhpm^2,\mhpm^2)}{}\\ \hline
\end{tabular}
\end{table}

\begin{table}[ht!]
\addtolength{\tabcolsep}{5pt}
\small
\centering
\caption{Standard Model contributions to the oblique parameters.}
\label{tabSM}
\begin{tabular}{|c|}\hline
 Contributions to $\Pi_{WW}^{SM}(m_W^2)$ and $\Pi_{ZZ}^{SM}(m_Z^2)$\T \Bot \\ \hline
\gaugeprop{W^+}{W^+}{W^+}{\phi}{=-\frac{g^2m_W^2}{16\pi^2}B_0(m_W^2;m_W^2,m_1^2)}\\
\loopgraph{W^+}{W^+}{G^+}{\phi}{=\gtwo B_{22}(m_W^2;m_W^2,m_1^2)}{}\\ 
\gaugeprop{Z}{Z}{Z}{\phi}{=-\frac{g^2m_Z^2}{16 \pi^2 c_W^2}B_0(m_Z^2;m_Z^2,m_1^2)}\\
\loopgraph{Z}{Z}{G^0}{\phi}{= \gc B_{22}(m_Z^2;m_Z^2,m_1^2)}{}\\ \hline
Contributions to $A_{WW}^{SM}(0)$ and $A_{ZZ}^{SM}(0)$\T \Bot\\ \hline
\gaugeprop{W^+}{W^+}{W^+}{\phi}{=-\frac{g^2m_W^2}{16\pi^2}B_0(0;m_W^2,m_1^2)}\\
\loopgraph{W^+}{W^+}{G^+}{\phi}{=\gtwo B_{22}(0;m_W^2,m_1^2)}{}\\
\gaugeprop{Z}{Z}{Z}{\phi}{=-\frac{g^2m_Z^2}{16 \pi^2 c_W^2}B_0(0;m_Z^2,m_1^2)}\\
\loopgraph{Z}{Z}{G^0}{\phi}{= \gc B_{22}(0;m_Z^2,m_1^2)}{}\\ \hline
\end{tabular}
\end{table}
\clearpage

\chapter{Derivation of Tree-Level Unitarity Limits}\label{app:deriv}
In section \ref{unit}, constraints from perturbative unitarity in the Standard Model were reviewed.  In particular,  an upper bound for the SM Higgs boson mass $m_h$ were derived from the scattering of $W^+ W^- \rightarrow W^+ W^-$. 
One can use a similar argument to put upper bounds on the magnitudes of the $Z_i$ parameters in the CP-violating 2HDM. The implications of unitarity for the 2HDM has been studied in the context of scattering of gauge bosons and the physical scalars  ~\cite{akeroyd,Ginzburg:2003fe,Huffel:1980sk,Kanemura:1993hm,Kanemura:2004mg,Weldon:1984wt}.  By putting an upper limit $\xi$ on the amplitude for a process $\varphi_A \varphi_B \rightarrow \varphi_C \varphi_D$, one can quantify the constraints from perturbative unitarity as follows:
\beq \left| \frac{g_{ABCD}}{16 \pi} \right| < \xi.\eeq
Here the parameter $\xi$ will be taken to be $\half$, as in section \ref{unit}.  We will only consider tree-level scattering here, so only quartic couplings will be involved, namely
$W^+ W^-W^+W^-$, $W^+W^-H^+H^-$, $(H^+e^{i\theta_{23}})(H^+e^{i\theta_{23}})W^-W^-+\rm{h.c.}$,~$Z^0Z^0Z^0h_m$,\\ $G^0 h_mG^-(H^+e^{i\theta_{23}}) +\rm{h.c.}$, $Z^0Z^0H^+H^-$, and $ Z^0 Z^0W^-(H^+e^{i\theta_{23}}).$

  The equivalence theorem (see section \ref{scsec}) allows one to equate a high energy scattering amplitude involving gauge bosons to the analogous amplitude involving Goldstone bosons\footnote{up to a sign, which will not be important here}, by making the replacements $W^{\pm}\rightarrow G^\pm$, $Z^0 \rightarrow G^0$.  Thus, one can translate limits on the gauge boson/Higgs couplings into limits on the Goldstone/Higgs couplings. The resulting constraints on $Z_1$, $Z_3$, $Z_3 + Z_4$, $\Re(\zfive)$, and $\Re(\zsix)$ can be read off directly from $\mathcal{V}_4$ [\eq{scalpot}], as shown in Table~\ref{unitarity}.

\begin{table}[ht!]
\centering
\caption{Calculation of tree-level unitarity limits on the CP-conserving quartic couplings. Combinatoric factors are included to take into account identical particles. }
\label{unitarity}
\begin{tabular}{|c|c|c|}\hline
Relevant Term in Scalar Potential& Amplitude \T \Bot&Unitarity bound\\ \hline
$\half Z_1 G^+ G^- G^+ G^- $ \T \Bot &$ \frac{1}{16\pi}(\half Z_1)\cdot 4 $&$ |Z_1|<4\pi $\\
$\half Z_3 G^0 G^0 H^+H^-$\T \Bot& $\frac{1}{16\pi}(\half Z_3)\cdot 2 $&$ |Z_3|<8\pi $\\
$(Z_3+Z_4)G^+ G^- H^+ H^- $\T \Bot& $\frac{1}{16\pi}(Z_3+Z_4) $&$ |Z_3+Z_4|<8\pi $\\
$\half \zfive H^+ H^+ e^{2i\theta_{23}} G^-G^-+{\rm h.c.}$\T \Bot& $\frac{1}{16\pi}\Re(\zfive)\cdot 4 $& $|\Re(\zfive)|<2\pi $\\
$\zsix G^0 G^0 G^- (H^+ e^{i\theta_{23}} ) +{\rm h.c.}$\T \Bot&$ \frac{1}{16\pi}\Re(\zsix)\cdot 4 $&$ |\Re(\zsix)|<2\pi$ \\ \hline
\end{tabular}
\end{table}
The CP-violating parameters $\Im(\zfive)$ and $\Im(\zsix)$ appear in a more convoluted form in the quartic scalar potential.  From the interaction
\beq\half \Im(q_{m2}Z_6\,e^{-i\theta_{23}})\,G^0 G^0 G^0 h_m\,,\eeq and Table~\ref{tab1}, one can write Feynman rules for $m = 1,2$:
\beqa
g_{G^0G^0G^0h_1}&=&\half(-s_{12} \Im[\zsix] - c_{12} s_{13} \Re[\zsix])\cdot 3!\,,\nonumber\\
g_{G^0G^0G^0h_2}&=&\half(c_{12} \Im[\zsix] - s_{12} s_{13} \Re[\zsix])\cdot 3!\,.\label{g}
\eeqa
Unitarity requires $|g_{G^0G^0G^0h_m}|< 8\pi$.  It is convenient to combine the two limits in quadrature to isolate $\Im(\zsix)$:
\beqa |g_{G^0G^0G^0h_1}|^2 + |g_{G^0G^0G^0h_2}|^2&<& 64\pi^2\,,\nonumber\\
\left[\Im[\zsix]\right]^2 + s_{13}^2 \left[\Re[\zsix]\right]^2&<& \frac{64\pi^2}{9}\,.\eeqa
Since $s_{13}^2 \left[\Re(\zsix)\right]^2$ is real and non-negative, it must be true that $|\Im(\zsix)|< \frac{8\pi}{3}$.  Similarly, one can use the term 
\beq \half i\,G^0 h_m \biggl\{G^-H^+e^{i\theta_{23}}\left[q^*_{m2}Z_4-q_{m2}Z_5 e^{-2i\theta_{23}}\right]+{\rm h.c.}\biggr\}\,,\eeq with $m=1,2$ to derive the following:
\beqa
g_{G^0G^-(H^+ e^{i\theta_{23}})h_1}&=&-c_{12}s_{13} Z_4 - s_{12}\Im(\zfive) - c_{12} s_{13} \Re(\zfive),\nonumber\\
g_{G^0G^-(H^+ e^{i\theta_{23}})h_2}&=&-s_{12}s_{13} Z_4 + c_{12}\Im(\zfive) - s_{12} s_{13} \Re(\zfive).\label{gg}
\eeqa
Adding in quadrature and applying the unitarity bound gives, after some simplification:
\beq s_{13}^2 \left[ Z_4 + \Re(\zfive)\right]^2 + \left[\Im(\zfive)\right]^2 < 64 \pi^2.\eeq
One can conclude that $|\Im(\zfive)|< 8\pi$.  

\nocite{*}
\bibliographystyle{plain}
\bibliography{diss}
\end{document}